\documentclass[structabstract]{aa}  
\usepackage{color}
\definecolor{colurl}{rgb}{0.,0.5,0.}
\definecolor{colcite}{rgb}{0.5,0.,0.}
\definecolor{collink}{rgb}{0.,0.,0.5}
\usepackage[colorlinks=true,
            breaklinks=true,
            pdfstartview=FitV, 
            linkcolor=collink, 
            citecolor=colcite, 
            urlcolor=colurl, 
            backref=true]{hyperref}
\usepackage{graphicx}
\usepackage{natbib}
\usepackage{txfonts}
\usepackage{tabularx}
\usepackage{float}
\DeclareGraphicsExtensions{.pdf}
\newcommand{\eg}{{\it e.g.}}
\newcommand{\ie}{{\it i.e.}}
\newcounter{textlistctr}
\newcommand{\thetextlist}{, }
\newcommand{\textlist}[1]
            {\setcounter{textlistctr}{1}
             \renewcommand{\thetextlist}
             {{\it (\roman{textlistctr})}\stepcounter{textlistctr}}#1
              }
\newcommand{\sms}[1]{{\mbox{{\scriptsize #1}}}}
\newcommand{\seppar}{\vspace*{10pt}}
\newcommand{\citengl}[1]{``{\it #1}''}
\newcommand{\refsec}[1]{Sect.~\ref{#1}}
\newcommand{\E}[1]{\times10^{#1}}
\newcommand{\refeqp}[1]{(Eq.~\ref{#1})}
\newcommand{\reftab}[1]{Table~\ref{#1}}
\newcommand{\refsecs}[1]{Sects.~\ref{#1}}
\newcommand{\ddiff}{{\;\rm d}}
\newcommand{\refapp}[1]{App.~\ref{#1}}
\newcommand{\reffig}[1]{Fig.~\ref{#1}}
\newcommand{\dd}{{\rm d}}
\newcommand{\refeqnp}[1]{Eq.~\ref{#1}}
\newcommand{\refeqs}[1]{Eqs.~(\ref{#1})}
\newcommand{\refeq}[1]{Eq.~(\ref{#1})}
\newcommand{\hersc}{{\it Herschel}}
\newcommand{\spitz}{{\it Spitzer}}
\newcommand{\SPIREiii}{SPIRE$_\sms{500$\mu m$}$}
\newcommand{\planck}{{\it Planck}}
\newcommand{\MIPSiii}{MIPS$_\sms{160$\mu m$}$}
\newcommand{\IRACiv}{IRAC$_\sms{8$\mu m$}$}
\newcommand{\IRACi}{IRAC$_\sms{3.6$\mu m$}$}
\newcommand{\IRACii}{IRAC$_\sms{4.5$\mu m$}$}
\newcommand{\IRACiii}{IRAC$_\sms{5.8$\mu m$}$}
\newcommand{\MIPSi}{MIPS$_\sms{24$\mu m$}$}
\newcommand{\MIPSii}{MIPS$_\sms{70$\mu m$}$}
\newcommand{\SPIREi}{SPIRE$_\sms{250$\mu m$}$}
\newcommand{\SPIREii}{SPIRE$_\sms{350$\mu m$}$}
\newcommand{\ngc}[1]{NGC$\;$#1}
\newcommand{\mic}{\;\mu {\rm m}}
\newcommand{\mmic}{$\mu {\rm m}$}
\newcommand{\hi}{H$\,${\sc i}}
\newcommand{\hii}{H$\,${\sc ii}}
\newcommand{\hmol}{H$_2$}
\newcommand{\hiline}{[\hi]$_{21\,\rm cm}$}
\newcommand{\COio}{$^{12}$CO(J$=$1$\rightarrow$0)$_{2.6\rm mm}$}
\newcommand{\msun}{\;M_\odot}
\newcommand{\cii}{C$\,${\sc ii}}
\newcommand{\ciiline}{[\cii]$_{158\mu\rm m}$}
\newcommand{\COvtoiv}{$^{12}$CO(J$=$5$\rightarrow$4)$_{520\mu\rm m}$}
\newcommand{\modif}[1]{#1}
\begin{document}

  \title{Non-Standard Grain Properties, Dark Gas Reservoir, and
         Extended Submillimeter Excess, Probed by \hersc\ in the 
         Large Magellanic Cloud}
  \titlerunning{Variations on the Dust Mass Estimate in the LMC}

  \author{Fr\'ed\'eric {\sc Galliano}\inst{1}
          \and
          Sacha {\sc Hony}\inst{1}
          \and
          Jean-Philippe {\sc Bernard}\inst{2}
          \and
          Caroline {\sc Bot}\inst{3}
          \and
          Suzanne C. {\sc Madden}\inst{1}
          \and
          Julia {\sc Roman-Duval}\inst{4}
          \and
          Maud {\sc Galametz}\inst{5}
          \and
          Aigen {\sc Li}\inst{6}
          \and
          Margaret {\sc Meixner}\inst{4}
          \and
          Charles W. {\sc Engelbracht}\inst{7}
          \and 
          Vianney {\sc Lebouteiller}\inst{1}
          \and
          Karl {\sc Misselt}\inst{7}
          \and 
          Edward {\sc Montiel}\inst{7}
          \and
          Pasquale {\sc Panuzzo}\inst{1}
          \and
          William T. {\sc Reach}\inst{8}
          \and
          Ramin {\sc Skibba}\inst{7}
          }

   \institute{
              AIM, CEA/Saclay, L'Orme des Merisiers, 
              91191 Gif-sur-Yvette, France\\
              \email{frederic.galliano@cea.fr}
              \and
              Centre d'\'Etude Spatiale des Rayonnements, CNRS, 9 Av.\ du 
              Colonel Roche, BP 4346, 31028 Toulouse, France
              \and
              Observatoire Astronomique de Strasbourg, 11 rue de lÕuniversit\'e, 
              67000 Strasbourg, France
              \and
              Space Telescope Science Institute, 3700 San Martin Drive, 
              Baltimore, MD 21218, USA
              \and
              Institute of Astronomy, University of Cambridge, Madingley Road, 
              Cambridge CB3 0HA, UK
              \and
              314 Physics Building, Department of Physics and Astronomy, 
              University of Missouri, Columbia, MO 65211, USA
              \and
              Steward Observatory, University of Arizona, 
              933 North Cherry Ave., Tucson, AZ 85721, USA
              \and
              Spitzer Science Center, California Institute of Technology, 
              MS 220-6, Pasadena, CA 91125, USA
             }

  \date{Received 25 August 2011 / Accepted 6 October 2011}

  \abstract
  {\hersc\ provides crucial constraints on the 
   IR \modif{SEDs} of galaxies, allowing unprecedented 
   accuracy on the dust mass estimates.
   However, these estimates rely on non-linear models and poorly-known optical 
   properties.
   }
  {In this paper, we \modif{perform detailed} modelling of the \spitz\ and 
   \hersc\ observations of the LMC, in order 
   to:\textlist{\thetextlist~systematically study the uncertainties and biases
   affecting dust mass estimates; and to
   \thetextlist~explore the peculiar ISM properties of the LMC.}
   }
  {To achieve these goals, we have modelled the spatially resolved SEDs with two 
   alternate grain compositions, to study the impact of different submillimetre 
   opacities on the dust mass.
   \modif{We have} rigorously propagated the observational errors (noise and 
   calibration) through the entire fitting process, in order to derive 
   consistent parameter uncertainties.}
  {First, we show that using the integrated SED leads to underestimating the 
   dust mass by $\simeq50\,\%$ compared to the value obtained with 
   sufficient spatial resolution, for the region we studied.
   This might be the case, in general, for unresolved galaxies.
   Second, we show that Milky Way type grains produce higher gas-to-dust mass 
   ratios than what seems possible according to the element abundances in the 
   LMC.
   A spatial analysis shows that this dilemma is the result of an exceptional 
   property: the grains of the LMC have on average a larger intrinsic 
   submm opacity (emissivity index $\beta\simeq1.7$ and opacity 
   $\kappa_\sms{abs}(160\mic)=1.6\;\rm m^2\,kg^{-1}$) than those of the Galaxy. 
   By studying the spatial distribution of the gas-to-dust mass ratio, we are
   able to constrain the fraction of unseen gas mass between $\simeq10$, and 
   $\simeq100\,\%$ and show that it is not sufficient to explain the gas-to-dust 
   mass ratio obtained with Milky Way type grains.
   Finally, we confirm the detection of a 500~\mmic\ extended emission excess 
   with an average relative amplitude of $\simeq15\,\%$, varying up to $40\,\%$.
   This excess anticorrelates well with the dust mass surface density.
   Although we do not know the origin of this excess, we show that it is
   unlikely the result of very cold dust, or CMB fluctuations.}
  {} 


   \maketitle



\section{Introduction}

The infrared (IR) spectral energy distribution (SED) is widely used to derive the global properties of a system, \modif{such as} its instantaneous star formation rate, its dust and eventually gas masses, and the compactness of the star forming region.
The advent of the {\it Herschel Space Observatory} has opened the most important
spectral window to perfect these diagnostics, by observing the far-IR to submillimeter (submm) wavelengths (60-600~\mmic).
Indeed, this regime samples the peak and Rayleigh-Jeans wing of the dust emission.
It consequently constrains the emission by grains in thermal equilibrium with the radiation field, present in the different phases of the interstellar medium (ISM), including the coldest, most massive components (down to dust temperatures of $T_\sms{dust}\gtrsim12$~K).
This spectral domain is therefore crucial to derive accurate dust masses, and physical conditions, and can be used as a powerful, unprecedented tool to probe interstellar matter in regions where no other counterpart is accessible.

Unfortunately, there are several fundamental systematic unknowns inherent to dust modelling, which are questioning the reliability of these diagnostics.
First, the microscopic properties of the grains are still poorly known. 
In the Milky Way, the most complete and accurate models are constrained by observations of high latitude cirrus clouds: their IR emission, ultraviolet (UV)-to-near-IR extinction, and elemental depletions \citep{zubko04,draine07,compiegne11}.
The authors performing these models derive the size distribution and abundance of the different grain species --~silicates, carbon grains (graphite or amorphous carbons) and polycyclic aromatic hydrocarbons (PAH). 
\citet{zubko04} demonstrated an important degeneracy by presenting complete fits of the same data set (Galactic emission, extinction and depletion), with five different dust compositions, alternating bare and coated grains, as well as crystalline and amorphous solids.
Thus, the derived dust properties depend on the assumed chemical composition of each species.
The UV to millimetre (mm) opacities are sensitive to the grain
composition.
They are derived from sparse constraints including astrophysical features, laboratory spectra of \modif{analogs of interstellar dust materials}, and theoretical solid state physics \citep[{\eg}][]{weingartner01,draine03}.
Their universality is doubtful.

The second major source of uncertainties concerns the macroscopic variations of 
these microscopic grain properties, as a function of the environment.
These variations are numerous; some are 
speculative:\textlist{\thetextlist~PAHs are known to be destroyed in 
\hii\ regions \citep[{\eg}][]{madden06};
\thetextlist~the variations in the $R_V$ parameter of the extinction curve is 
interpreted as variations of the grain size distribution \citep{draine03c,fitzpatrick05};
\thetextlist~coagulation occurs in dense regions \citep[{\eg}][]{stepnik03,berne07};
\thetextlist~blast waves are responsible for grain fragmentation and erosion in the low-velocity phase
\citep{jones96} and destruction close to the remnant \citep{reach02};
\thetextlist~the dust abundances and properties are thought to evolve with the metallicity of the ISM \citep{galliano03,galliano05,galliano08a};
\thetextlist~a transition from amorphous to crystalline silicates is observed
in protostellar objects \citep[{e.g.}][]{hallenbeck00,poteet11}.}\modif{This
list is not exhaustive.}
Moreover, when considering the SED of a given region, it is possible to \modif{confuse} variations of the physical properties of the grains (\eg\ their optical
properties) with variations of their physical conditions (\eg\ the starlight intensity to which they are exposed).
This problem becomes even more intricate, when considering the integrated SED of a galaxy.

The various processes controlling the lifecycle of dust throughout the ISM are
not known with enough precision to break \modif{these kinds} of degeneracies.
Even the origin of interstellar dust is uncertain.
The contribution of supernovae (SN) and asymptotic giant branch (AGB) stars to the observed content of ISM dust, and the dust growth in interstellar cloud is still debated \citep[{\eg}][]{galliano08a,draine09}.
Dust is believed to constantly evolve throughout the ISM, being photoprocessed, altered by cosmic rays, accreting atoms in dense regions, and being shattered in shocks \citep[][for a review]{jones04}.
We are compelled to find observational cases where there will be enough redundancy in the data to isolate one of these processes.
This is the goal of this paper.

\seppar
The present article scrutinizes the different methodological biases, as well as the fundamental physical processes affecting the dust mass estimate in galaxies. 
Our demonstration is performed on the \hersc\ and \spitz\ observations of the 
Large Magellanic Cloud \citep[LMC; $d=50\pm 2.5$~kpc;][]{schaefer08}.
Due to its proximity, it is an ideal laboratory to study the variations of the far-IR properties, down to spatial scales of $\simeq 10$~pc (\SPIREiii\ angular resolution of 36\arcsec).
Moreover, it offers an environment containing massive star clusters, allowing us to study the impact of intense star formation on the surrounding ISM.
Finally, its metallicity is moderately sub-solar, with 
$\rm(O/H)_\sms{LMC}\simeq 0.5\times(O/H)_\odot$ and 
$\rm(C/H)_\sms{LMC}\simeq 0.3\times(C/H)_\odot$ \citep{pagel03}.
The comparison of its dust properties with those of the Galaxy therefore provides insights on cosmic dust evolution.

\seppar
In general, low-metallicity dwarf galaxies, a category to which the LMC belongs, have peculiar dust properties.
They exhibit a deficit of PAH strength, that appears to be correlated with the metallicity of their ISM.
The origin of this trend is still 
debated:\textlist{\thetextlist~PAHs could be more massively destroyed by permeating hard radiation, in sub-solar ISM \citep[{\eg}][]{madden06};
\thetextlist~the delayed injection of PAHs by AGB stars could explain their lower intrinsic abundance, in young systems \citep{galliano08a};
\thetextlist~the PAHs could form in molecular clouds, which have a lower filling factor, in low-metallicity environments \citep[{\eg}][]{sandstrom10}.}

The IR SED of dwarf galaxies usually peaks at shorter wavelengths, indicating
hotter equilibrium grains, on average.
In addition, the mid-IR continuum is steeply rising, similarly to what is observed in Galactic compact \hii\ regions \citep[{\eg}][]{peeters02}.
This peculiar mid-IR continuum, was modelled by \citet{galliano03,galliano05}
with an increase of the very small grain abundances, which could be the consequence of the high number density of shock waves.
It is consistent with the peculiar shape of the extinction curve, in the Magellanic clouds:
their lower 2175~\AA\ bump, and their steeper near-UV rise could be the result of an excess of small grains \citep{weingartner01,galliano03,galliano05}.
This typical continumm shape was reported as a \citengl{mid-IR excess} compared to the Galactic SED by \citet{bernard08}, in the LMC.

The dust-to-gas mass ratio increases with metallicity \citep{lisenfeld98,draine07b,galliano08a,engelbracht08}.
At first order, it can be understood since dust is made out of the metals
synthesized by the various stellar populations.
However, the detailed dependency of the dust-to-gas mass ratio with the metallicity remains unknown.
In addition, the gas mass, especially the molecular phase, is uncertain, because
the CO-to-\hmol\ conversion factor is a strong unknown function of the environment --~in particular, it is a function of the metallicity. 
A given CO line intensity will translate in a larger molecular gas mass in a lower metallicity environments 
\citep[{\eg}][]{madden97,israel97,leroy07,leroy09,leroy11}.
Moreover, we can expect the presence of large quantities of \hmol\ in regions where no CO at all is detected \citep{madden00}.
\citet{bernard08} unveiled a \citengl{far-IR excess}, compared to the gas column density, in the LMC, which is likely the evidence of such a gas reservoir.

Finally, at submm wavelengths, the SED of dwarf galaxies differs significantly from solar metallicity systems.
\citet{galliano03,galliano05} reported in four blue compact dwarf galaxies an excess emission at 850~\mmic\ (SCUBA) and 1.2~mm (MAMBO). 
Such an excess was then confirmed in other similar systems \citep{dumke04,galametz09,galametz10,bot10,grossi10}.
It extends up to centimetre (cm) wavelengths in the LMC and SMC \citep{israel10,planck-collaboration11}.
Although, the COBE data of our Galaxy presented a submm excess \citep{reach95}, the intensity of this excess is much more pronounced in low-metallicity systems.
Several explanations are in competition, for the origin of this 
excess:\textlist{\thetextlist~very cold dust, in dense clumps, accounting for a large fraction of the dust budget of the galaxy \citep{galliano03,galliano05,galametz09,ohalloran10};
\thetextlist~temperature dependent submm emissivity \citep{meny07};
\thetextlist~rapidly spinning grains in addition to another component \citep{bot10,planck-collaboration11}.}

The first \hersc\ observations of the LMC showed that the slope of the submm SED was flatter than in the Galaxy \citep{gordon10}.
\citet{meixner10} showed that modelling this SED with standard Galactic grain properties required too much mass, and therefore concluded that it 
required modified grain optical properties.
\citet{roman-duval10} confirmed the \citet{bernard08} \citengl{far-IR excess}
toward several molecular clouds.
\citet{hony10} demonstrated the complex structure of two massive starforming regions.
\citet{gordon10} reported a \SPIREiii\ excess, which is likely the rise of the
submm excess previously discussed.

\seppar
The unprecedented sensitivity and wavelength coverage of \hersc, at far-IR/submm wavelengths, allow us, for the first time, to study in \modif{detail} processes that were previously glimpsed at.
With a rigorous method, accounting for the different sources of error, it is 
now possible to unveil the systematic effects inherent to SED modelling.
The common assumptions, concerning the universality of dust properties, the accuracy of gas mass estimates, and the homogeneity of gas-to-dust mass ratios can be confronted by data.
From a technical point of view, all the processes that have been \modif{previously described} here define the required model parameters, as well as the unknowns when interpreting the IR/submm emission of the 
LMC.

For that purpose, the present paper is organized as follows.
\refsec{sec:data} presents the data set upon which we have based our analysis,
and discusses the reference observational quantities we consider.
In \refsec{sec:model}, we present our SED model, and the way we rigorously propagate the various sources of observational errors through the entire fitting procedure, in order to provide reliable errors on the derived physical parameters.
\refsec{sec:discussion} attempts to reconcile different interpretations of the
peculiar far-IR properties of the LMC: modified grain composition and/or undetected gas reservoir.
We end by a discussion on the origin of the \SPIREiii\ excess.
\refsec{sec:conclusion} synthesizes the paper and emphasizes the consequences of our findings.
The various appendices give details on technical points that would otherwise alter the flow of the discussion, if they were included in the main text.


\section{The Data Set}
\label{sec:data}

The data set we are using are the science demonstration (SD)
\hersc/SPIRE observations of the LMC \citep{meixner10}, together with its \spitz/IRAC and \spitz/MIPS data \citep{meixner06}, and some ancillary data.
These SD \hersc\ data cover only one fourth of the LMC.
Although the complete PACS and SPIRE maps have now been obtained, we have
performed our analysis on the sole SD strip, since we want to demonstrate general effects on the dust estimate, that do not require the totality of the LMC.

  \subsection{\hersc\ Observations and Data Reduction}
  \label{sec:herschel}

We use the SPIRE maps presented by \citet{meixner10}.
We point the reader to this paper for the detailed description of the data reduction.
The SD SPIRE data of the LMC cover a $2\degr\times 8\degr$, at 250, 350 and 500~\mmic.
The extended source calibration was performed assuming SPIRE beam areas of 395, 740 and 1517$\arcsec^2$, at 250, 350 and 500~\mmic, respectively.

A background was subtracted, taking as a reference the two outer edges of the strip. 
Those edges are supposed to be out of the LMC.
The same regions are considered for the background subtraction at other wavelengths, and for the gas maps.

First, as discussed by \citet{bernard08}, there are residual foreground \modif{Galactic} filamentary structures. 
We used the \hi\ map, whose velocity range corresponds to the Galaxy, in order to quantify the contribution of these fluctuations.
Converting the Galactic \hi\ column density into Galactic IR emission \citep[using the][model]{zubko04}, we find that the (non-subtracted) foreground accounts for $\simeq15\,\%$ of the IR power of the strip.
When this foreground is subtracted, the remaining fluctuations are on average $\simeq1\,\%$ of the IR power.
Therefore, this contamination is smaller than our uncertainties.

Second, we note that the larger scale \planck\ images \citep{planck-collaboration11} show
that \modif{the edges of our maps are not completely beyond the LMC.} 
The \planck\ images show that \modif{there is outer emission}, which is colder than that of the rest of the LMC, in particular to the South of the strip. 
This oversubtraction may bias the dust temperatures, that we will derive in \refsec{sec:discussion}, toward hotter emission.
However, as will be discussed in \refsec{sec:discussion}, our results are based on an excess of cold emission in the SPIRE bands.
Therefore, this potential oversubtraction is conservative.

We have compared the data of the SD paper, which were a single scan, with the more recent, accurately calibrated data, which also include a cross scan.
The relative difference between the integrated SPIRE fluxes of the two sets
is less than $2\,\%$.

  \subsection{\spitz\ Data}

The \spitz\ data include the four IRAC bands (3.6, 4.5, 5.8, 8.0~\mmic)
and the three MIPS bands (24, 70 and 160~\mmic).
They have been presented by \citet{meixner06}.

  \subsection{Gas Tracers}
  \label{sec:gas}

To compare the gas and dust contents in the LMC, we have completed our data set with observations of the atomic and molecular gas phases.
These maps were presented by \citet{roman-duval10}.

We use the \hiline\ map observed by \citet{kim03}. 
Their original beam size is $1.0\arcmin$.
The total atomic gas mass in the strip is 
$M_\sms{gas}^\sms{\hi}=7.14\E{7}\;M_\odot$.
The noise at the original resolution is 
$\sigma_\sms{\hi}(1.0\arcmin)\simeq 1.07\;M_\odot\,\rm pc^{-2}$.

We use the \COio\ map observed by \citet[][with the NANTEN telescope]{fukui08}.
The beam size is $2.6\arcmin$.
We assume a constant CO-to-\hmol\ conversion factor of 
$X_\sms{CO}=7\E{20}\;{\rm H\,cm^{-2}(K\,km\,s^{-1})^{-1}}$, based
on the virial estimate of \citet{fukui08}.
It accounts for the variation of the $X_\sms{CO}$ factor with metallicity, due to less efficient shielding of CO by dust and to a lower intrinsic C and O abundance
We emphasize that this value of $X_\sms{CO}$ is larger than in the Milky Way.
It accounts for discrepancies of the CO-to-\hmol\ conversion factor in regions where CO is detected.
However, it does not take into account potential regions where large envelopes of \hmol\ could be present, but where the CO would be massively photodissociated, and therefore not detected in emission by ground based radio telescopes.
With this conversion, the total molecular gas mass is 
$M_\sms{gas}^\sms{\hmol}=2.23\E{7}\; M_\odot$.
The noise at the original resolution is $6.05\;M_\odot\,\rm pc^{-2}$.
The total uncertainty in $M_\sms{gas}^\sms{\hmol}$ is dominated by the 
uncertainty in $X_\sms{CO}$ itself.
We will discuss that point in \refsec{sec:darkCO}.

The gas masses above include the mass of Helium and heavier elements.
The mean atomic weight used is:
\begin{equation}
  \mu_\sms{LMC} = \frac{1}{1-Y_\odot-Z_\sms{LMC}} = 1.34,
  \label{eq:mu}
\end{equation}
where the mass fractions of Helium and heavy elements are $Y_\odot=0.248$ and
$Z_\sms{LMC}=0.5\times Z_\odot=8.5\E{-3}$, respectively 
\citep{grevesse98,pagel03}.
The fraction of molecular gas in the strip is 
$M_\sms{gas}^\sms{\hmol}/(M_\sms{gas}^\sms{\hi}+M_\sms{gas}^\sms{\hmol})\simeq 24\;\%$.
It is higher than integrated over the entire LMC \citep[$\simeq10\;\%$;][]{bernard08}, since the strip includes a large number of molecular clouds.

  \subsection{Exploring the Effects of Spatial Resolution}
  \label{obs:res}

\begin{table*}[h!tbp]
  \centering
  \begin{tabular}{lll|rrr}
    \hline\hline
      Label         & Angular resolution  
                    & Linear \modif{resolution} ($l_\sms{pix}$)
                    & \multicolumn{3}{c}{Number of pixels}     \\
                    & 
                    &
                    & Total  & Dust modelling  
                    & Comparison to \hi\ \&\ \hmol\ \\
      \hline
      {\tt R1}      & $42\arcsec\times42\arcsec$ 
                    & $10\times10$~pc        & $193\times794$   
                    & 89245  & \ldots \\
      {\tt R2}      & $56\arcsec\times56\arcsec$ 
                    & $14\times14$~pc        & $144\times535$   
                    & 50367  & \ldots \\
      {\tt R3}      & $112\arcsec\times112\arcsec$ 
                    & $27\times27$~pc        & $72\times267$   
                    & 12750  & \ldots \\
      {\tt R4}      & $3.7\arcmin\times3.7\arcmin$ 
                    & $54\times54$~pc        & $36\times133$   
                    & 3278  & 2442 \\
      {\tt R5}      & $7.5\arcmin\times7.5\arcmin$ 
                    & $109\times109$~pc        & $18\times66$   
                    & 852  & 645 \\
      {\tt R6}      & $14.9\arcmin\times14.9\arcmin$ 
                    & $217\times217$~pc        & $9\times33$   
                    & 225  & 169 \\
      {\tt R7}      & $29.9\arcmin\times29.9\arcmin$ 
                    & $434\times434$~pc        & $4\times16$   
                    & 64  & 52 \\
      {\tt R8}      & $1\degr\times1\degr$ 
                    & $869\times869$~pc        & $2\times8$ 
                    & 16  & 15 \\
      {\tt R9}      & $2\degr\times2\degr$ 
                    & $1738\times1738$~pc        & $1\times4$ 
                    & 4  & 4 \\
      {\tt R10}     & $2.17\degr\times8.7\degr\Leftrightarrow4.3\degr$ 
                    & $1890\times7560\;\rm pc\Leftrightarrow 3780\;\rm pc$
                    & $1\times1$ 
                    & 1  &  1 \\
    \hline
  \end{tabular}
  \caption{{\sl Characteristics of the maps modelled in the present paper.}
           \modif{Each label (R1, R2, etc.) corresponds to the full data set (\spitz, \hersc\ and gas maps), but with different pixel sizes.}
           The resolution R10 corresponds to the integrated strip;
           it is a rectangle. 
           That is the reason why there are two dimensions.
           In the text, we will refer to the geometric mean of these two 
           dimensions.
           \modif{The three columns on the right side of the table list the 
           number of pixels.
           The first (\citengl{total}) is the total number of pixels in the map, 
           at a given resolution.
           The second (\citengl{dust modelling}) is the number of pixels used 
           for the dust modelling.
           It is the total number of pixels in the maps minus the pixels which 
           give bad fits at R1. 
           When building lower resolution maps, we exclude these bad pixels.
           Consequently, for a given waveband, the sum of the pixel fluxes is 
           rigorously the same at all spatial resolutions.
           The third number (\citengl{comparison to \hi\ and \hmol}) is the 
           number of pixels used when comparing the dust and gas properties.
           Since the \COio\ map is not defined on the entire strip, it 
           corresponds to a smaller number of pixels.}}
  \label{tab:dolls}
\end{table*}
All our maps are regridded and reprojected on a common frame.
They have been convolved with various kernels, in order to match the
spatial resolution of \MIPSiii\ ($38\arcsec$).
The full process is described in \citet{gordon10}.

We aim at studying the systematic effects that would bias the dust mass estimates.
The spatial resolution is one of these effects.
Indeed, SED models are highly non-linear, since the power emitted by 
a grain at thermal equilibrium with the radiation 
field (temperature $T_\sms{dust}$) is proportional to $T_\sms{dust}^{4+\beta}\simeq T_\sms{dust}^6$, where
$\beta\simeq2$ is the standard \citengl{emissivity index} \refeqp{eq:beta}.
Therefore, the sum of the modelled dust masses of $N$ regions is likely to be different than the modelled dust mass of the sum of the emissions of these $N$ regions.
In order to study this effect, we will model several maps of the same region, but with different pixel sizes.

\reftab{tab:dolls} lists the different resolutions.
The highest spatial resolution (R1, $42\arcsec$, 10~pc) is slightly larger than the largest beam size (\MIPSiii).
\modif{We then construct each set of maps (\spitz, \hersc\ and gas) by summing the pixels in a $2\times 2$ pixel window.
We repeat this process until we reach the size of the full integrated strip (R10).}
The resolution of the combined gas maps is lower than the dust maps.
We will therefore not study the spatial distribution of gas-to-dust mass ratios
at resolutions higher than R4.
From a technical point of view, the \COio\ map does not cover the entire strip.
Consequently, we define a subsample of pixels where both the gas
and dust maps are defined.


\section{A Phenomenological Dust SED Model}
\label{sec:model}

  \subsection{Motivations}

We have developed a model aimed at accurately fitting the observed mid-IR to mm SEDs of various regions of the ISM of the LMC.
At the spatial resolutions we consider here ($\gtrsim 10$~pc), the SEDs will 
likely be the combination of several regions with different physical conditions
--~photodissociation regions (PDR), diffuse ISM, etc.
In principle, we should perform a radiative transfer model, in order to determine the irradiation of each mass element of the ISM, then compute its spectrum, and transfer the IR radiation to the observer. 
However, we do not have the necessary information on the detailed matter and stellar 3D distributions, at these spatial scales, to constrain this type of model.
Moreover, such an analysis is unnecessary in our case.
Indeed, we are interested in the dust mass estimate.
The mass is dominated by large grains, at thermal equilibrium with the radiation field. 
The spectrum of these grains depends only on the stellar power they absorb (or on their equilibrium temperature), and not on the details of the stellar spectrum and spatial distribution.

This is not going to be true for grains which are out of thermal equilibrium with the radiation field (with typical radius $a\lesssim10$~nm), especially PAHs.
The spectrum of these grains depends on the hardness of the radiation field which determines the maximum temperature up to which the grain is fluctuating.
The fact is that most of the emission of these grains arises at short wavelengths \modif{($\lambda\lesssim50\mic$)}, and is not contaminating the 
far-IR-to-submm SED.

Finally, the regions we are considering here are optically thin in the IR.
Some compact sources show signs of absorption in the mid-IR \citep[9.8 and 18~\mmic\ silicate features, 15.2~\mmic\ CO$_2$ ice feature, etc.;][]{kemper10,hony11}, but the bulk of the grain emission, at longer wavelengths is unaffected.

Considering the previously exposed arguments, we could derive reliable dust masses by simply fitting the observed SEDs with a combination of several modified black bodies.
Nonetheless, we still choose to fit a combination of realistic dust models,
even if the very small grain (VSG) and PAH spectra will not be perfectly accurate, due to the lack of constraints on the radiation field hardness.
This approach is not providing significantly better mass estimates, but is
providing more reliable estimates of the physical conditions of the hottest equilibrium grains.
This is crucial for the interpretation, as will be demonstrated in \refsecs{sec:darkCO} and \ref{sec:r500}.
In particular, our approach allows us to avoid unphysical fits where a hot equilibrium component will be fit in place of the PAH emission.

Our phenomenological dust SED model can be decomposed in two levels:
\begin{enumerate}
  \item the dust SED of a mass element of the ISM, which is controlled by
        the microscopic grain properties;
  \item the synthesis of several mass elements, to account for the 
        macroscopic variations of the illumination conditions.
\end{enumerate}
\refsecs{sec:ISMdust} and \ref{sec:dale} details these two levels.
This model was previously used, in particular, by \citet{galametz09,galametz10}, \citet{ohalloran10}, \citet{cormier10}, \citet{hony10} and \citet{meixner10}.

  \subsection{Dust SED of a Mass Element of the ISM}
  \label{sec:ISMdust}

\begin{table*}[h!tbp]
  \centering
  \begin{tabular}{lrrr}
    \hline\hline
      Component  & Dielectric function & Heat capacity 
                 & Mass density \\
    \hline
      Smoothed UV astronomical silicates & \citet{weingartner01} 
        & \citet{draine85}
        & $3.50\E{3}\;\rm kg\,m^{-3}$ \\
      Graphite & \citet{laor93} 
        & \citet{dwek97}
        & $2.26\E{3}\;\rm kg\,m^{-3}$ \\      
      ACAR amorphous carbon & \citet{zubko96} 
        & \citet[][graphite]{dwek97} 
        & $1.85\E{3}\;\rm kg\,m^{-3}$ \\      
      Neutral PAHs & \citet{draine07} 
        & \citet{dwek97}
        & $2.24\E{3}\;\rm kg\,m^{-3}$ \\      
      Ionized PAHs & \citet{draine07} 
        & \citet{dwek97}
        & $2.24\E{3}\;\rm kg\,m^{-3}$ \\      
    \hline
  \end{tabular}
  \caption{{\sl Grain composition of our dust model.}
            We give the references where the dielectric functions and heat
            capacities can be found, as well as the mass density of the 
            material.}
  \label{tab:grains}
\end{table*}
Let's consider the SED emitted by an element of mass of the ISM, where we can assume that the starlight intensity is uniform.
For simplicity, we assume that the starlight intensity heating the grains has the spectral shape of the interstellar radiation field (ISRF) of the 
solar neighborhood \citep[noted $U_\lambda^\odot(\lambda)$;][]{mathis83}.
We parametrize its integrated intensity by:
\begin{equation}
  U = \frac{\displaystyle
            \int_{0.0912\mic}^{8\mic} U_\lambda^\odot(\lambda)\ddiff\lambda}
           {2.2\E{-5}\rm\; W\,m^{-2}}.
  \label{eq:U}
\end{equation}
The value $U=1$ corresponds to the intensity of the solar neighborhood.
In these conditions, large interstellar silicates have an equilibrium 
temperature of $\simeq17.5$~K.
It is possible that the ISRF of the LMC differs from $U_\lambda^\odot$, since it has younger stellar populations.
This spectral shape is also likely to vary spatially within the LMC, being harder in transparent regions, and redder in dense regions.
However, as explained previously, this will impact only the PAH and VSG spectra, which are not determinant to estimate the total dust mass.

In the same way, we are adopting the Galactic framework,
by using the grain properties of \citet{zubko04}.
We choose the bare grain model, with solar abundance constraints (BARE-GR-S).
The abundance and size distribution of each grain species was constrained by
fitting the IR emission, UV-visible extinction and elemental depletions of the
Galactic diffuse ISM.
We have updated this model with the new \citet{draine07} PAH optical 
properties, that includes more accurate band profiles, based on \spitz\ spectra.

The optical properties and enthalpies considered here are summarized in 
\reftab{tab:grains}.
The grain cross-sections are computed using a Mie code, and following 
the method of \citet[][Sect.~2.2; \refapp{ap:mie} of the present paper]{laor93}.
The temperature fluctuations are computed for each grain size of each 
component, and for each starlight intensity, using the transition matrix 
method \citep{guhathakurta89}.

The specific monochromatic luminosity of a mass element of ISM exposed
to the starlight intensity $U$ is:
\begin{equation}
  l_\nu^\sms{dust}(U,\lambda) = f_\sms{PAH} l_\nu^\sms{PAH}(U,\lambda)
                              + f_\sms{carb} l_\nu^\sms{carb}(U,\lambda)
                              + f_\sms{sil} l_\nu^\sms{sil}(U,\lambda),
  \label{eq:lnu}
\end{equation}
where $f_\sms{PAH}$, $f_\sms{carb}$ and $f_\sms{sil}$ are the mass fractions
of PAHs (charge fraction of 1/2), graphite and silicate ($f_\sms{PAH}+f_\sms{carb}+f_\sms{sil}=1$), and $l_\nu^\sms{PAH}(U,\lambda)$, $l_\nu^\sms{carb}(U,\lambda)$, $l_\nu^\sms{sil}(U,\lambda)$ are the corresponding size distribution integrated specific 
monochromatic luminosities.
In this paper, we keep the mass fractions to the Galactic values \citep{zubko04}, except $f_\sms{PAH}$ that we vary to fit the observed 
\IRACiv\ band.

\begin{figure*}[h!tbp]
  \centering
  \includegraphics[width=0.95\linewidth]{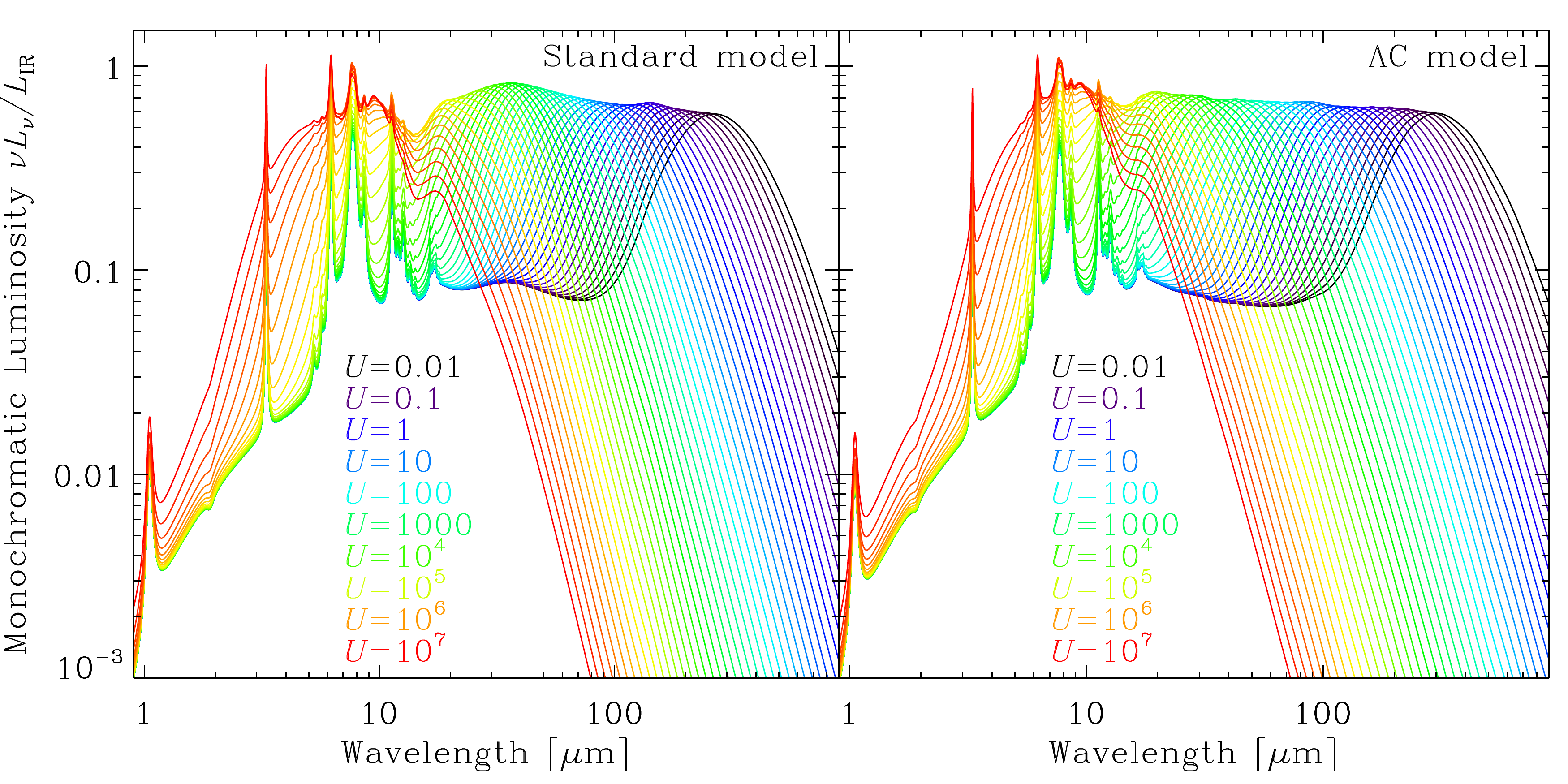}
  \caption{{\sl
           Uniformly illuminated dust SEDs, exposed to various radiation field
           intensities $U$.}
           Each curve represents the sum of the emission by PAH$^0$, PAH$^+$,
           carbon grains and silicates, exposed to $U$, in units of 
           $2.2\E{-5}\;\rm W\,m^{-2}$. 
           The monochromatic luminosity is normalised by its integrated 
           luminosity $L_\sms{IR}$.}
  \label{fig:demoSED}
\end{figure*}
Throughout this paper, we will systematically compare the two following 
dust compositions, in order to study a possible evolution of composition between the Milky Way and the LMC.
\begin{enumerate}
  \item \citengl{The standard model} (hereafter labeled {\sl Std}) is the 
        original \citet{zubko04} Galactic grain composition, made of PAHs, 
        graphite and silicates.
        The effective submillimeter opacity of this dust corresponds to 
        an emissivity index $\beta\simeq2$ (\refapp{ap:mie}).
  \item \citengl{The AC model} (hereafter labeled {\sl AC}) is the 
        \citengl{standard model}, 
        but replacing graphite by amorphous carbons \citep[ACAR;][]{zubko96}.
        This substitution is arbitrary.
        The purpose of this model is to test realistic compositions having
        a higher effective submillimeter opacity ($\beta\simeq1.7$; 
        \refapp{ap:mie}), without violating the elemental abundances.
\end{enumerate}
We note that the heat capacities of amorphous carbons are unknown.
We adopt those of graphite in replacement.
It might be a crude approximation.
However, this inconsistency will affect only the stochastically
heated grains, which do not contribute significantly to the dust mass.

The individual SEDs are shown in \reffig{fig:demoSED}.
The two models show similar features.
\begin{enumerate}
  \item The far-IR peak is dominated by grains in thermal equilibrium with the 
        radiation field. 
        Their spectrum is roughly a modified black body. 
        The peak wavelength of the emission shifts to shorter wavelengths when 
        $U$ rises, as the equilibrium temperature rises.
  \item The mid-IR continuum, for $U\lesssim 10^4$, is dominated by small grains
        (radius $a\lesssim 0.01\mic$) and PAHs (prominent emission bands at 3.3, 
        6.2, 7.7, 8.6, 11.3~\mmic), both out of equilibrium with the 
        radiation field.
        These grains \modif{are being heated by single photon events and} the 
        spectral shape of 
        their emission is independent of $U$.
        Their spectrum normalized to $L_\sms{IR}$ (or $U$) is constant.
        The change of shape of the mid-IR spectrum with $U$ is only due to the
        contribution of equilibrium grains at these wavelengths, when their
        temperature reaches $T_\sms{eq}\gtrsim80$~K ($U\gtrsim10^4$).
        \modif{In particular, the prominent $9.7\mic$ silicate feature in 
        emission dominates the mid-IR wavelengths, in this temperature regime.}
\end{enumerate}

  \subsection{Synthetic Multi-Environment SED}
  \label{sec:dale}

Each SED we model in this paper is likely to be the combination of the emission from regions with different physical conditions.
To account for this diversity of conditions, we make the following assumptions.
\begin{enumerate}
  \item We assume that the dust properties are uniform within the modelled  
        region: the size distribution, and mass fractions are constant.
        Only the starlight intensity varies.
        This is an approximation.
        We will discuss in \refsec{sec:interpretation} potential local 
        variations of the grain properties.
  \item The distribution of starlight intensities per unit dust mass,
        throughout the region, can be approximated by a power-law
        \citep{dale01}:
        \begin{equation}
          \frac{\dd M_\sms{dust}}{\dd U} \propto U^{-\alpha} 
          \mbox{ with } U_\sms{min}\leq U \leq U_\sms{min}+\Delta U.  
          \label{eq:dale}
        \end{equation}
        \modif{This is an empirical prescription.
        \citet[][Sect.~5.5]{dale01} provide a physical justification of this 
        formulation.
        However, its main advantage is that it allows for flexible parametrizing 
        of the physical conditions.}
        A more complex formulation is discussed in 
        \refapp{ap:compD07}.
\end{enumerate}
The total dust mass of each \modif{modelled region} is therefore:
\begin{equation}
  M_\sms{dust} = \int_{U_\sms{min}}^{U_\sms{min}+\Delta U} 
                 \frac{\dd M_\sms{dust}}{\dd U}\ddiff U.
  \label{eq:Mdust}
\end{equation}

In addition, to subtract the stellar contribution \modif{from} the mid-IR bands, we
add a stellar continuum, parametrized by the stellar mass in the region 
$M_\star$:
\begin{equation}
  L_\nu^\star(\lambda) = M_\star\times l_\nu^\star(\lambda),
\end{equation}
where $l_\nu^\star(\lambda)$ is the specific monochromatic luminosity of 
a 1~Gyr stellar population, synthesized with the model PEGASE \citep{fioc97}.
Since, this population is constrained mainly by the \IRACi\ and \IRACii\ bands, the age of the populations do not have a big effect. 
On the other \modif{hand}, this stellar mass is poorly determined and should not be trusted.
The only purpose of this component is to give a better $\chi^2$. 
To accurately determine the mass of the stellar populations, we would need to take into account shorter wavelengths.
This is not the purpose of this paper.

In summary, the total monochromatic luminosity of the model is:
\begin{eqnarray}
  L_\nu^\sms{mod}(\lambda) & = & \int_{U_\sms{min}}^{U_\sms{min}+\Delta U}
    l_\nu^\sms{dust}(U,\lambda)\times \frac{\dd M_\sms{dust}}{\dd U}\ddiff U \nonumber\\
    & + & M_\star\times l_\nu^\star(\lambda).
    \label{eq:model}
\end{eqnarray}
For each waveband $\lambda_i$, where the observed monochromatic luminosity is
$L_\nu^\sms{obs}(\lambda_i)$, we compute the synthetic photometry, $L_\nu^\sms{mod}(\lambda_i)$, by convolving the model with \modif{the} instrumental spectral response, using the appropriate conventions provided by the user's manuals of each instrument.
We minimize the $\chi^2$, using a Levenberg-Marquart algorithm \citep{press92}.
The $\chi^2$ is weighted as follows:
\begin{equation}
  \chi^2 = \sum_{i=1}^{n} \frac{\left(L_\nu^\sms{obs}(\lambda_i)-L_\nu^\sms{mod}(\lambda_i)\right)^2}{\left(\Delta L_\nu^\sms{RMS}(\lambda_i)\right)^2
  +\left(\Delta L_\nu^\sms{cal}(\lambda_i)\right)^2},
  \label{eq:chi2}
\end{equation}
where $\Delta L_\nu^\sms{RMS}(\lambda_i)$ and $\Delta L_\nu^\sms{cal}(\lambda_i)$ are respectively the RMS and calibration errors of the waveband centered at 
wavelength $\lambda_i$ (see \refsec{sec:error}).
We do not use the \SPIREiii\ flux as a constraint, because of its excess relative to the model, as discussed by \citet{gordon10}.
Instead, we will study the behaviour of this excess in \refsec{sec:r500}, in order to attempt to decipher its origin.

\begin{figure*}[h!tbp]
  \centering
  \includegraphics[width=0.95\linewidth]{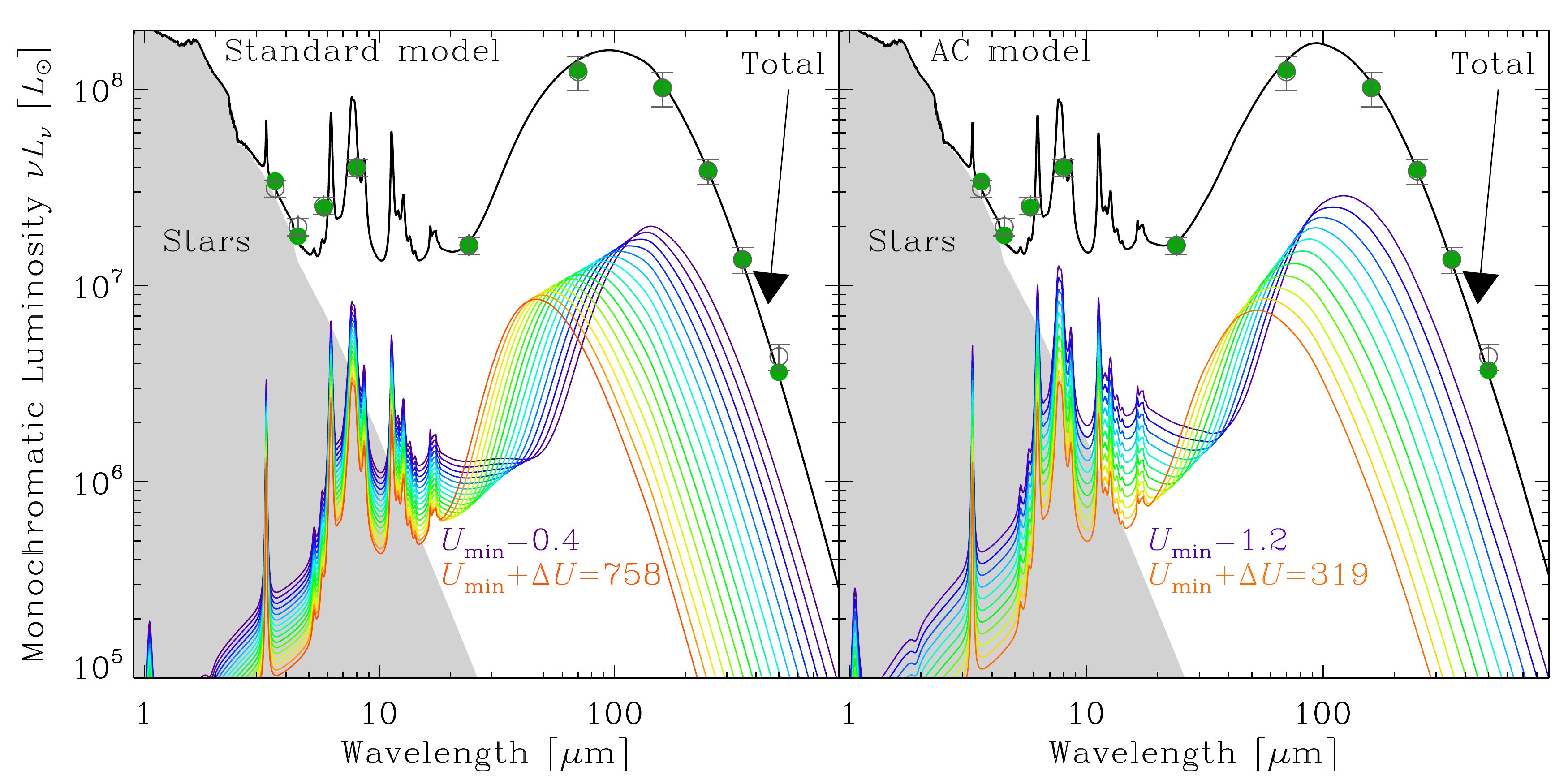}
  \caption{{\sl Decomposition of the integrated strip SED into the individual
           uniformly illuminated SEDs.}
           The \modif{grey} circles and error bars are the integrated observed 
           fluxes 
           of R10 (\reftab{tab:dolls}). 
           The total model (black line; \refeqnp{eq:model}) is the sum of 
           the independent stellar component 
           (grey filled area), and of the integral of uniformly illuminated 
           dust SEDs (in colors; \refeqnp{eq:dale}).
           There is linear gradation in $U$ between colors for the uniformly 
           illuminated SEDs.
           The starlight intensity $U$ is in units of $2.2\E{-5}\;\rm W
           \,m^{-2}$.
           The sum of these components is the black line.
           The green dots are the synthetic photometry (\ie\ the model 
           integrated in each instrumental filter).
           The {\it left panel} shows the \citengl{standard model}, while
           the {\it right panel} shows the \citengl{AC model}.
           To quantify the quality of the fits, the reduced chi square
           is $\bar{\chi}^2_\sms{Std}=1.07$ for the \citengl{standard model}
           and $\bar{\chi}^2_\sms{AC}=0.89$ for the \citengl{AC model}.}
  \label{fig:demoMix}
\end{figure*}
\reffig{fig:demoMix} demonstrates this model on the integrated strip SED (R10).
The two panels highlight the degeneracy between the starlight intensity distribution and the submillimeter opacities, by comparing the fit of the
two compositions to the same observations.
Having a flatter submillimeter opacity gives a fit with less massive, 
hotter dust populations.

\reftab{tab:parm} summarizes the parameters of the model.
In particular, the last column of \reftab{tab:parm} describes the behaviour of the free parameters.
When an SED fit is performed, a slight variation of one parameter value,
due to observational errors, may systematically be compensated by the variation 
of another parameter.
Although the free parameters are rigorously independent, this effect may induce
a correlation between these parameters.
The Monte-Carlo error analysis that will be discussed in \refsec{sec:MC} is a good way to quantify these correlations.
The most striking example, in our case, is the correlation between the three parameters controlling the starlight intensity distribution ($\alpha$, $U_\sms{min}$ and $\Delta U$; \refeqnp{eq:dale}).
Moreover, these parameters do not have a physical meaning.
Rather than discussing their values, it is more convenient to consider the first 
two moments of the starlight intensity distribution:
\begin{eqnarray}
  \langle U\rangle 
  & = & 
  \frac{1}{M_\sms{dust}}
  \int_{U_\sms{min}}^{U_\sms{min}+\Delta U} U\times \frac{\dd M_\sms{dust}}{\dd U}\ddiff U \\
  \sigma^2(U)
  & = & 
  \frac{1}{M_\sms{dust}}
  \int_{U_\sms{min}}^{U_\sms{min}+\Delta U}\left(U-\langle U\rangle\right)^2
  \times \frac{\dd M_\sms{dust}}{\dd U}\ddiff U,
\end{eqnarray}
which develop into:
\begin{equation}
  \langle U\rangle = 
  \left\{
  \begin{array}{ll}
    \displaystyle
    \frac{1-\alpha}{2-\alpha}
    \frac{\left(U_\sms{min}+\Delta U\right)^{2-\alpha}-U_\sms{min}^{2-\alpha}}
         {\left(U_\sms{min}+\Delta U\right)^{1-\alpha}-U_\sms{min}^{1-\alpha}}
    & \mbox{ if } \alpha\neq1 \;\&\; \alpha\neq2 \\
    & \\
    \displaystyle
    \frac{\Delta U}{\ln\left(U_\sms{min}+\Delta U\right)-\ln U_\sms{min}}
    & \mbox{ if } \alpha = 1 \\
    & \\
    \displaystyle
    \frac{\ln\left(U_\sms{min}+\Delta U\right)-\ln U_\sms{min}}
         {U_\sms{min}^{-1}-\left(U_\sms{min}+\Delta U\right)^{-1}}
    & \mbox{ if } \alpha = 2, \\
  \end{array}
  \right.
  \label{eq:avU}
\end{equation}
and:
\begin{equation}
  \sigma^2(U) = 
  \left\{
  \begin{array}{ll}
    \displaystyle
    \frac{1-\alpha}{3-\alpha}
    \frac{\left(U_\sms{min}+\Delta U\right)^{3-\alpha}-U_\sms{min}^{3-\alpha}}
         {\left(U_\sms{min}+\Delta U\right)^{1-\alpha}-U_\sms{min}^{1-\alpha}}
    - \langle U\rangle^2
    & \mbox{ if } \alpha\neq1 \\
    & \\
    \displaystyle
    \frac{1}{2}
    \frac{\left(U_\sms{min}+\Delta U\right)^2-U_\sms{min}^2}
         {\ln\left(U_\sms{min}+\Delta U\right)-\ln U_\sms{min}}
    - \langle U\rangle^2
    & \mbox{ if } \alpha = 1. \\
  \end{array}
  \right.
  \label{eq:sigU}
\end{equation}
We define the infrared luminosity as the power emitted by the dust:
\begin{equation}
  L_\sms{IR} = \int_0^\infty L_\nu^\sms{dust}(\nu)\ddiff\nu.
  \label{eq:LIR}
\end{equation}
Finally, the \citengl{gas-to-dust mass ratio} is the ratio of the total mass of gas (\hi\ and \hmol, with Helium and heavy elements) to the total dust mass, in the same region:
\begin{equation}
  G_\sms{dust} = \frac{M_\sms{gas}^\sms{\hi}+M_\sms{gas}^\sms{\hmol}}{M_\sms{dust}}.
  \label{eq:G2D}
\end{equation}
For comparison, the Galactic value is \citep{zubko04}:
\begin{equation}
  G_\sms{dust}^\odot\simeq158.
  \label{eq:Gsun}
\end{equation}
It is important to note that, for the discussion in \refsec{sec:discussion}, this value is consistent with the dust properties we use.
Moreover, this model is consistent with the elemental depletion constraints.
\begin{table*}[h!tbp]
  \centering
  \begin{tabularx}{\linewidth}{lllXX}
    \hline\hline
      Parameter       & Units \&\ normalization  & Range of values
        & Definition  
        & Comments \\
    \hline
      \multicolumn{5}{c}{{\it 1.\ Free parameters}} \\
    \hline
      $M_\sms{dust}$  & $[M_\odot]$  & $[0,\infty[$ 
        & Dust mass (all components; \refeqnp{eq:Mdust})    
        & Linearly correlated with $\langle U\rangle$  \\
      $f_\sms{PAH}$   & $[4.6\;\%]$  & $[0,1/0.046]$   
        & PAH-to-dust mass ratio (relative to Galactic)
        & Degenerate with the arbitrarily fixed PAH charge fraction \\
      $\alpha$        & \ldots  & $[1,2.5]$
        & Index of the power-law distribution of starlight intensities 
        \refeqp{eq:dale} 
        & Correlated with $U_\sms{min}$ and $\Delta U$ \\
      $U_\sms{min}$        & $[2.2\E{-5}\;\rm W\,m^{-2}]$  & $[10^{-2},10^7]$
        & Lower cut-off of the power-law distribution of starlight intensities 
        \refeqp{eq:dale} 
        & Correlated with $\alpha$ and $\Delta U$ \\
      $\Delta U$        & $[2.2\E{-5}\;\rm W\,m^{-2}]$  & $[1,10^7-U_\sms{min}]$
        & Range of starlight intensities 
          (\refeqnp{eq:dale})
        & Correlated with $\alpha$ and $U_\sms{min}$ \\
      $M_\star$        & $[M_\odot]$  & $[0,\infty[$
        & Mass of the old stellar population
        & Poorly constrained; this parameter is used only to subtract the 
        stellar continuum to the mid-IR bands;
        this component has no relation to the starlight intensity \\
    \hline
      \multicolumn{5}{c}{{\it 2.\ Derived parameters}} \\
    \hline
      $L_\sms{IR}$        & $[L_\odot]$                  & $[0,\infty[$                
      & Infrared luminosity (total dust power; \refeqnp{eq:LIR})
      & Almost model independent \\
      $\langle U\rangle$  & $[2.2\E{-5}\;\rm W\,m^{-2}]$ & $[10^{-2},10^7]$
      & Mass averaged starlight intensity \refeqp{eq:avU}
      & \ldots \\
      $\sigma(U)$   & $[2.2\E{-5}\;\rm W\,m^{-2}]$ & $[1,10^7-\langle U\rangle]$
      & Second moment of the starlight intensity distribution \refeqp{eq:sigU}
      & \ldots \\
      $G_\sms{dust}$ & \ldots   & $[0,\infty[$ 
      & Gas-to-dust mass ratio \refeqp{eq:G2D}
      & The Galactic value is $G_\sms{dust}^\odot\simeq 158$ \citep{zubko04} \\
    \hline
  \end{tabularx}
  \caption{{\sl Parameters of the model.}
           The first category lists the independent free parameters.
           The second category lists the quantities derived from these 
           parameters.
           The last column (\citengl{comments}) describes the behaviour
           of the parameters when fitting an SED.
           In absolute, $M_\sms{dust}$ and $\langle U\rangle$ are 
           independent.
           However, when fitting an observed SED, a slight variation of 
           $\langle U\rangle$ will have \emph{systematic} consequences on 
           $M_\sms{dust}$, to compensate the variation and minimize the 
           $\chi^2$.
           The same comment applies to the other dependent parameters.}
  \label{tab:parm}
\end{table*}

  \subsection{Rigorous Error Propagation}
  \label{sec:error}

Since our dust model is highly non-linear, it is crucial to rigorously propagate the observational errors through the entire fitting procedure, taking into 
account the fact that some errors are independent and others are correlated.
In this way, we will be able to quote consistent errors on the parameters.
We first need to identify the various sources of error

\subsubsection{Sources of Observational Error}
\label{sec:errobs}

\begin{table*}[h!tbp]
\centering
\begin{tabular}{l*{10}{r}}
\hline\hline
Filter & \multicolumn{10}{c}{$F_\nu^\sms{RMS}\;\rm[10^{-3}\, MJy\,sr^{-1}]$} \\
\hline
 & R1 & R2 & R3 & R4 & R5 & R6 & R7 & R8 & R9 & R10 \\
\hline
\IRACi & $19.8$ & $14.9$ & $7.4$ & $3.7$ & $1.86$ & $0.92$ & $0.40$ & $0.206$ & $0.104$ & $0.201$ \\
\IRACii & $2.61$ & $1.96$ & $0.98$ & $0.49$ & $0.244$ & $0.121$ & $0.053$ & $0.0271$ & $0.0137$ & $0.142$ \\
\IRACiii & $3.8$ & $2.82$ & $1.41$ & $0.70$ & $0.35$ & $0.174$ & $0.076$ & $0.039$ & $0.0197$ & $0.239$ \\
\IRACiv & $3.7$ & $2.76$ & $1.38$ & $0.69$ & $0.34$ & $0.170$ & $0.074$ & $0.038$ & $0.0192$ & $0.45$ \\
\MIPSi & $5.0$ & $3.7$ & $1.86$ & $0.93$ & $0.47$ & $0.230$ & $0.100$ & $0.052$ & $0.0260$ & $1.31$ \\
\MIPSii & $57$ & $42$ & $21.2$ & $10.6$ & $5.3$ & $2.62$ & $1.14$ & $0.59$ & $0.296$ & $16.0$ \\
\MIPSiii & $207$ & $155$ & $77$ & $39$ & $19.4$ & $9.6$ & $4.2$ & $2.14$ & $1.08$ & $19.8$ \\
\SPIREi & $248$ & $186$ & $93$ & $47$ & $23.3$ & $11.5$ & $5.0$ & $2.58$ & $1.30$ & $13.8$ \\
\SPIREii & $180$ & $135$ & $68$ & $34$ & $16.9$ & $8.3$ & $3.6$ & $1.87$ & $0.94$ & $6.3$ \\
\SPIREiii & $66$ & $49$ & $24.7$ & $12.3$ & $6.2$ & $3.05$ & $1.33$ & $0.68$ & $0.34$ & $2.73$ \\
\hline
\end{tabular}
\caption{{\sl RMS values for each filter and each spatial resolution.}
         These values come from the standard deviation of the upper and lower
         borders of the maps.
         This RMS value represents the $1\times\sigma$ of a normally distributed
         random 
         variable.}
\label{tab:rms}
\end{table*}
For each wavelength, at each spatial resolution, we measure the noise of the map by taking the standard deviation of the pixel values in what is considered to be the background, \ie\ the upper and lower ends of the strip.
These values are given in \reftab{tab:rms}.
These errors are independent from one wavelength to the other, and from one
pixel to the other.
A simple check of the distribution of pixel values shows that the uncertainty is 
well described by a Gaussian.

The calibration error is the error on the flux conversion factor.
This error is therefore correlated between each pixel.
It can be synthesized as follows.
\begin{description}
  \item[IRAC:] the $1\sigma$ calibration uncertainty is 
    $\sigma_\sms{cal}(\mbox{IRAC})=2\;\%$ \citep{reach05}.
    The different wavelengths are correlated.
  \item[\MIPSi:] \citet{engelbracht07} quote a $1\sigma$ calibration error of 
    $\sigma_\sms{cal}(\mbox{\MIPSi})=4\;\%$.
    It is independent of the other wavebands.
  \item[\MIPSii:] \citet{gordon07} quote a $1\sigma$ calibration error of
    $\sigma_\sms{cal}(\mbox{\MIPSii})=5\;\%$ for the coarse scale mapping used 
    for the SAGE observations.
  \item[\MIPSiii:] \citet{stansberry07} report a $1\sigma$ calibration error
    of $\sigma_\sms{cal}(\mbox{\MIPSiii})=12\;\%$.
    This error is correlated with the \MIPSii\ error.
  \item[SPIRE:] Although \citet{swinyard10} report a calibration error of 
    $\sigma_\sms{cal}(\mbox{SPIRE})=15\;\%$, it is necessary to decompose 
    this error into its components \citep{spire-consortium10}, as the SPIRE fluxes are the most crucial 
    constraints on the dust mass and emissivity.
    \begin{enumerate}
      \item The $3\sigma$ error on the calibration model is 
        $3\times\sigma_\sms{cal}^\sms{model}(\mbox{SPIRE}) = 5\;\%$, 
        each waveband being correlated.
      \item The noise in the calibration observations (Ceres) are:
        \begin{itemize}
          \item $\sigma_\sms{cal}^\sms{noise}(\mbox{\SPIREi}) = 7\;\%$;
          \item $\sigma_\sms{cal}^\sms{noise}(\mbox{\SPIREii}) = 12\;\%$;
          \item $\sigma_\sms{cal}^\sms{noise}(\mbox{\SPIREiii}) = 6\;\%$.
        \end{itemize}
        These errors are independent.
      \item The error on the beam area are 
        $\sigma_\sms{cal}^\sms{beam}(\mbox{SPIRE}) = {-2\;\%}/{+5\;\%}$ 
        and they are independent.
    \end{enumerate}
\end{description}
The latter SPIRE calibration errors correspond to the initial calibration performed during the SD phase.
We kept this error, in order to remain consistent with our data set.
In any case, this estimate is conservative, the new calibration errors being smaller.

  \subsubsection{Monte-Carlo Iterations}
  \label{sec:MC}

\begin{figure*}[h!tbp]
  \centering
  \includegraphics[width=0.95\linewidth]{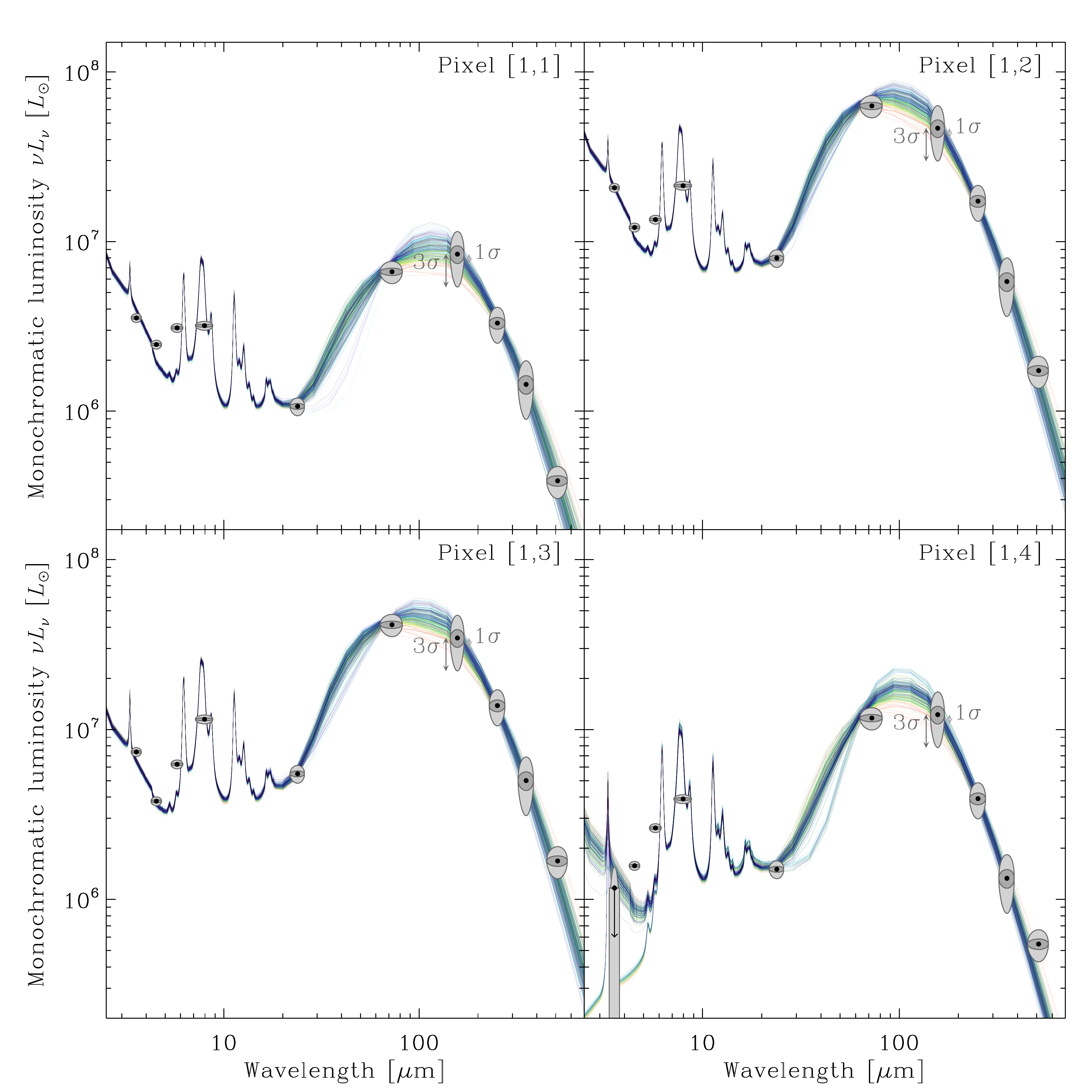}
  \caption{{\sl Demonstration of the Monte-Carlo method on the 4 pixels of the
            R9 map.}
           The grey ellipses represent the unperturbed observations and 
           their uncertainties: the filter band widths on the $x$ direction; 
           the $\pm1\sigma$ error bars on the $y$ direction (dark grey); 
           and the $\pm3\sigma$ error on the $y$ direction (concentric 
           light grey ellipse).
           The color lines show the $N_\sms{MC}=300$ model fits to the
           perturbed fluxes.
           We used the \citengl{standard model} for the demonstration.
           Some points (like \MIPSii) appears shifted from the models because
           of the color correction.
           The two shortest wavelengths of the $[1,4]$ pixel are poorly fitted
           since the \IRACi\ is only an upper limit.
           This might be the result of oversubtraction of point sources in this
           low surface brightness region.
           However, this discrepancy affects only the level of the independent 
           stellar contribution.
           It does not affect the longer wavelength fit.}
  \label{fig:demoMC}
\end{figure*}
We propagate the observational errors detailed in \refsec{sec:errobs}, by performing Monte-Carlo iterations of each fit.
More precisely, for each observed SED, we perform a large number 
($N_\sms{MC}\simeq300$) \modif{of} fits of the SED with additional random perturbations.
\reffig{fig:demoMC} demonstrates the fits of the perturbed SEDs of the 4 pixels of the R9 map.
Each model corresponds to one particular set of random perturbations.
These perturbations take into account the two main sources of errors, as follows.
\begin{enumerate}
  \item The pixel noise at each wavelength (\refsec{tab:rms}) is assumed to be  
        a normal random independent variable.
        The noise is independent from one pixel to the other.
  \item The calibration error is assumed to be a normal random variable, with 
        standard deviation and correlation between wavelengths as described in 
        \refsec{sec:errobs}.
        The calibration error from one pixel to the other is correlated.
\end{enumerate}

From a technical point of view, we generate the complete set of independent random variables necessary for the calibration errors, and keep them for our entire analysis.
Indeed, one of the advantages of the Monte-Carlo technique is that it allows us to account for complex correlations between variables. 
As will be demonstrated in \refsec{sec:resolution}, the error on the ratio of two quantities depending on the calibration error is often lower than the errors on each individual \modif{quantity}.
It is due to the fact that the calibration error cancels when considering a relative quantity.
It is even possible to take this effect into account when comparing the results of two different models, as will be shown in \refsec{sec:compmod}.

\begin{figure*}[h!tbp]
  \centering
  \begin{tabular}{cc}
    \includegraphics[width=0.48\textwidth]{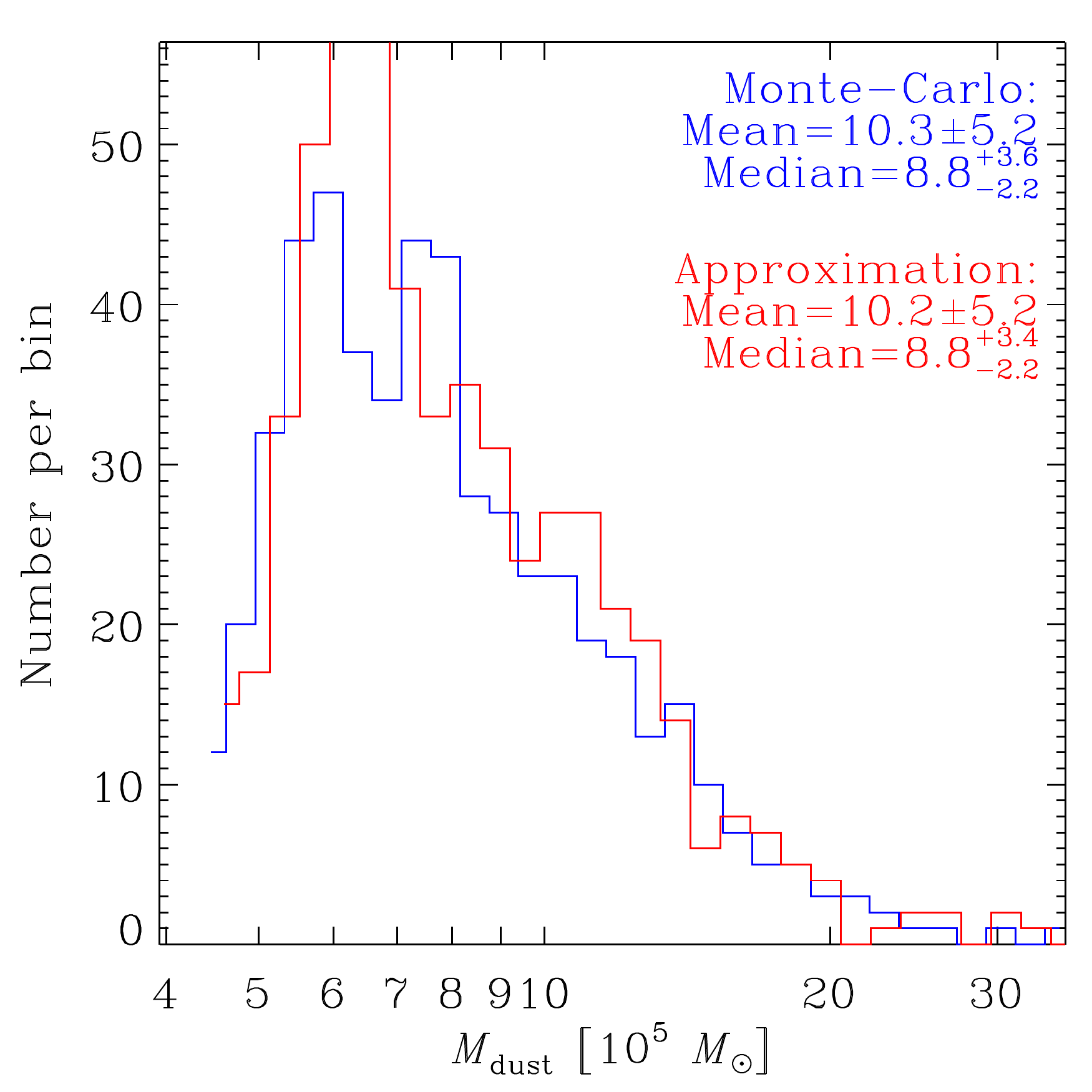} & 
    \includegraphics[width=0.48\textwidth]{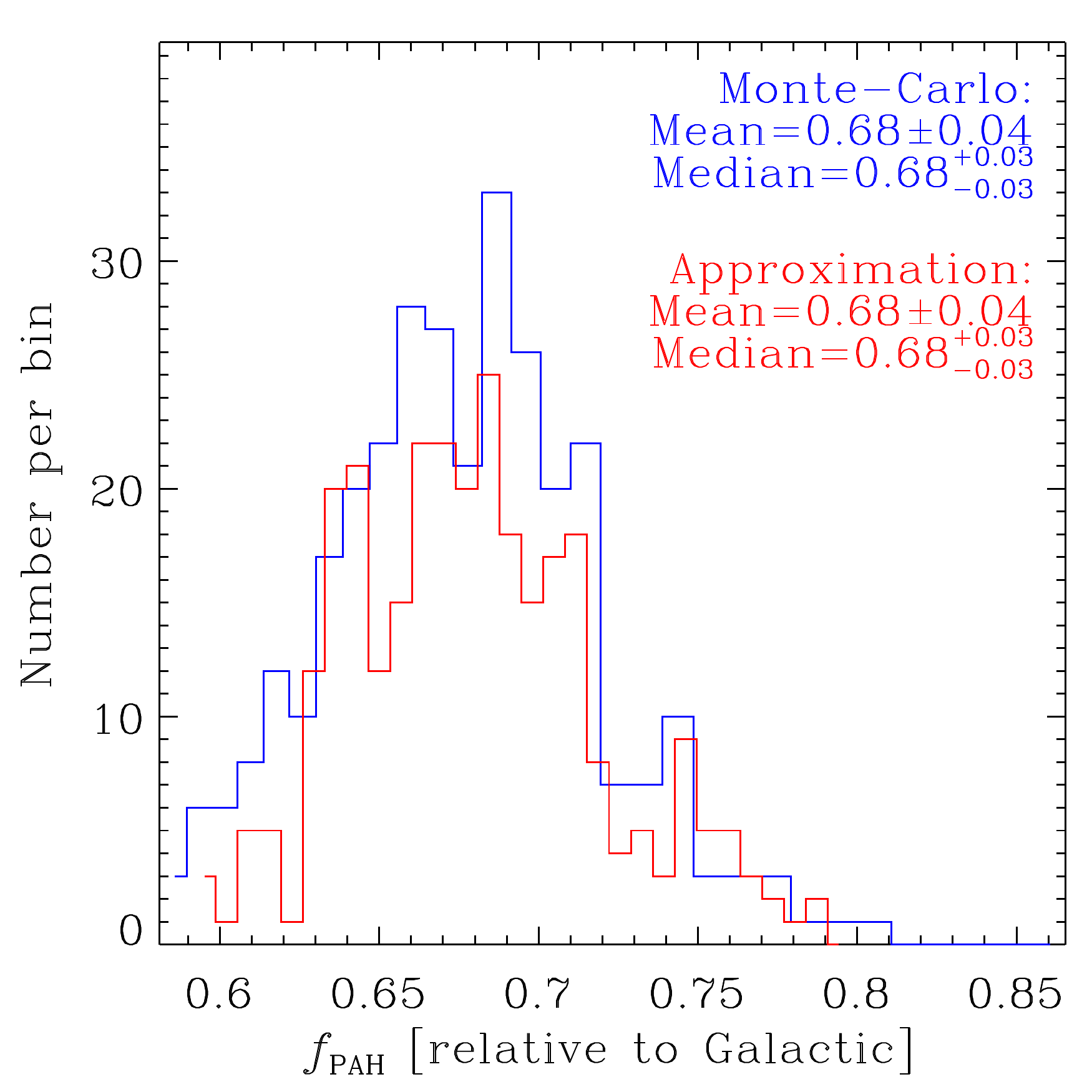} \\
    \includegraphics[width=0.48\textwidth]{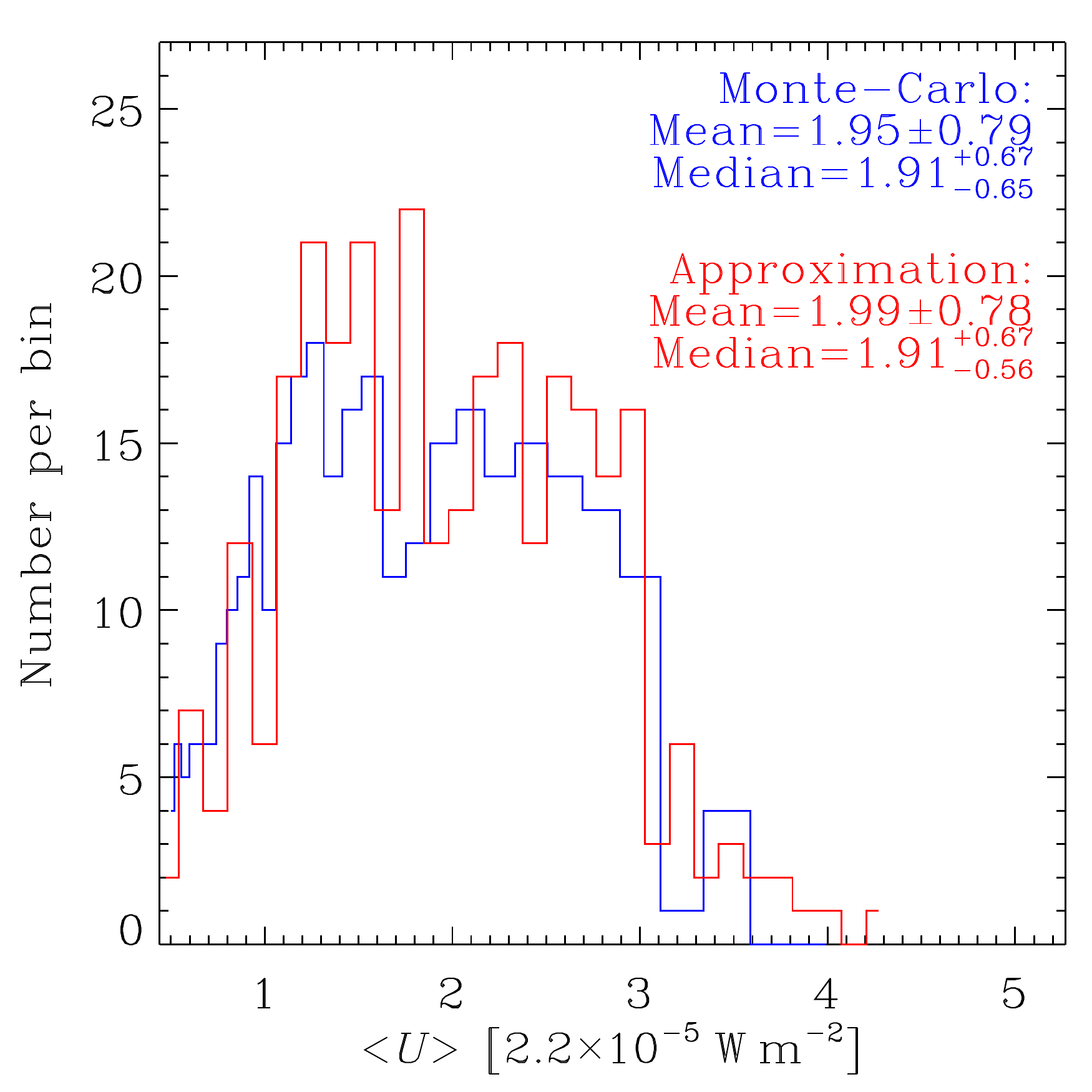} & 
    \includegraphics[width=0.48\textwidth]{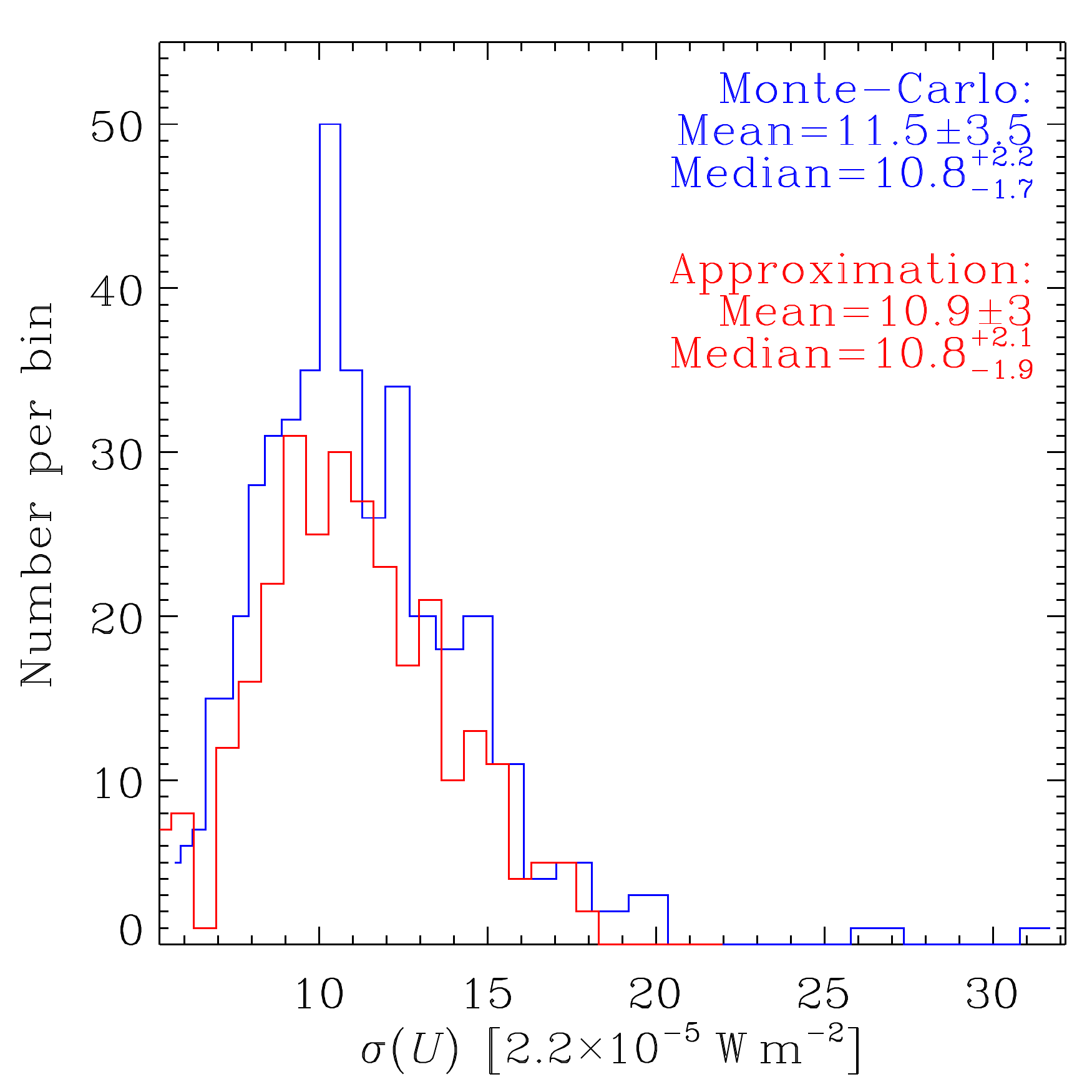} \\
  \end{tabular}
  \caption{{\sl Distribution of the main parameters of the perturbed SED 
            fits of the four R9 pixels (\reffig{fig:demoMC}).}
           For each parameter, we show the number of Monte-Carlo fits
           to the perturbed SED per bin of parameter value.
           The value of each parameter, for a given iteration is the combination 
           of the
           values of the parameter for each of the 4 pixels.
           We quote both the mean and the median.
           The blue distributions correspond to the rigorous Monte-Carlo 
           statistics, while the red distributions are from our reconstruction
           method (\refapp{ap:MC}).}
  \label{fig:dist_MC}
\end{figure*}
\reffig{fig:dist_MC} shows the statistical distributions of the main model parameters, corresponding to the fits of \reffig{fig:demoMC}.
The parameter value of each Monte-Carlo iteration is the sum of the parameter value of each pixel.
The first panel of this figure demonstrates, in particular, that the dust mass has a $\simeq50\,\%$
uncertainty, even for very high signal-to-noise ratio SEDs.
In addition it clearly shows the asymmetry of the distribution, which is a result of the non-linearity of the model.
The upper end of the statistical distribution of dust masses is less constrained than the lower end, because of the non-linear dependence of the dust mass with the temperature.

Even with a fast running model, computing $N_\sms{MC}\simeq300$ fits for each pixel is 
CPU intensive, and not necessary.
Instead, we use an approximation to reconstruct the probability distributions
of each parameter.
This method is detailed in \refapp{ap:MC}.
It consists of interpolating the pre-computed errors of 30 classes of SEDs, parametrized by their specific power ($L_\sms{IR}/M_\sms{dust}$) and the RMS level.
\reffig{fig:dist_MC} compares this reconstruction method to the rigorous Monte-Carlo results.
It succeeds \modif{in reproducing} the correct central value and errors.
In particular, it reproduces accurately the skewness of the probability distribution.

  \subsubsection{Error display}
  
Throughout this paper, each time a numerical quantity is reported as $X\simeq a\pm b$,
the two quantities $a$ and $b$ will refer to the mean and standard deviation, 
or in other words:
\begin{equation}
  X\simeq\langle X\rangle\pm\sigma_X.
  \label{eq:mean}
\end{equation}
On the other \modif{hand}, for numerical quantities having a strongly asymmetric
distribution, we will quote the error as $X\simeq a_{-b}^{+c}$. 
Only in this case will the quantities refer to the three quartiles, 
$Q_1(X)$, $Q_2(X)$, $Q_3(X)$, corresponding to values of the repartition function of 1/4, 1/2 and 3/4, respectively:
\begin{equation}
   X\simeq Q_2(X)_{\displaystyle-[Q_2(X)-Q_1(X)]}^{\displaystyle+[Q_3(X)-Q_2(X)]}.
   \label{eq:median}
\end{equation}

The latter error \refeqp{eq:median} corresponds to a confidence level of 
$50\;\%$, by definition.
It is also interesting to consider a higher confidence interval.
Most of our error estimates are based on $N_\sms{MC}=300$ Monte-Carlo 
iterations.
It would therefore be meaningless to go down to less than 
$1/\sqrt{N_\sms{MC}}\simeq6\;\%$ error tolerance.
We therefore choose to quote the $90\;\%$ confidence level, defined by the range of the parameter values between 0.05 and 0.95 of the repartition function.
From a technical point of view, with $N_\sms{MC}=300$, the limits of this interval are simply the 15$^{th}$ and 286$^{th}$ ordered Monte-Carlo parameter values. 
We will note this interval: 
\begin{equation}
  X \simeq [X_\sms{inf},X_\sms{sup}]_{90\;\%}.
  \label{eq:sup}
\end{equation}
On figures, the $50\;\%$ error bars will be displayed with a solid line, and 
the $90\;\%$ interval with a dashed line.

We note that taking the median \refeqp{eq:median} gives a central value very close to the best fit, while taking the mean \refeqp{eq:mean} gives a central
value systematically shifted from the best fit.
\reffig{fig:dist_MC} demonstrates this effect: the best fit value of the dust mass is $8.62\E{5}\msun$, very close to the median of the Monte-Carlo iterations 
($8.76\E{5}\msun$).
On the contrary, the mean, $1.03\E{6}\msun$, is significantly higher.
This is due to the skewness of the probability distribution of the parameter values.


\section{The Dust Mass Estimate and its Uncertainties}
\label{sec:discussion}

  \subsection{Systematic Discrepancies Between the Two Models}
  \label{sec:compmod}

\begin{figure*}[h!tbp]
  \centering
  \begin{tabular}{cc}
    \includegraphics[width=0.45\linewidth]{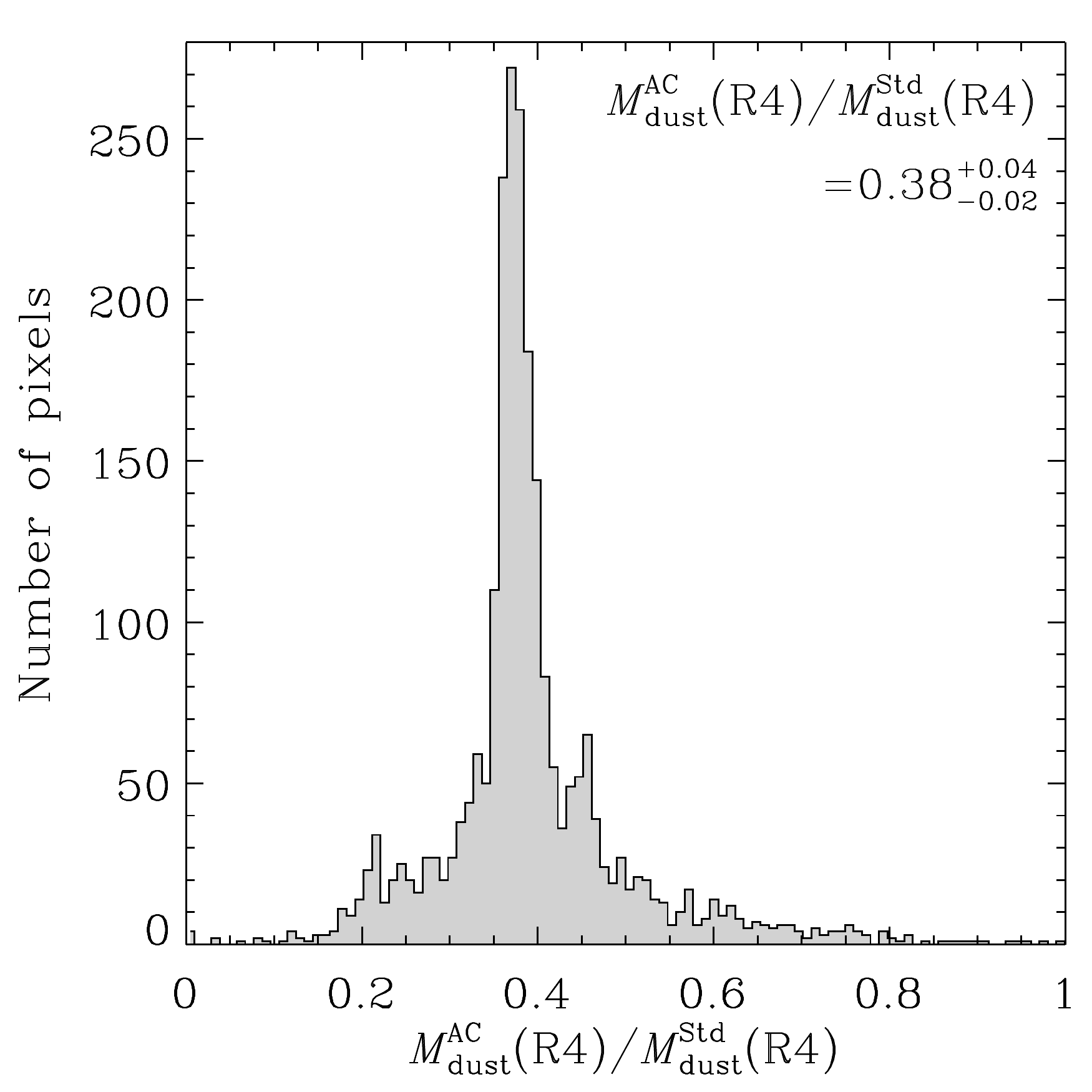} &
    \includegraphics[width=0.45\linewidth]{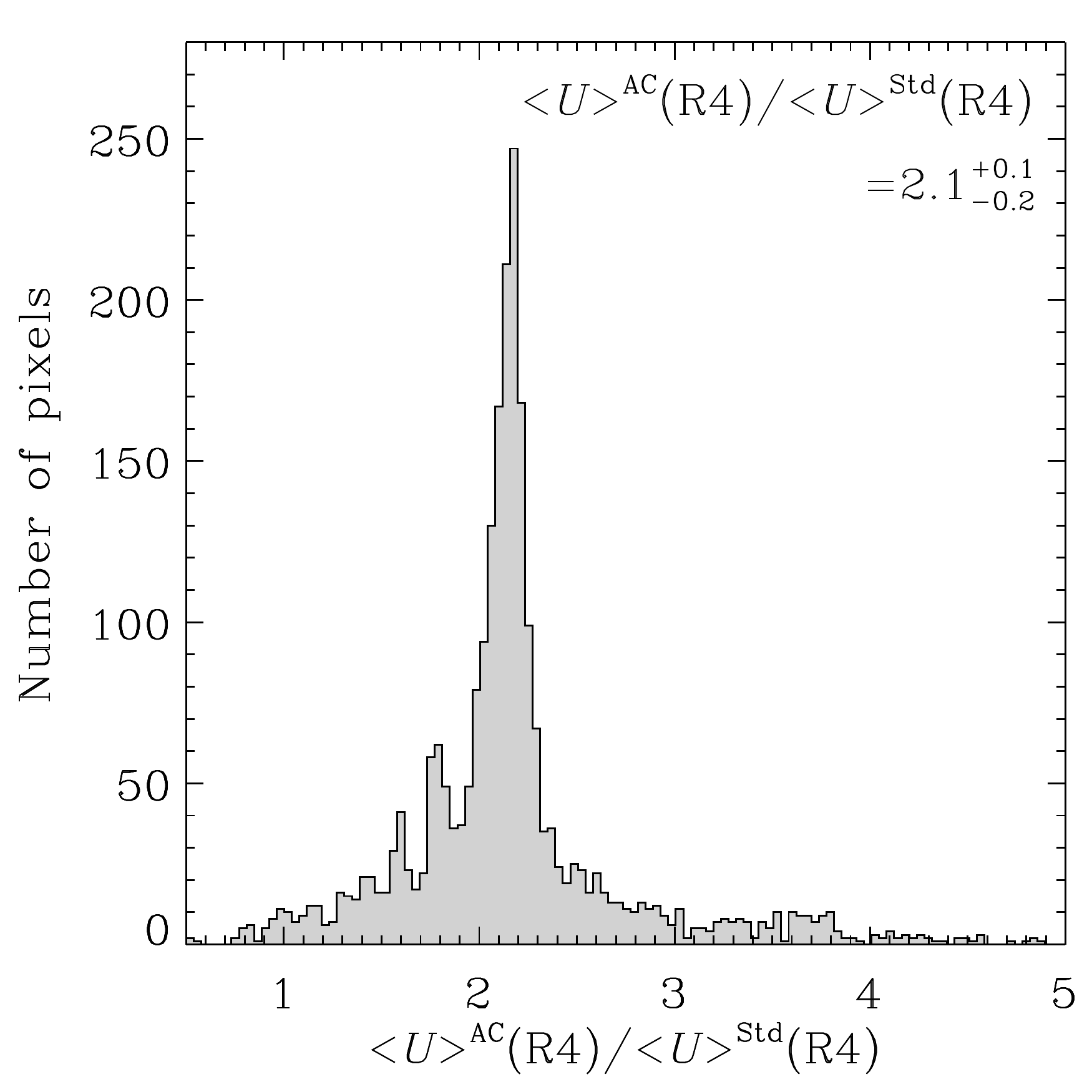} \\
  \end{tabular}
  \caption{{\sl Comparison between the parameters obtained with the two 
            models}.
            The {\it left panel}
            shows the statistical distribution of the pixel-to-pixel
            dust mass ratio between the \citengl{AC model} and the 
            \citengl{standard model}
            at spatial resolution R4.
            The {\it right panel} shows a similar ratio for the mass averaged
            starlight intensity $\langle U\rangle$.}
  \label{fig:comparison}
\end{figure*}
We have applied the model presented in \refsec{sec:model}, with the two compositions (\citengl{standard} and \citengl{AC}), to the maps listed in \reftab{tab:dolls}.
The dust mass spatial distribution obtained with the two models are very similar.
\reffig{fig:comparison} shows the distribution of the pixel-to-pixel ratio 
of the dust masses (\citengl{standard} over \citengl{AC} models), and starlight intensities,
at spatial resolution R4.
These distributions are very tight.
It shows, that the dust masses obtained with the \citengl{AC model} are systematically
lower by a factor of $0.38_{-0.02}^{+0.04}$ than dust masses obtained with the
\citengl{standard model}, while the starlight intensities are systematically higher
by a factor of $2.1_{-0.2}^{+0.1}$.
These parameters are tied together.
Grains with the \citengl{AC model} absorb more light than grains of the \citengl{standard model} (\refapp{ap:mie}). 
They therefore require less mass to account for the observed IR luminosity.

Thus, the results derived with the two compositions give similar trends, but the absolute value of their parameters systematically differ.

  \subsection{Bias Originating in the Lack of Spatial Resolution}
  \label{sec:resolution}

\begin{figure*}[h!tbp]
  \centering
  \begin{tabular}{cc}
    \includegraphics[width=0.47\linewidth]{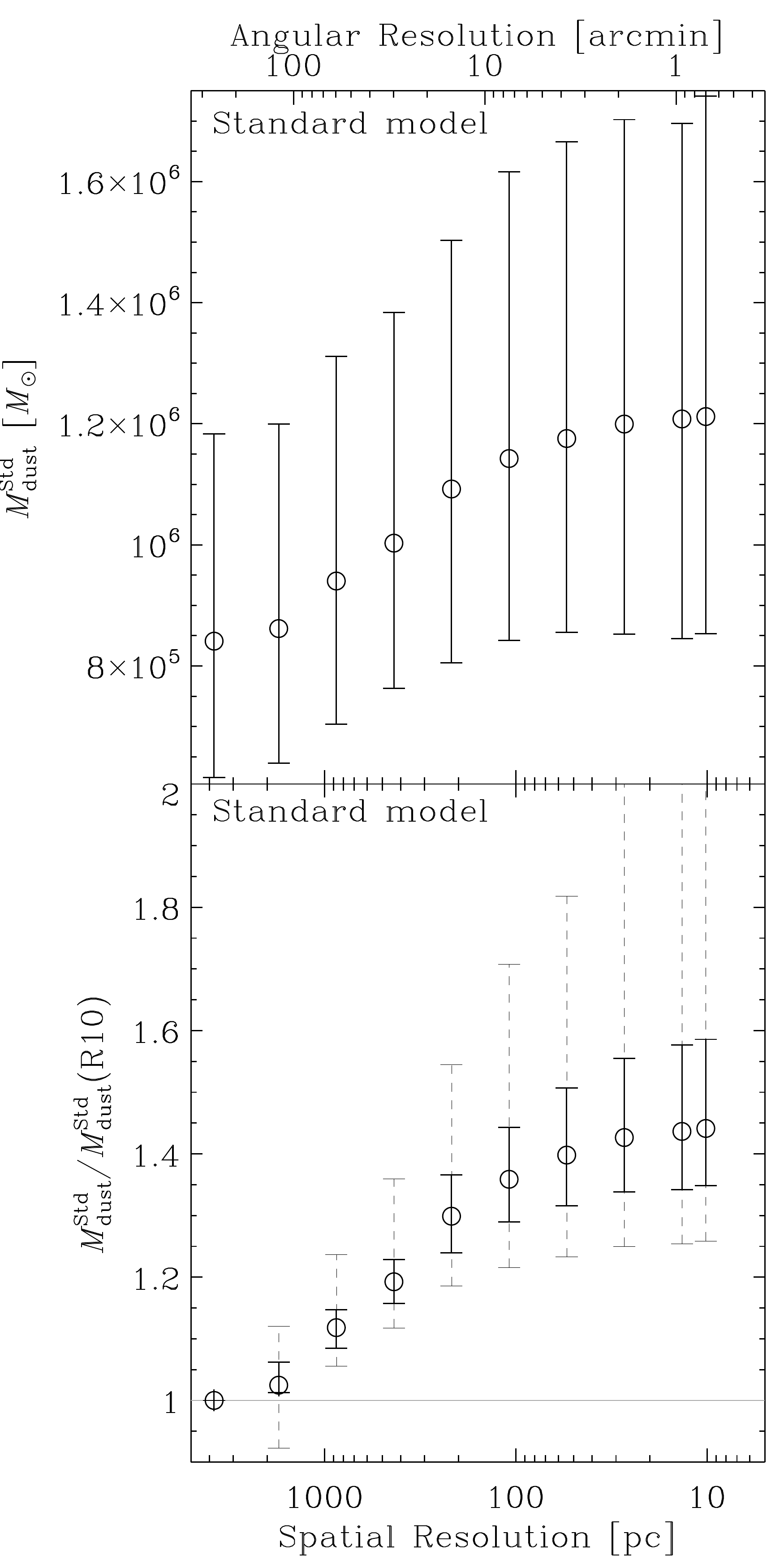} &
    \includegraphics[width=0.47\linewidth]{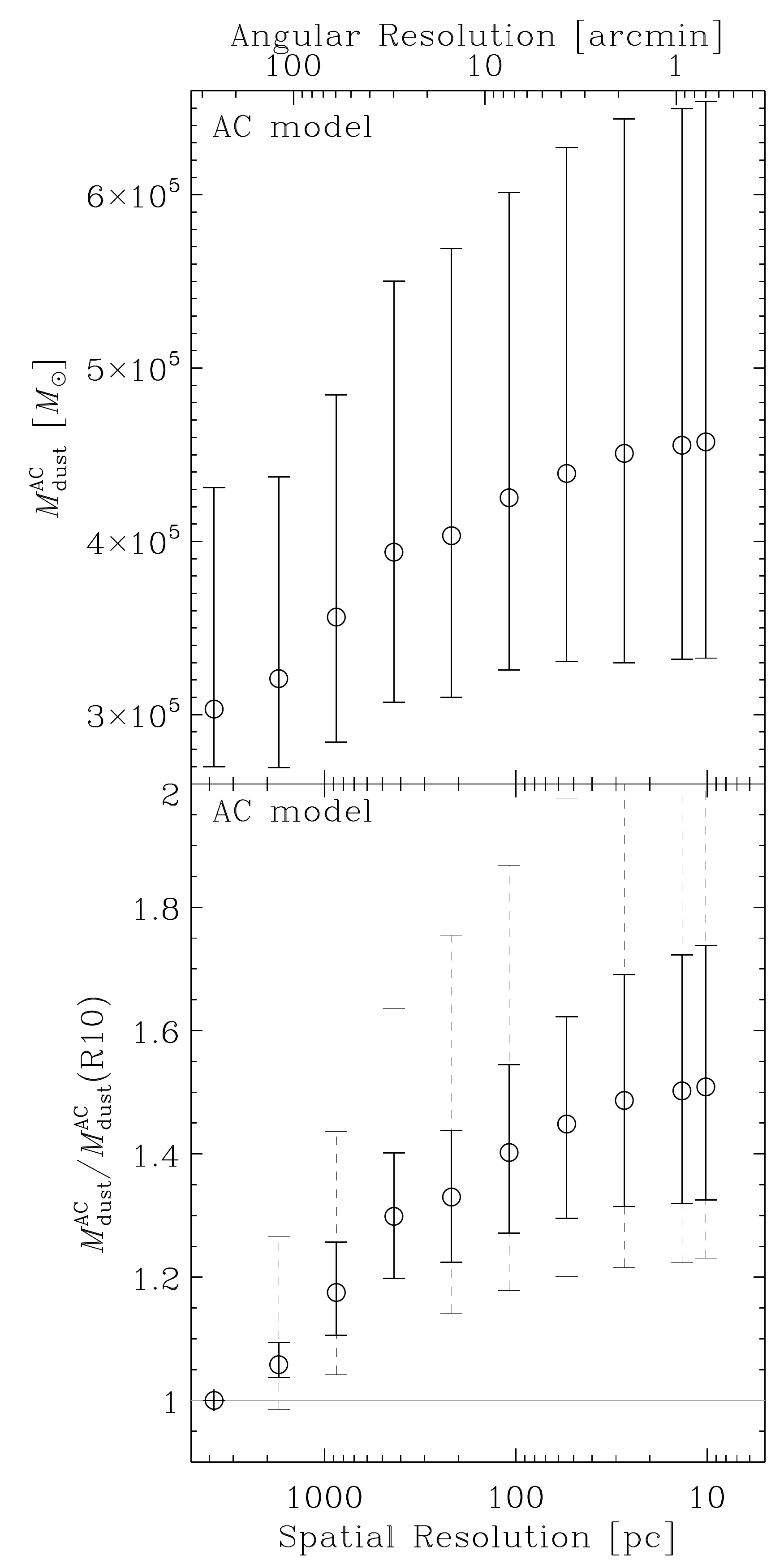} \\
  \end{tabular}
  \caption{{\sl Trend of the total dust mass with spatial resolution.}
            The {\it left panels} show the trends for the \citengl{standard 
            model}, while 
            the {\it right panels} show the trends for the \citengl{AC model}. 
            For each panel, the $x$-axis is the spatial resolution of the maps
            used to derive the dust mass.
            Each point of the trends \modif{corresponds} to one of the maps listed 
            in \reftab{tab:dolls}.
            For the two {\it top panels}, the $y$-axis is the total dust mass.
            This mass is the sum of the dust mass of each pixel.
            For each resolution, there are as many SED fits as the number of 
            pixels listed in \reftab{tab:dolls}.
            The two {\it bottom panels} show the relative dust mass variation.
            It is normalized by the integrated strip (R10).
            In that way, the calibration errors cancel, and the trend has 
            smaller error bars.
            The dashed error bars display the $90\;\%$ confidence interval.}
  \label{fig:Md_doll}
\end{figure*}
In order to quantify the effect of the non-linearity of our SED model, we
compute the total dust mass for each map listed in \reftab{tab:dolls},
with our two grain compositions.
\reffig{fig:Md_doll} shows the resulting trends of dust mass with spatial
resolution.
These trends are shown for both models.
The mass at each resolution is the sum of the masses of all defined pixels.
The top panels show the trend of the absolute value of the dust mass.
The trends look similar for both models.
They appear systematically shifted as discussed in \refsec{sec:compmod}.
The error bar on each value is large and covers roughly the range of the 
trend.
However, the relative variation of the dust mass (bottom panels of \reffig{fig:Md_doll}) has significantly smaller error bars.
These trends are obtained by normalizing each one of the $N_\sms{MC}$ Monte-Carlo results of a given spatial resolution, by the corresponding mass for the integrated strip (R10).
In that way, most of the calibration error cancels.
The only remaining source of uncertainty is the intercalibration error between the various instruments and the RMS noise.
The dynamics of the trend is unchanged.

The relative trends of \reffig{fig:Md_doll} show that measuring the dust mass of an integrated galaxy can give significantly lower values than performing 
fits of its spatially resolved regions, providing that the spatial resolution is fine enough.
For the LMC, the spatially resolved dust mass estimate is $\simeq50\,\%$
higher than the integrated SED fit.
This variation is not due to the noise, since the error bars are much smaller than the spread of the trend.
Thus, the true mass is probably closer to the high spatial resolution value (R1) than to the integrated flux value (R10).
It appears that there is a spatial scale where the trend stabilizes.
This transition is an optimal resolution for our model, as
it \modif{probably provides} a correct dust mass, and the noise is lower than at R1.
This optimal spatial scale corresponds to R3--R4 ($\simeq27-54$~pc; 
\reftab{tab:dolls}).

The origin of this trend likely lies in the morphology of the ISM.
It could be the result of the dilution of cold massive regions in
hotter regions.
It is possible that there is a typical spatial scale below which most of the cold regions dominate the SED of the pixels where they lie. 
It may correspond to the typical scale of molecular complexes.
With only a few far-IR/submm constraints, it is difficult to account for 
these regions when modelling the integrated SED.

\begin{figure}[h!tbp]
  \centering
  \includegraphics[width=0.95\linewidth]{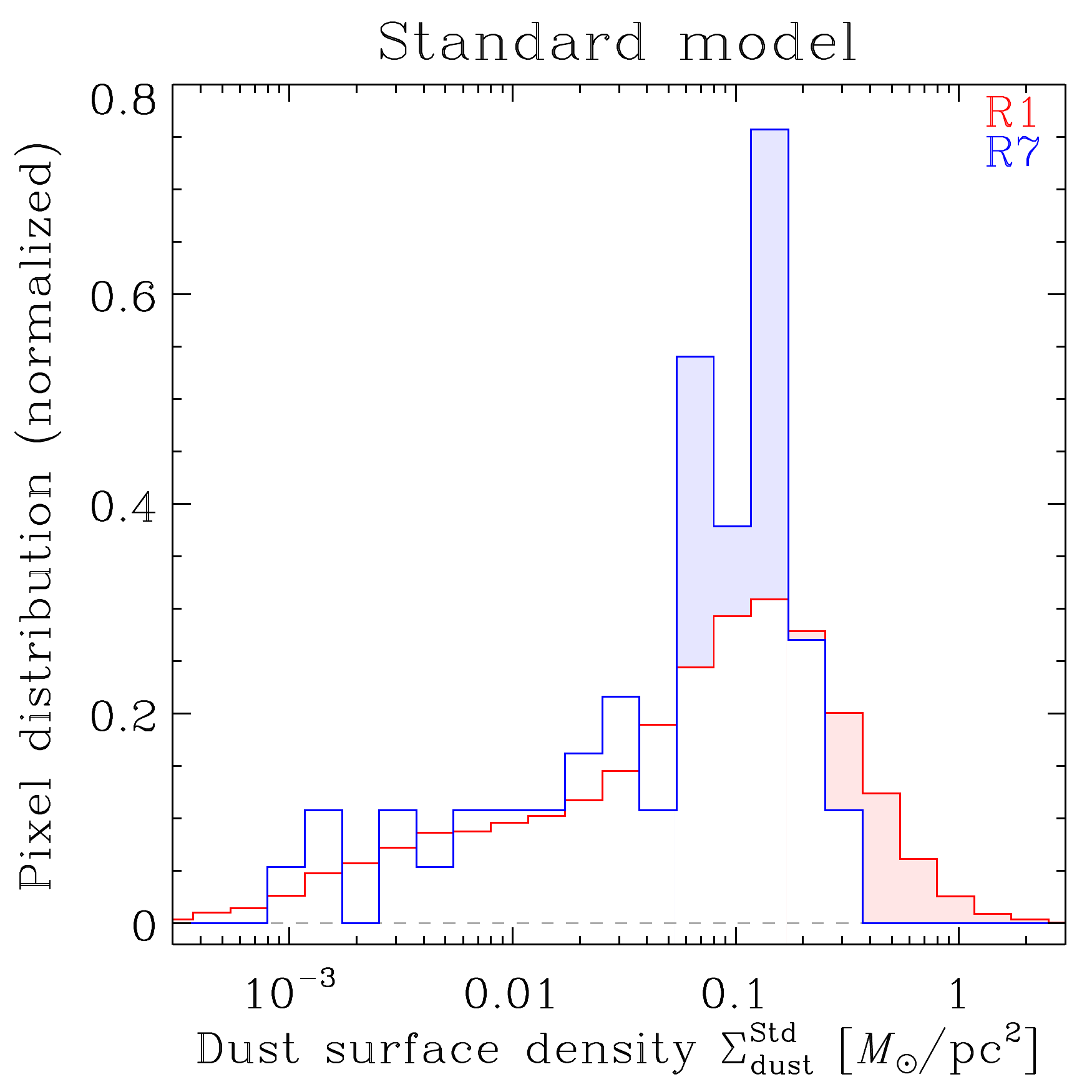}
  \caption{{\sl Pixel distribution of the dust mass surface density for two
            spatial resolutions.}
            We plot only the \citengl{standard model}.
            The figure is qualitatively similar for the \citengl{AC model}.
            The distributions are normalized.
            It shows that, at lower spatial resolution, the very high surface 
            densities are missed (red filled area), and there is an excess 
            of intermediate surface densities (blue filled area).}
  \label{fig:compdistdoll}
\end{figure}
To demonstrate the origin of this trend, \reffig{fig:compdistdoll} compares
the pixel distribution of the dust mass surface density, at two spatial resolutions. 
Taking into account the statistical fluctuations, due to the low pixel number of R7, the two distributions (R1 and R7) agree at low dust mass surface densities 
($\Sigma_\sms{dust}^\sms{Std}\lesssim0.06\msun\,\rm pc^{-2}$).
However, at high dust mass surface densities ($\Sigma_\sms{dust}^\sms{Std}\gtrsim0.2\msun\,\rm pc^{-2}$), there is a higher fraction of pixels for the high resolution map.
This fraction is compensated by an excess of intermediate surface density pixels at low spatial resolution ($0.06\msun\,{\rm pc}^{-2}\lesssim\Sigma_\sms{dust}^\sms{Std}\lesssim0.2\msun\,\rm pc^{-2}$).
Thus, the origin of the dust mass underestimation is due to the inability of our model to probe dense regions, at low spatial resolutions.
\modif{At these spatial resolutions, the mass and average temperature of cold regions are biased by the contribution from hot regions present in the same beam, as the latter are more emissive.
On the other hand, at high spatial resolution, cold regions tend to be better separated from hot regions and can therefore be modelled more accurately.}

  \subsection{The Gas-to-Dust Mass Ratio Crisis: Several Competitive 
              Scenarios}
  \label{sec:darkCO}

  \subsubsection{Metal Abundance Constraints}
  
To be rigorous, we first need to estimate the uncertainty on the metallicity of the LMC.
\citet{pagel03} compiles the literature for numerous \modif{elemental} abundances in different regions of the LMC and of the Galaxy.
We estimate the error on each element of Table~1 of \citet{pagel03}, by taking the dispersion of the different measures, excluding cepheids which are very dispersed and not available for all elements.
These uncertainties are summarized in \reftab{tab:Z}.
\begin{table}[h!tbp]
  \centering
  \begin{tabular}{ll*{3}{r}}
    \hline\hline
      & & \multicolumn{1}{c}{O} & \multicolumn{1}{c}{C} & \multicolumn{1}{c}{N} \\
    \hline
      $12+\log\left(X/H\right)$ & Galaxy
        & $8.6-8.7$ & $8.4-8.6$ & $7.7-7.9$ \\
      & LMC
        & $8.4-8.5$ & $7.1-8.1$ & $6.9-7.7$ \\
    \hline
  \end{tabular}
  \caption{{\sl Literature compilation for the elemental abundances.}
           The range of values reflects the dispersion of the observations
           listed in Table~1 of \citet{pagel03}.}
  \label{tab:Z}
\end{table}

We assume that we can reliably derive the metallicity by scaling the mass of O, C and N.
In particular, this is supported by the fact that O and C are the major dust constituents.
The metallicity of the LMC we adopt is thus:
\begin{eqnarray}
  \frac{Z_\sms{LMC}}{Z_\odot} &
  = & \frac{16\left(\frac{\mbox{O}}{\mbox{H}}\right)_\sms{LMC} 
           +12\left(\frac{\mbox{C}}{\mbox{H}}\right)_\sms{LMC}
           +14\left(\frac{\mbox{N}}{\mbox{H}}\right)_\sms{LMC}}
           {16\left(\frac{\mbox{O}}{\mbox{H}}\right)_\odot 
           +12\left(\frac{\mbox{C}}{\mbox{H}}\right)_\odot
           +14\left(\frac{\mbox{N}}{\mbox{H}}\right)_\odot} \nonumber\\
    & = & 0.47_{-0.07}^{+0.06}.
  \label{eq:Z}
\end{eqnarray}
This estimate is more accurate than simply scaling the oxygen abundance, as it is usually done.

\begin{figure}[h!tbp]
  \centering
  \includegraphics[width=0.95\linewidth]{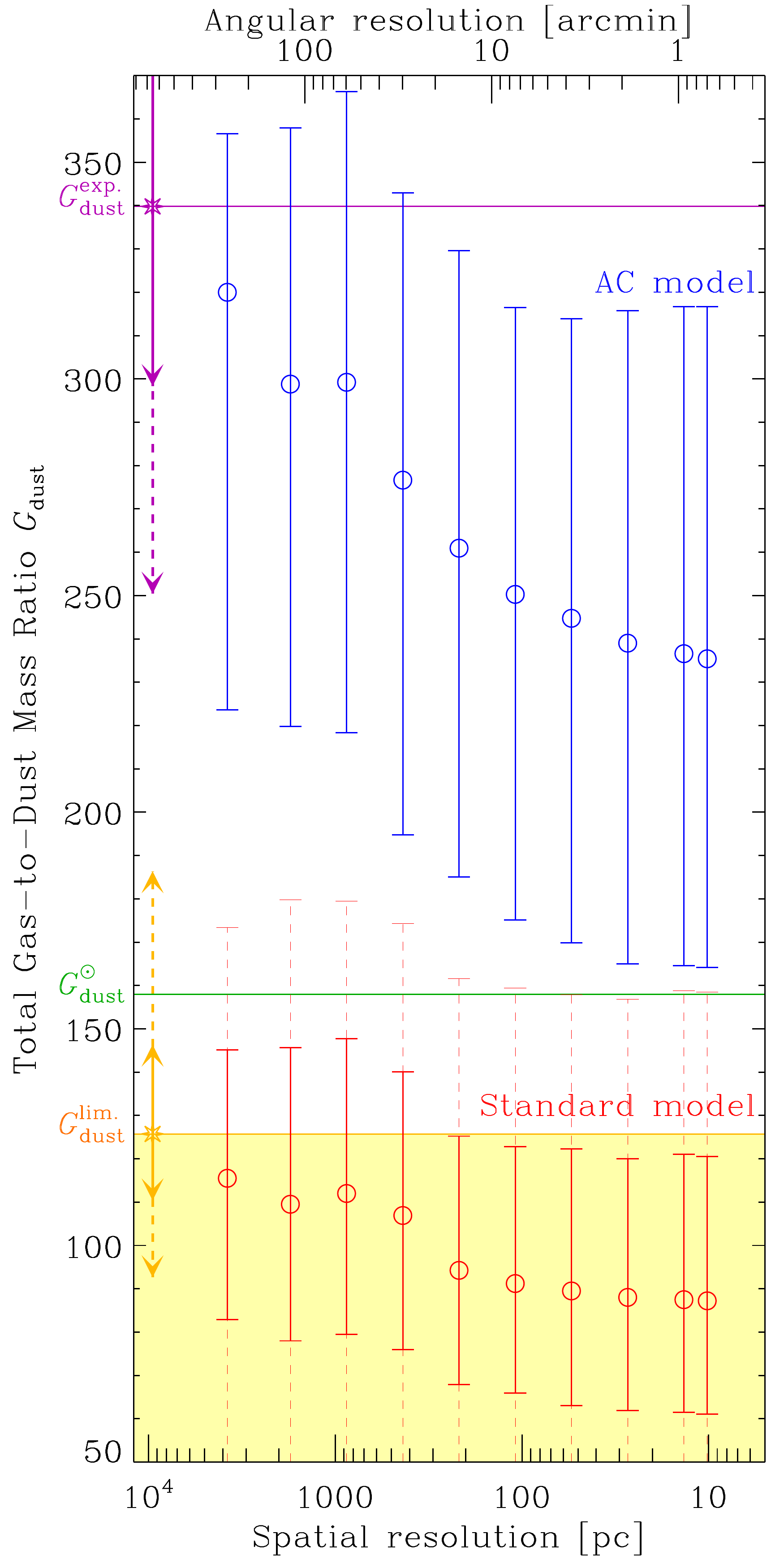}
  \caption{{\sl Consistency of the gas-to-dust mass ratios.}
           The two trends show the gas-to-dust mass ratio as a function of the
           spatial resolution, for each model.
           We also display the $90\;\%$ confidence interval for the 
           \citengl{standard model}, with dashed lines.
           The yellow and purple error bars, with the star symbol, represent 
           the uncertainities on 
           $G_\sms{dust}^\sms{lim.}$ and $G_\sms{dust}^\sms{exp.}$, 
           respectively (\reftab{tab:Z}; solid line: $50\,\%$; 
           dashed line: $90\,\%$).
           The purple, green and yellow solid lines show the central values of 
           $G_\sms{dust}^\sms{exp.}$, $G_\sms{dust}^\odot$
           and $G_\sms{dust}^\sms{lim.}$, respectively.
           The yellow filled area represents the gas-to-dust mass ratio range
           below
           the central value of the limit ratio $G_\sms{dust}^\sms{lim.}$.}
  \label{fig:G2D}
\end{figure}
Let's check the physical consistency of our dust masses.
The Galactic gas-to-dust mass ratio is $G_\sms{dust}^\odot\simeq158$ 
\citep{zubko04}.
We emphasize that this value is consistent with the elemental depletion patterns.
Assuming that the dust-to-gas mass ratio scales with metal abundance (\ie\
the dust-to-metal mass ratio is constant), the expected gas-to-dust mass ratio 
for the LMC is:
\begin{equation}
  G_\sms{dust}^\sms{exp.}\simeq \frac{G_\sms{dust}^\odot}{Z_\sms{LMC}/Z_\odot}
    \simeq339_{-41}^{+55}.
  \label{eq:Gexp}
\end{equation}
Assuming that the mass fraction of gaseous heavy elements in the Galaxy is 
$Z_\odot\simeq0.017$ \citep{grevesse98}, 
the solar metallicity dust-to-metal mass ratio is:
\begin{equation}
  \mathcal{D}_\odot\simeq \frac{1}{G_\sms{dust}^\odot Z_\odot}
    \simeq 0.37.
  \label{eq:D2M}
\end{equation}
It is therefore difficult to understand how the gas-to-dust mass ratio in the 
LMC could be lower than:
\begin{equation}
  G_\sms{dust}^\sms{lim.}\simeq G_\sms{dust}^\sms{exp.}\times\mathcal{D}_\odot
    \simeq125_{-15}^{+20},
  \label{eq:Glim}
\end{equation}
without requiring a larger amount of metals locked-up in grains than what is
available in the ISM.
These values are summarized in \reftab{tab:refG2D}.
\begin{table}[h!tbp]
  \centering
  \begin{tabular}{lrr}
    \hline\hline
      Quantity & \multicolumn{2}{c}{Value \&\ Uncertainties}  \\
    \hline
      $Z_\sms{LMC}/Z_\odot$ 
        & $0.47_{-0.07}^{+0.06}$ & $[0.31,0.63]_{90\;\%}$ \\
      $G_\sms{dust}^\sms{exp.}$ 
        & $339_{-41}^{+55}$ & $[250,503]_{90\;\%}$ \\
      $G_\sms{dust}^\sms{lim.}$
        & $125_{-15}^{+20}$ & $[92,186]_{90\;\%}$ \\
    \hline
  \end{tabular}
  \caption{{\sl Reference metallicity and gas-to-dust mass ratios.}
           These values are made explicit in \refeqs{eq:Z}, (\ref{eq:Gexp}) 
           and (\ref{eq:Glim}).
           The convention for error display is defined in \refeqs{eq:median} 
           and (\ref{eq:sup}).}
  \label{tab:refG2D}
\end{table}

  \subsubsection{Preliminary: Global Analysis}
  \label{sec:global}

\begin{table}[h!tbp]
  \centering
  \begin{tabular}{l|rr|rr}
    \hline\hline
      Resolution 
        & \multicolumn{4}{c}{Gas-to-Dust Mass Ratio $G_\sms{dust}$} \\
        & \multicolumn{2}{c}{\citengl{Standard model}}
        & \multicolumn{2}{c}{\citengl{AC model}} \\
    \hline
    R1   & $87_{-26}^{+33}$ & $[$$38$,$159$$]_{90\;\%}$ & $235_{-71}^{+81}$ & $[$$86$,$410$$]_{90\;\%}$ \\
    R2   & $87_{-26}^{+34}$ & $[$$39$,$159$$]_{90\;\%}$ & $237_{-72}^{+80}$ & $[$$86$,$410$$]_{90\;\%}$ \\
    R3   & $88_{-26}^{+32}$ & $[$$38$,$157$$]_{90\;\%}$ & $239_{-74}^{+77}$ & $[$$84$,$410$$]_{90\;\%}$ \\
    \hline
    R4   & $89_{-26}^{+33}$ & $[$$39$,$158$$]_{90\;\%}$ & $245_{-75}^{+69}$ & $[$$78$,$400$$]_{90\;\%}$ \\
    R5   & $91_{-25}^{+32}$ & $[$$40$,$160$$]_{90\;\%}$ & $250_{-75}^{+66}$ & $[$$78$,$400$$]_{90\;\%}$ \\
    R6   & $94_{-26}^{+31}$ & $[$$42$,$162$$]_{90\;\%}$ & $261_{-76}^{+68}$ & $[$$81$,$410$$]_{90\;\%}$ \\
    R7   & $107_{-31}^{+33}$ & $[$$47$,$174$$]_{90\;\%}$ & $277_{-82}^{+66}$ & $[$$85$,$420$$]_{90\;\%}$ \\
    R8   & $112_{-33}^{+36}$ & $[$$49$,$179$$]_{90\;\%}$ & $299_{-81}^{+67}$ & $[$$90$,$440$$]_{90\;\%}$ \\
    R9   & $110_{-32}^{+36}$ & $[$$47$,$180$$]_{90\;\%}$ & $299_{-79}^{+59}$ & $[$$88$,$430$$]_{90\;\%}$ \\
    R10   & $115_{-33}^{+30}$ & $[$$50$,$173$$]_{90\;\%}$ & $320_{-100}^{+40}$ & $[$$91$,$430$$]_{90\;\%}$ \\
    \hline
  \end{tabular}
  \caption{{\sl Total gas-to-dust mass ratio, as a function of the spatial 
            resolution, for the two models.}
            This gas-to-dust mass ratio is the ratio of the total gas mass
            over the total dust mass, and not the averaged $G_\sms{dust}$
            over the map.
            That is the reason why the high spatial resolutions are defined.
            The errors account for the uncertainties on both the dust 
            and the gas masses.
            We give also the $90\;\%$ confidence interval.
            The convention for error display is defined in \refeqs{eq:median} 
            and (\ref{eq:sup}).}
  \label{tab:G2D}
\end{table}
\reftab{tab:G2D} shows the gas-to-dust mass ratio at each spatial resolution
for the two models.
These ratios are displayed \modif{in} \reffig{fig:G2D}.
The total gas-to-dust mass ratios given by the \citengl{standard model} 
are too low by a factor of 
$G_\sms{dust}^\sms{exp.}/G_\sms{dust}^\sms{Std}({\rm R1})\simeq3.8_{-1.0}^{+1.7}\simeq[2.1,9.7]_{90\;\%}$.
They are even lower than the physical limit by a factor of 
$G_\sms{dust}^\sms{lim.}/G_\sms{dust}^\sms{Std}({\rm R1})\simeq1.4_{-0.4}^{+0.6}\simeq[0.8,3.6]_{90\;\%}$.
We emphasize here that there is no spatial correlation between the foreground
Galactic \hi\ column density (\refsec{sec:herschel}) and the gas-to-dust mass
ratio deficit of the \citengl{standard model}.
Therefore, the residual cirrus emission is not responsible for this deficit.
Statistically, the \citengl{standard model} violates the elemental abundances with a probability of $80\;\%$, while the \citengl{AC model} is consistent.
Therefore, if we assume that our gas mass is correct, then we can conclude that the properties of the \citengl{standard model} do not apply to the LMC.
This is \modif{one} scenario.

However, there is a second scenario:
the discrepant gas-to-dust mass ratio obtained with the 
\citengl{standard model} could 
result from the underestimate of the total gas mass.
In particular, the mass of molecular gas could have been underestimated.
It is known that in low-metallicity environments, the \hmol\ gas is not properly traced by \COio.
In these environments, the CO cores are thought to be much smaller relative to their \hmol\
envelope (\hmol\ being more efficiently self-shielded).
This scenario is supported by the exceptionally high observed \ciiline/\COio\ luminosity ratio in dwarf galaxies \citep[e.g.\ $\mbox{\ciiline/\COio}\simeq 20\,000$ in the LMC, compared to $\mbox{\ciiline/\COio}\simeq4\,000$ in normal metallicity galaxies;][]{poglitsch95,israel97,madden97,madden00}.
In principle, the underestimation of the molecular gas mass, using \modif{the} CO line, could be a factor of $\simeq10-100$, in these environments \citep{madden11}.

Let's assume that our dust properties are correct, and that 
$G_\sms{dust}^\sms{Std}$ is very low due to having underestimated the molecular gas mass.
Noting $\mathcal{C}_\sms{\hmol}$ the correction factor accounting for the hypothetical molecular gas not traced by CO and for the fact that our 
$X_\sms{CO}$ conversion factor (\refsec{sec:gas}) might be wrong, the total gas-to-dust mass ratio would have to be:
\begin{eqnarray}
  \frac{M_\sms{gas}^\sms{\hi}+\mathcal{C}_\sms{\hmol}M_\sms{gas}^\sms{\hmol}}{M_\sms{dust}^\sms{Std}({\rm R1})}
  & = & G_\sms{dust}^\sms{exp.} \\
  \Rightarrow
  \mathcal{C}_\sms{\hmol} & \simeq & 10.1_{-3.6}^{+5.9} = [4.0,30.6]_{90\;\%}. \nonumber
\end{eqnarray}
In other words, to explain the discrepant gas-to-dust mass ratio of the \citengl{standard model}, we would have to conclude that the molecular gas mass would have been globally underestimated by at least one order of magnitude.

These two alternative scenarios are degenerate, when considering only global values.
We therefore need to take into account the redundancy provided by the spatial distribution of the gas-to-dust mass ratio, in order to sort these scenarios out.

\begin{figure}[h!tbp]
  \centering
  \begin{tabular}{cc}
     \includegraphics[width=0.95\linewidth]{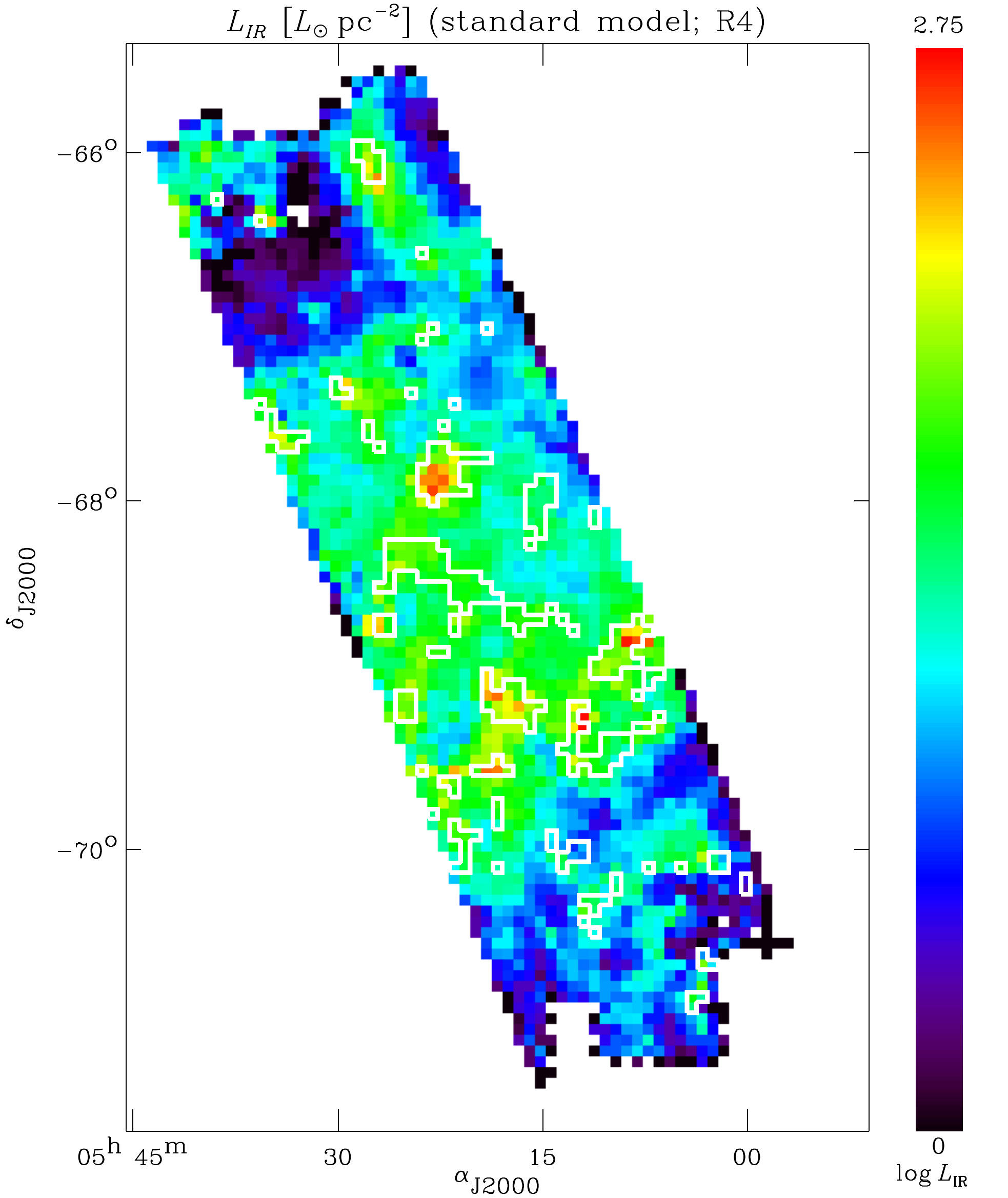}
  \end{tabular}
  \caption{{\sl Spatial distribution of the IR luminosity, 
            for the \citengl{standard model},} at resolution R4 (54~pc).
            The color image represents the IR luminosity map.
            The color scale is logarithmic.
            The white contours show the main CO concentrations from the Nanten 
            map \citep{fukui08}.
            This contour level is chosen so that $90\;\%$ of the CO mass
            has a higher column density than this level.
            In other words, $90\;\%$ of the CO mass is in these concentrations.
            The map is almost rigorously identical with the \citengl{AC model}.}
  \label{fig:mapLIR}
\end{figure}

  \subsubsection{Spatial Distribution of the Gas-to-Dust Mass Ratio}

\begin{figure*}[h!tbp]
  \centering
  \begin{tabular}{cc}
    \includegraphics[width=0.47\linewidth]{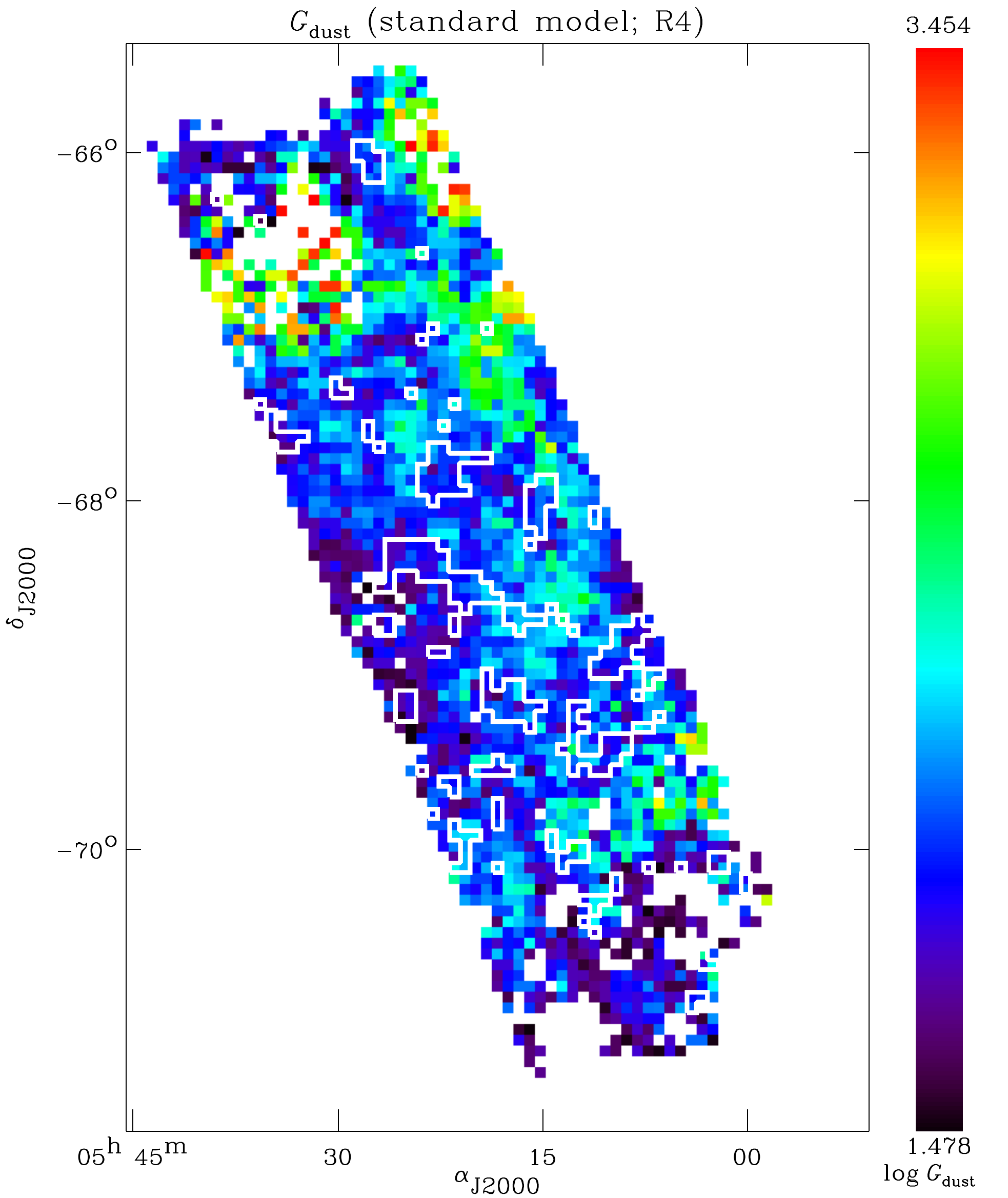} &
    \includegraphics[width=0.47\linewidth]{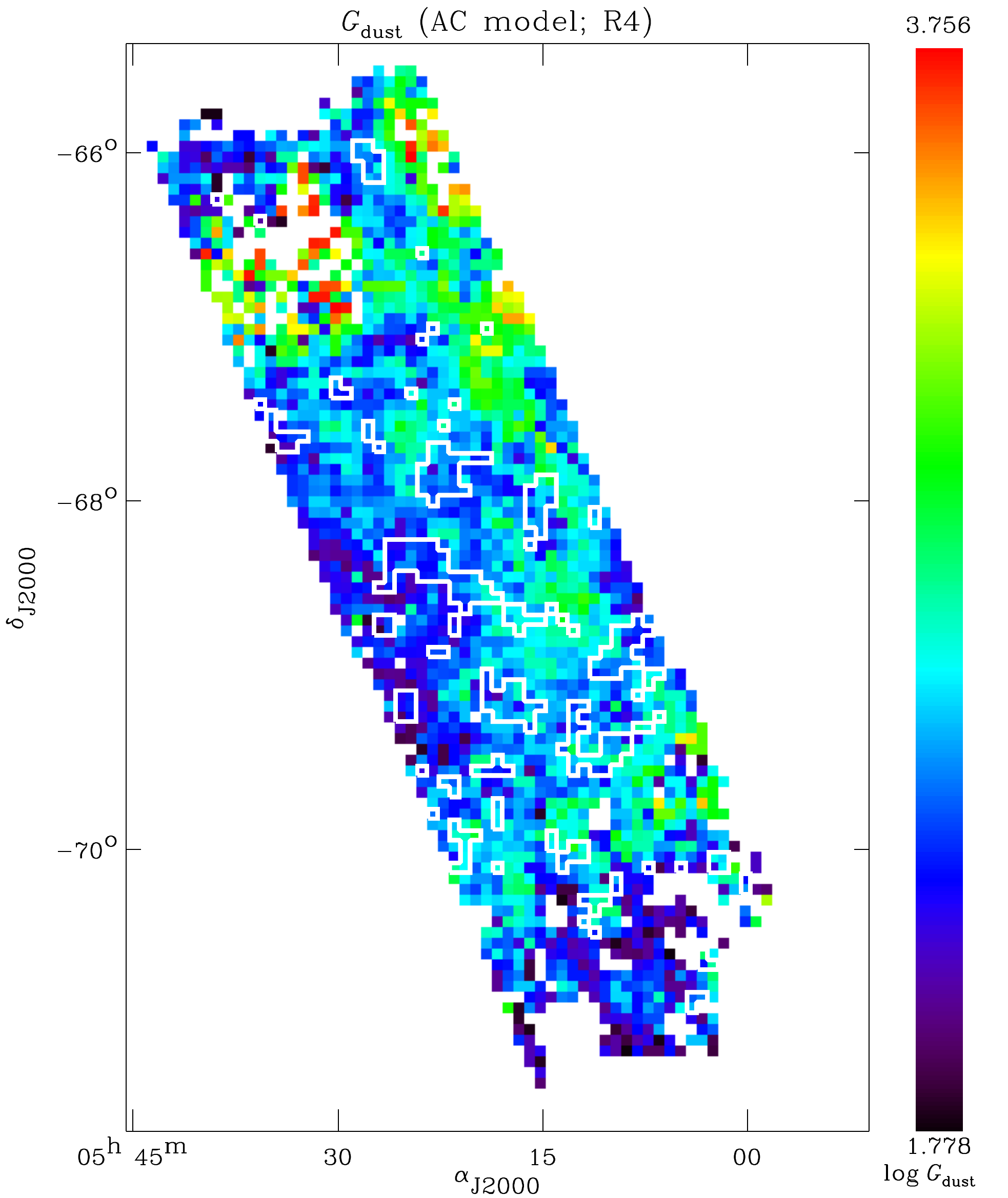} \\
    \includegraphics[width=0.47\linewidth]{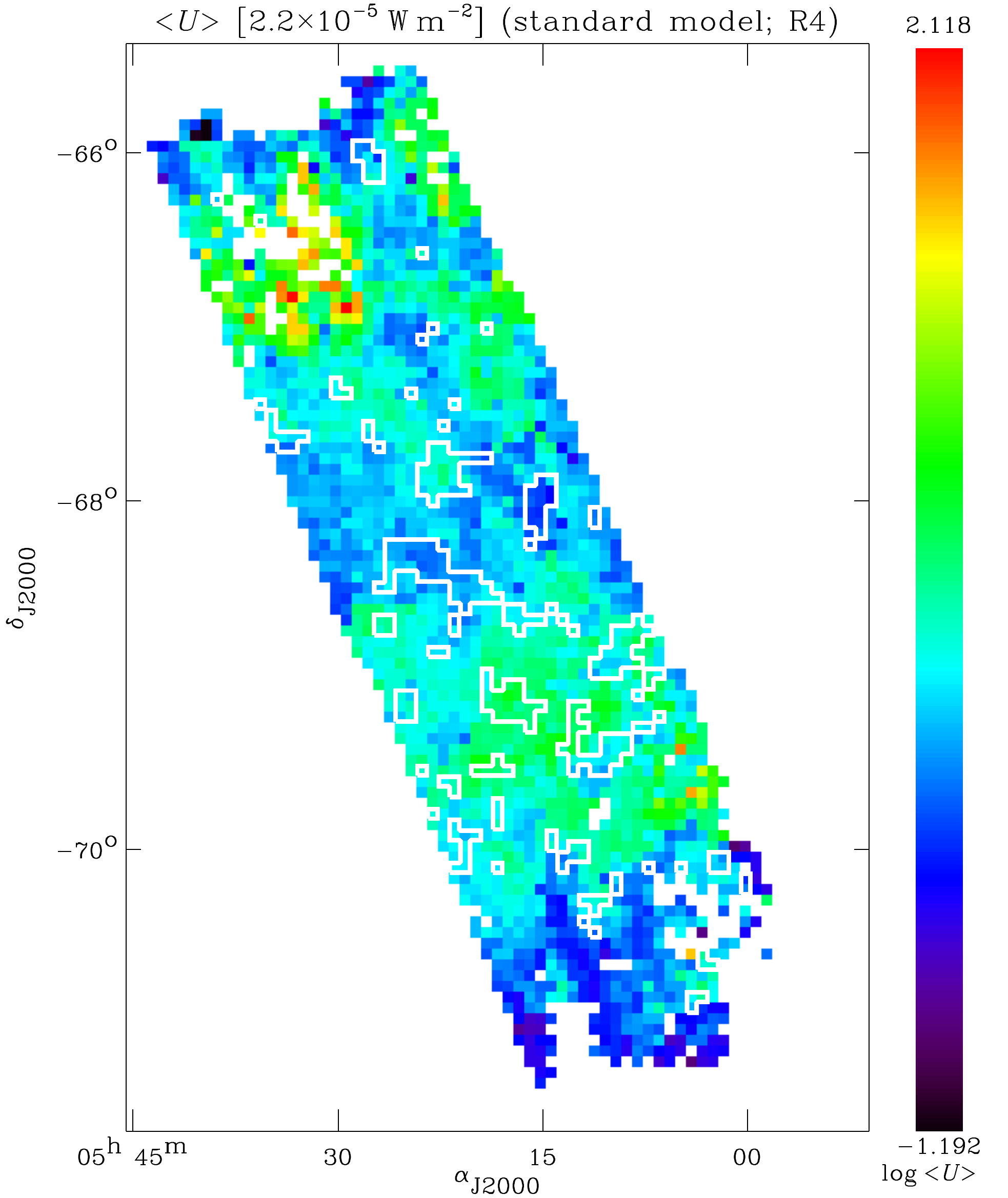} &
    \includegraphics[width=0.47\linewidth]{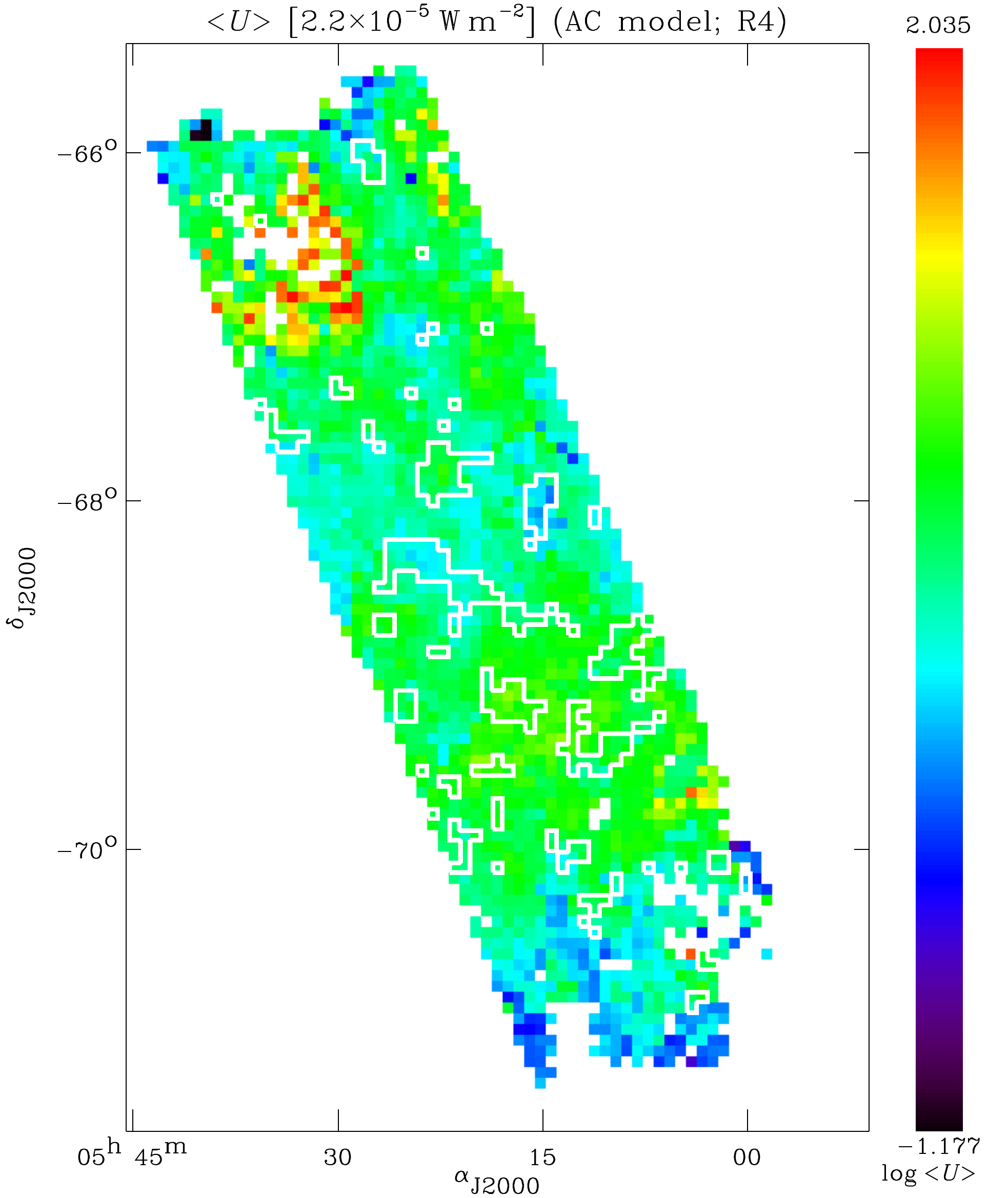} \\
  \end{tabular}
  \caption{{\sl Spatial distribution of the main dust parameters, 
            for the two models,} at resolution R4 (54~pc).
            The color images of the two upper panels represent 
            the gas-to-dust mass ratio map, for each model.
            The color images of the two lower panels represent 
            the mass averaged starlight intensity map, for each model.
            The color scale is logarithmic.
            The white contours show the main CO concentrations from the Nanten 
            map \citep{fukui08}.
            This contour level is chosen so that $90\;\%$ of the CO mass
            has a higher column density than this level.
            In other words, $90\;\%$ of the CO mass is in these concentrations.}
  \label{fig:mapG2D}
\end{figure*}
\reffig{fig:mapLIR} shows the map of IR luminosity for the \citengl{standard model}.
Since this quantity is the integration of the interpolated observed SED, it 
depends very little on the model; \modif{$L_\sms{IR}$ distribution of the \citengl{standard model} looks almost identical to the \citengl{AC model}.}
That is the reason why we displayed it only for the \citengl{standard model}.
The upper panels of \reffig{fig:mapG2D} show the spatial distribution of the gas-to-dust mass ratio with the two models.
A few regions at the two ends of the strip exhibit noisy pixels. 
They correspond to very low surface densities.
In general, there is no particular spatial correlation between the gas-to-dust 
mass ratio and the CO concentrations.
The lower panels of \reffig{fig:mapG2D} show the corresponding map of the mass averaged starlight intensity $\langle U\rangle$, for both models.

\begin{figure*}[h!tbp]
  \centering
  \includegraphics[width=0.95\linewidth]{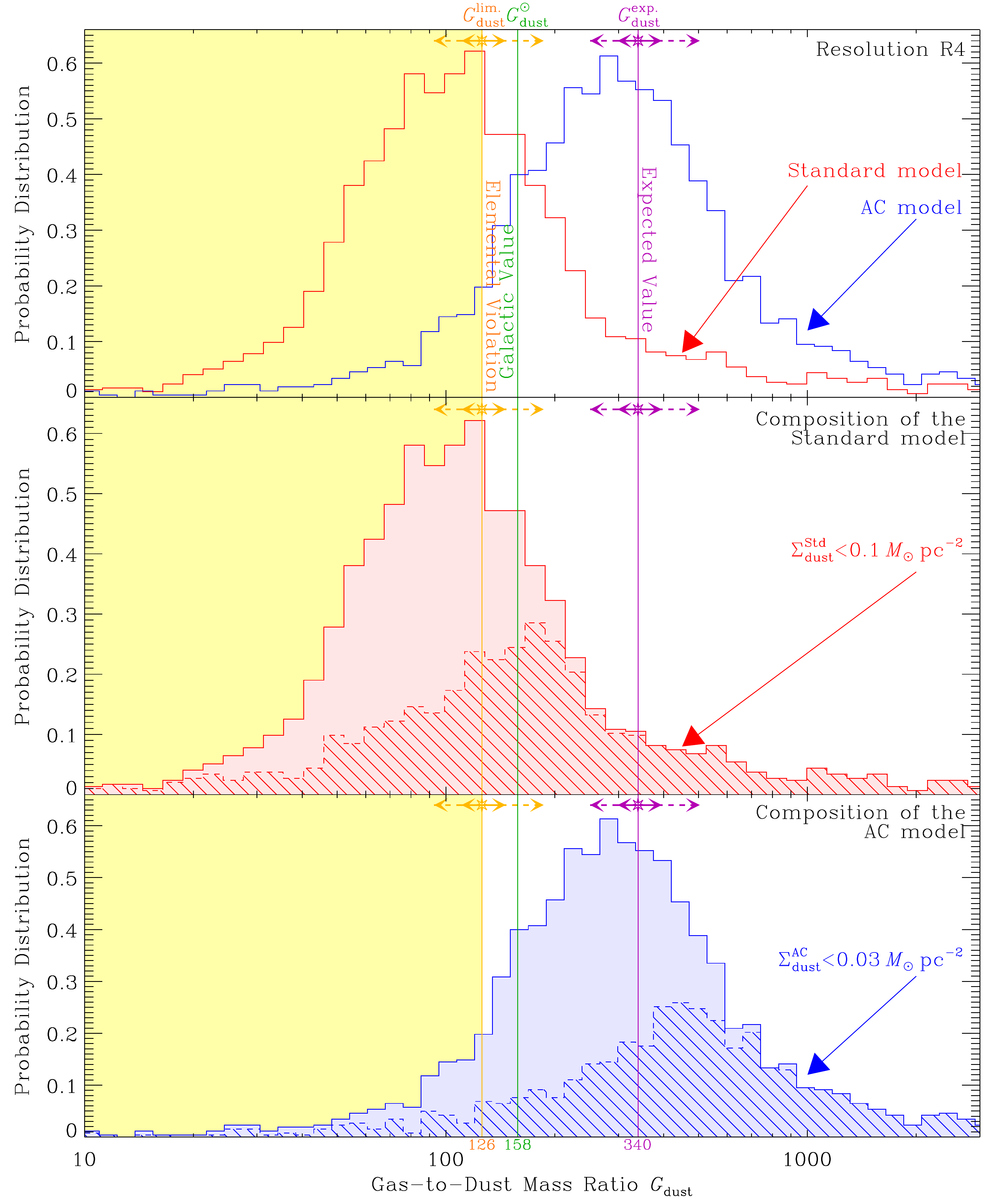}
  \caption{{\sl Pixel-to-pixel distribution of the gas-to-dust mass ratio for 
            R4 (54~pc).}
           The two histograms of the {\it top panel} are the pixel probability 
           distribution of the
           gas-to-dust mass ratio for the two models.
           The spread of the distribution reflects the pixel to pixel spread
           fluctuations of the ratio.
           The actual error on the total $G_\sms{dust}$ is smaller 
           than this spread (\reftab{tab:G2D}).
           For comparison, we show the values of $G_\sms{dust}^\sms{exp.}$ 
           \refeqp{eq:Gexp},
           $G_\sms{dust}^\odot$ \refeqp{eq:Gsun}, and
           $G_\sms{dust}^\sms{lim.}$ \refeqp{eq:Glim}.
           In the {\it two lower panels}, each histogram \modif{is repeated and} 
           is decomposed according to its dust mass surface density. 
           The hatched components correspond to lower surface densities (below
           the arbitrary limit indicated in each panel).
           The yellow and purple error bars, with the star symbol, represent 
           the uncertainities on 
           $G_\sms{dust}^\sms{lim.}$ and $G_\sms{dust}^\sms{exp.}$, 
           respectively (\reftab{tab:Z}; solid line: $50\,\%$; 
           dashed line: $90\,\%$).}
  \label{fig:dist_G2D}
\end{figure*}
\reffig{fig:dist_G2D} shows the pixel-to-pixel distribution of 
$G_\sms{dust}$ for the highest spatial resolution where the gas maps are defined 
(R4).
It appears that the distribution for the \citengl{standard model} is shifted to lower values, the pixels are systematically too low, compared to the expected value.
In addition, we have built a histogram of the pixels below an arbitrary column density ($\Sigma_\sms{dust}^\sms{Std}<0.1\;M_\odot\,\rm pc^{-2}$ and $\Sigma_\sms{dust}^\sms{AC}<0.03\;M_\odot\,\rm pc^{-2}$), in the two lower panels of \reffig{fig:dist_G2D}.
We have defined the surface densities at a given spatial resolution by dividing the mass in the pixel by the area of this pixel (\reftab{tab:dolls}): 
$\Sigma_X=M_X/l_\sms{pix}^2$.
The two lower panels of \reffig{fig:dist_G2D} demonstrate that
most of the pixels exhibiting high gas-to-dust mass ratios are located in
regions of low dust column density, independently of the model used.

\modif{Regarding} this property, we need to study the variations of the gas-to-dust mass ratio as a function of surface density.

  \subsubsection{The Correlation Between Gas and Dust Column Densities}
  \label{sec:correlin}

\begin{figure}[h!tbp]
  \centering
  \includegraphics[width=0.95\linewidth]{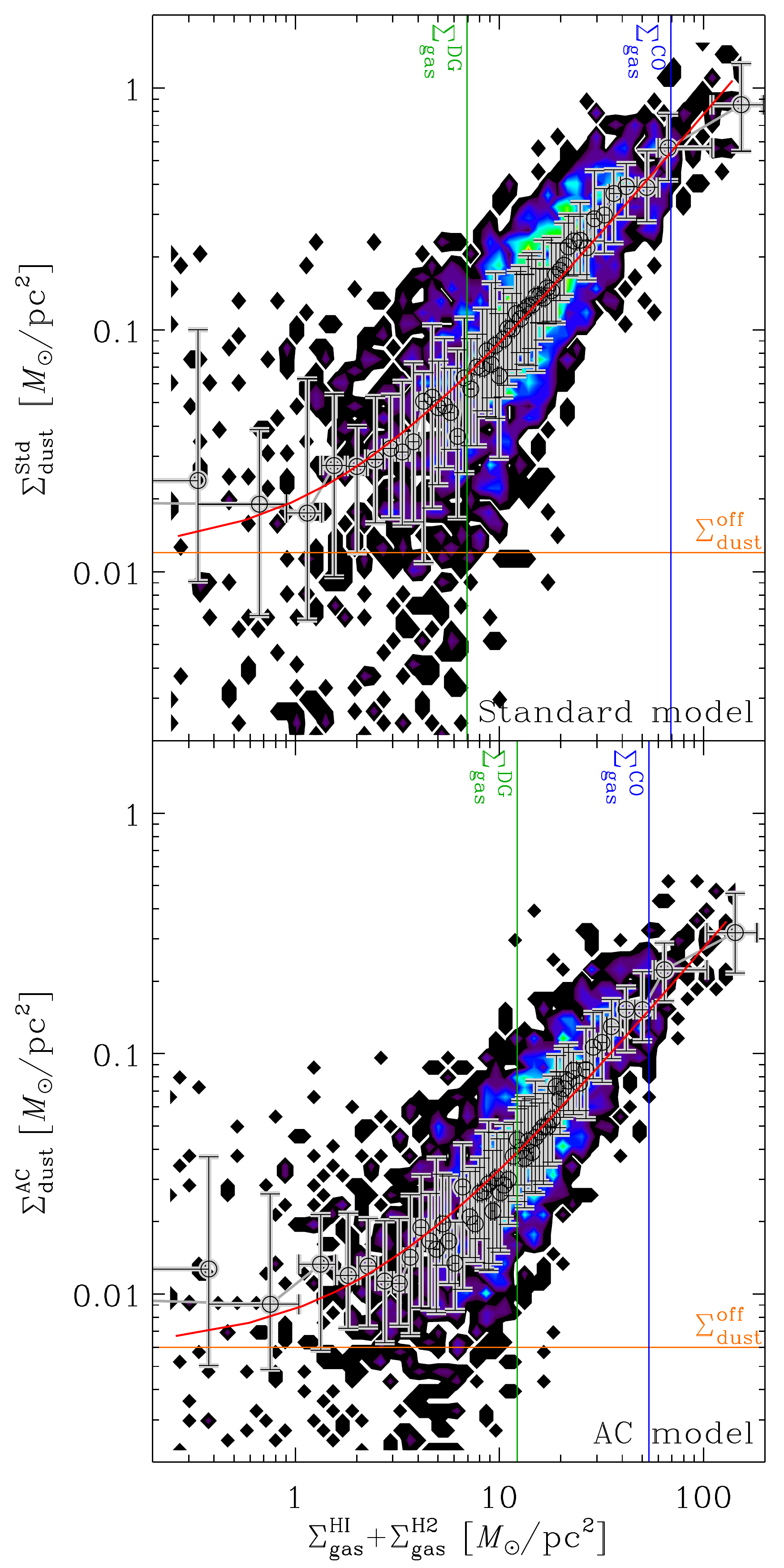}
  \caption{{\sl Correlation between the gas and dust mass column densities.}
           Each panel corresponds to a model.
           The spatial resolution is R4 (54~pc).
           \modif{The colored area is the pixel density. 
           The number density of pixels for each 
           $[\Sigma_\sms{gas},\Sigma_\sms{dust}]$ values is coded with the same 
           color scale as in \reffig{fig:correldarkgas}.}
           The circles with error bars show the binned trend. 
           The horizontal error bar displays the $\Sigma_\sms{gas}$ bin width.
           The vertical error bar displays the dispersion in $\Sigma_\sms{dust}$ 
           within each $\Sigma_\sms{gas}$ bin.
           The $\Sigma_\sms{gas}$ bin sizes are chosen so that each bin 
           contains the same number of pixels.
           The red line is the best fit of \refeq{eq:correlin} to the binned 
           trend.
           The vertical lines show the best fit values 
           $\Sigma_\sms{gas}^\sms{DG}$ and 
           $\Sigma_\sms{gas}^\sms{CO}$.
           The horizontal line shows the best fit value of
           of $\Sigma_\sms{dust}^\sms{off}
               = \Sigma_\sms{gas}^\sms{off}/G_\sms{dust}^\sms{fit}$.}
  \label{fig:correlin}
\end{figure}

In this section, we analyze the correlation between gas and dust 
using a method similar to that of \citet{planck-collaboration11b}.
It is aimed at identifying the presence of \citengl{dark gas} 
(hereafter DG)\footnote{In this paper, we choose the \citengl{dark gas} 
terminology, to remain more general.
However, we acknowledge that CO-free molecular gas is not dark, since its dust grains are radiating and it likely emits copious amounts of \ciiline.}, as a departure from the linear correlation between gas and dust.
\modif{
We have defined bins of gas mass surface density ($\Sigma_\sms{gas}^\sms{\hi}+\Sigma_\sms{gas}^\sms{\hmol}$), so that the same number of pixels falls within 
each bin. 
We have then computed the average value and the scatter of the dust mass of the pixels within each bin.
\reffig{fig:correlin} shows this binned trend on top of the pixel density plot.}
The binned dust mass surface density is fit with the following function:
\begin{eqnarray}
  \Sigma_\sms{dust}^\sms{fit} &
    = & \frac{\Sigma_\sms{gas}^\sms{\hi} 
           + \mathcal{C}_\sms{CO}^\sms{fit}\Sigma_\sms{gas}^\sms{\hmol}
           + \Sigma_\sms{gas}^\sms{off}}{G_\sms{dust}^\sms{fit}}   \label{eq:correlin}
\\
  & \mbox{for} & \Sigma_\sms{gas}^\sms{\hi} 
               + \mathcal{C}_\sms{CO}^\sms{fit}\Sigma_\sms{gas}^\sms{\hmol}
               \leq\Sigma_\sms{gas}^\sms{DG} 
 \;\mbox{ and }\;              
   \Sigma_\sms{gas}^\sms{\hi} 
               + \mathcal{C}_\sms{CO}^\sms{fit}\Sigma_\sms{gas}^\sms{\hmol}
               \geq\Sigma_\sms{gas}^\sms{CO}, \nonumber  
\end{eqnarray}
where the free parameters are the following.
\begin{itemize}
  \item $G_\sms{dust}^\sms{fit}$ is the fit gas-to-dust mass ratio.
    It assumes that the actual gas-to-dust mass ratio is the same everywhere in 
    the LMC.
  \item $\mathcal{C}_\sms{CO}^\sms{fit}$ is the fit correction factor of the 
    assumed $X_\sms{CO}$. 
    Since we have assumed 
    $X_\sms{CO}=7\E{20}\;{\rm H\,cm^{-2}(K\,km\,s^{-1})^{-1}}$ 
    (\refsec{sec:gas}), then the value of the fit conversion factor is
    $X_\sms{CO}^\sms{fit}=\mathcal{C}_\sms{CO}^\sms{fit}\,7\E{20}\;{\rm H\,cm^{-2}(K\,km\,s^{-1})^{-1}}$.
  \item $\Sigma_\sms{gas}^\sms{off}$ accounts for a possible offset in the gas 
    mass surface density compared to the dust mass surface density.
  \item $\Sigma_\sms{gas}^\sms{DG}$ is the gas mass surface density above which 
    the dark gas contributes.
  \item $\Sigma_\sms{gas}^\sms{CO}$ is the gas mass surface density above which 
    the molecular gas is reliably traced by \COio, and the dark gas therefore
    \modif{does} not contribute anymore.
    Above this value, the molecular phase is not \citengl{dark} anymore.
\end{itemize}
\refeq{eq:correlin} is not fit between $\Sigma_\sms{gas}^\sms{DG}$ and 
$\Sigma_\sms{gas}^\sms{CO}$, where the dark gas is assumed to contribute.
The total dark gas mass is computed as the difference between the binned trend and the fit of \refeq{eq:correlin}.
It is noted $M_\sms{gas}^\sms{DG}$.

\begin{table*}[h!tbp]
  \centering
  \begin{tabular}{l|rr|rr}
    \hline\hline
      & \multicolumn{2}{|c}{\citengl{Standard model}}
      & \multicolumn{2}{|c}{\citengl{AC model}} \\
    \hline
      $G_\sms{dust}^\sms{fit}$ 
        & $131_{-42}^{+39}$ & $\left[55,229\right]_{90\,\%}$
        & $370_{-90}^{+80}$ & $\left[103,590\right]_{90\,\%}$ \\
      $\Sigma_\sms{gas}^\sms{off}\;[M_\odot\,{\rm pc^{-2}}]$
        & $1.57_{-0.36}^{+0.45}$ & $\left[0.8,2.96\right]_{90\,\%}$
        & $2.23_{-0.36}^{+0.34}$ & $\left[1.34,3.2\right]_{90\,\%}$ \\
      $\Sigma_\sms{gas}^\sms{DG}\;[M_\odot\,{\rm pc^{-2}}]$
        & $6.9_{-3.2}^{+1.4}$ & $\left[3.5,18.2\right]_{90\,\%}$
        & $12.2_{-1.0}^{+0.9}$ & $\left[9.6,15.2\right]_{90\,\%}$ \\
      $\Sigma_\sms{gas}^\sms{CO}\;[M_\odot\,{\rm pc^{-2}}]$
        & $69_{-16}^{+23}$ & $\left[53,98\right]_{90\,\%}$
        & $54_{-0}^{+3}$ & $\left[53,98\right]_{90\,\%}$ \\
      $\mathcal{C}_\sms{CO}^\sms{fit}$
        & $1.42_{-0.47}^{+0.41}$ & $\left[0.79,3.16\right]_{90\,\%}$
        & $1.29_{-0.26}^{+0.34}$ & $\left[0.72,2.23\right]_{90\,\%}$ \\
      $M_\sms{gas}^\sms{DG}/(M_\sms{gas}^\sms{\hi}+M_\sms{gas}^\sms{\hmol})$
        & $11.1_{-3.9}^{+6.2}\,\%$ & $\left[0.45\,\%,24.5\,\%\right]_{90\,\%}$
        & $13.6_{-3.5}^{+2.9}\,\%$ & $\left[3.8\,\%,25.9\,\%\right]_{90\,\%}$ \\
    \hline
  \end{tabular}
  \caption{{\sl Parameters of the fit of $\Sigma_\sms{dust}$ as a
           function of $\Sigma_\sms{gas}$ (\reffig{fig:correlin}; 
           \refeqnp{eq:correlin}).}
           The spatial resolution is R4 (54~pc).
           The uncertainties come from the Monte-Carlo analysis.
           The convention for error display is defined in \refeqs{eq:median} 
           and (\ref{eq:sup}).}
  \label{tab:correlin}
\end{table*}
\reffig{fig:correlin} shows the trend and the best fit for each model.
Although the departure between the line fit and the binned trend, in the $[\Sigma_\sms{gas}^\sms{DG},\Sigma_\sms{gas}^\sms{CO}]$ is visible, its deviation
is lower than the typical dispersion of the correlation.
However, the parameter values and their uncertainties, given in
\reftab{tab:correlin}, show this departure is significant.
It appears that the gas-to-dust mass ratios derived from these simple fits are consistent with the pixel-to-pixel values (\reftab{tab:G2D}).
Therefore, this analysis tends to confirm that the \citengl{standard model}
violates elemental abundances in the LMC, while the \citengl{AC model}
is physically valid.
The fact that the parameter $\mathcal{C}_\sms{CO}^\sms{fit}$ is consistent with 
unity indicates that our  adopted $X_\sms{CO}$ conversion factor  
($X_\sms{CO}=7\E{20}\;{\rm H\,cm^{-2}(K\,km\,s^{-1})^{-1}}$; \refsec{sec:gas})
is probably not \modif{incorrect.}
Finally, the dark gas mass fraction, around $10\,\%$, seems to be moderate.

There are several hypotheses entering into \refeq{eq:correlin}.
First, its interpretation implicitly relies on the translation between $\Sigma_\sms{dust}$ and $A_V$, the extinction magnitude in $V$ band, because it defines the limits on the dark gas regime ($\Sigma_\sms{gas}^\sms{DG}$ and $\Sigma_\sms{gas}^\sms{CO}$; \refeqnp{eq:correlin}).
However, at the spatial scales considered here (54~pc), molecular clouds are not resolved, and several phases are mixed within each pixel. 
Therefore, the dust mass surface density is in principle a biased estimator 
of the $A_V$. 
The value of $A_V$ derived from the dust mass surface density, assuming a uniform dust distribution, is actually always going to be lower than the $A_V$ of a molecular cloud that would lie in the pixel.
Second, the approach of \refeq{eq:correlin} is a perturbative method.
It is correct only if the fraction of dark gas is small.
In particular, this method would fail if dark gas was present at low $\Sigma_\sms{gas}$.
Finally, it relies on the assumption of a uniform gas-to-dust mass ratio throughout the entire galaxy.
In fact, our $G_\sms{dust}$ spatial distributions (\reffig{fig:mapG2D}) are not noise maps, they contain clear structures.
Moreover, our error analysis demonstrates that the amplitude of these structures 
is larger than the typical error bar on an individual pixel value.
And the possible offset in background subtraction between the gas and dust maps
(estimated in \reftab{tab:correlin}) is small enough to affect only the low surface brightness pixels.
Consequently, this prompts us to further scrutinize the 
observed variations of $G_\sms{dust}$ as a function of the physical conditions.

  \subsubsection{Variations of the Gas-to-Dust Mass Ratio with Physical 
                 Conditions}
  \label{sec:correldarkgas}

\begin{figure*}[h!tbp]
  \centering
  \includegraphics[width=1.15\linewidth,angle=-90]{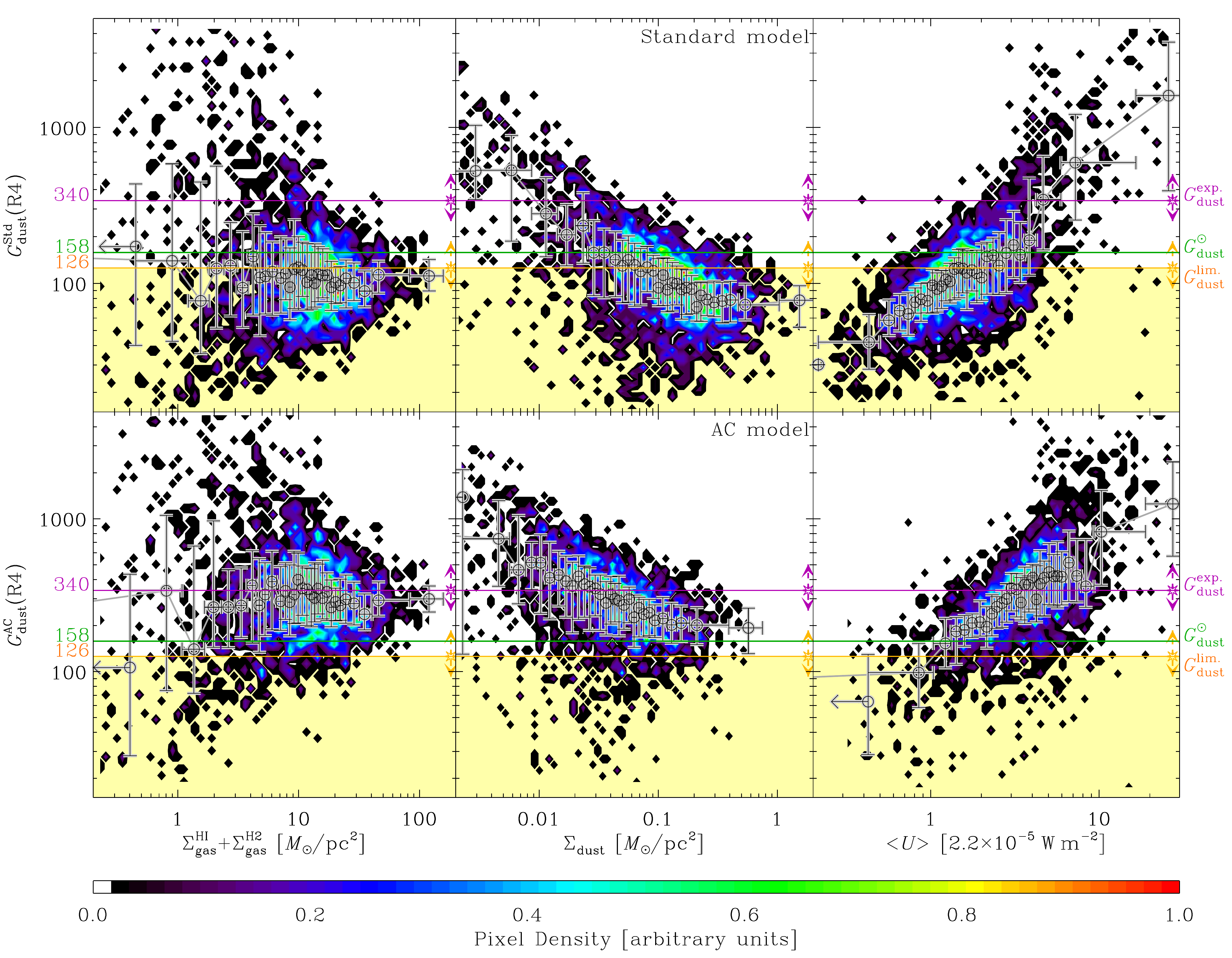}
  \caption{{\sl Pixel-to-pixel correlations between various tracers of the 
            physical conditions and the gas-to-dust mass ratio.} 
            \modif{Results are shown} for the two models, at spatial resolution 
            R4.
            The color scale represents the density of pixels in various bins
            of the two parameters.
            The overplotted grey error bars are the trends binned over the
            $x$-axis parameter.
            The bin size is defined so that the same number of pixels
            falls within each bin. 
            That is the reason why the bins are very narrow in the center, and
            there are only a few points in the outer parts where the pixel 
            density is low.
            The central value is the median gas-to-dust mass ratio, and
            the error bars account for the pixel-to-pixel 
            scatter, which is similar \modif{to}
            or larger than the typical intrinsic error 
            bar of a single pixel.
            The yellow and purple dashed error bars, with the star symbol,
            represent the $90\;\%$ confidence uncertainities on 
            $G_\sms{dust}^\sms{lim.}$ and $G_\sms{dust}^\sms{exp.}$, 
            respectively (\reftab{tab:Z}).
            This complex figure is explained in details in 
            \refsec{sec:correldarkgas}.}
  \label{fig:correldarkgas}
\end{figure*}
In this section, we describe the observed trends of the gas-to-dust mass ratio as a function of several parameters.
\reffig{fig:correldarkgas} shows the pixel-to-pixel variations of the
observed gas-to-dust mass ratio, as a function of several tracers of the physical conditions, for the two models.
We note that the trends are similar with the two models, but the quantities proportional to the dust mass appear to be scaled down by a factor of 
$\simeq 0.38$ with the \citengl{AC model}.
On the contrary, $\langle U\rangle$ appears to be scaled up by a factor of $\simeq2.1$ (see also \reffig{fig:comparison}).

The two left panels of \reffig{fig:correldarkgas} show the variation
of $G_\sms{dust}$ as a function of the gas mass surface density $\Sigma_\sms{gas}$.
There is a large scatter within each bin of $\Sigma_\sms{gas}$, and the scatter
is higher at low surface densities, since the signal-to-noise ratio is lower and offsets in the zero level of the gas and dust masses can impact the ratio.
The gas-to-dust mass ratio at high gas mass column densities ($\Sigma_\sms{gas}\gtrsim10\;M_\odot\,\rm pc^{-2}$) appears to deviate slighty from a constant value, as discussed in \refsec{sec:correlin}.
However, this deviation is smaller than the dispersion.
Overall, this trend does not show any general correlation between the gas-to-dust mass ratio and the gas mass surface density.
The central values of $G_\sms{dust}$ are on average below the limit value with the \citengl{standard model}, but are found around the expected value with the 
\citengl{AC model}.

The two central panels of \reffig{fig:correldarkgas} show the variation of 
$G_\sms{dust}$ with the dust mass surface density $\Sigma_\sms{dust}$.
There is a clear anticorrelation between the two quantities.
Although, the trend at low surface densities ($\Sigma_\sms{dust}\lesssim0.01\;M_\odot\,\rm pc^{-2}$) is very scattered and sensitive to the offset 
$\Sigma_\sms{gas}^\sms{off}$ (\reftab{tab:correlin}), the rest of the trend
is clear.
The fact that the gas-to-dust mass ratio correlates better with the dust mass column density than with the gas mass column density is the sign that the origin of the variations of $G_\sms{dust}$ is not directly linked with the {\it observed} gas content.
Most of the trend is below the limit value, with the \citengl{standard model}.
On the contrary the values for the \citengl{AC model} are all realistic.

Finally, the two right panels of \reffig{fig:correldarkgas} show the variation
of $G_\sms{dust}$ with the mass averaged starlight intensity $\langle U\rangle$.
There is a clear relation between the two quantities.
Contrary to the central panels ($G_\sms{dust}$ vs. $\Sigma_\sms{dust}$), these
two quantities are independent.
Moreover, $\langle U\rangle$ is a specific quantity, while $\Sigma_\sms{dust}$
is integrated over the line of sight.
It appears that $\langle U\rangle$ is the parameter giving the best trend with $G_\sms{dust}$, exhibiting larger dynamics and smaller scatter than the trends with $\Sigma_\sms{gas}$ and $\Sigma_\sms{dust}$.
We note that, with both models, the slope of the correlation (in bi-logarithmic representation) is not constant.
Schematically, for $\langle U\rangle^\sms{Std}\lesssim1$  (or $\langle U\rangle^\sms{AC}\lesssim2$), $G_\sms{dust}\propto \langle U\rangle$.
Then, for $1\lesssim\langle U\rangle^\sms{Std}\lesssim4$ 
(or $2\lesssim\langle U\rangle^\sms{AC}\lesssim8$), the trend flattens.
There is no significant variation in this range.
At larger values, $G_\sms{dust}\propto \langle U\rangle$ again, but the pixels are very dispersed in this range, \modif{and the statistics are limited.}

In summary, the observed variations of $G_\sms{dust}$ do not appear to be
significantly traced by the observed gas content.
On the contrary, these variations are well correlated with tracers of the dust physical conditions, in particular their irradiation conditions.
It suggests that the origin of these variations is linked in some way with the 
radiation energy density.
We are now going to explore the possible interpretations of these variations.

  \subsubsection{Physical Interpretation: Dark Gas, Modified Submm Opacities or 
                 Enhanced Dust Condensation?}
  \label{sec:interpretation}

To interpret the trends of \reffig{fig:correldarkgas}, we need to list the possible physical processes responsible for variations of the observed 
gas-to-dust mass ratio, as well as the known biases affecting these trends.
The known biases of our trends are the following.
\begin{description}
  \item[{\bf Bias 1.}] 
    There is a potential offset between the gas and dust mass zero level, 
    due to differences in background subtractions.
    This offset has an impact on low surface density pixels 
    (or high $\langle U\rangle$).
    It is quantified in \reftab{tab:correlin}.
  \item[{\bf Bias 2.}] 
    There is a known bias of our SED fitting. 
    At high average starlight intensities ($\langle U\rangle\gtrsim10$), 
    our model tends to underestimate the dust mass.
    This effect is demonstrated in \reffig{fig:bias_Md} of \refapp{ap:MC}.
    An artificial correlation between $\langle U\rangle$ and $G_\sms{dust}$ 
    is expected in this regime.
\end{description}
The possible causes of variations of the {\it observed} gas-to-dust mass ratio are the following.
\begin{description}
  \item[{\bf Dark gas.}]
        As discussed in \refsec{sec:global}, low $G_\sms{dust}$ could be the
        sign of massive \citengl{dark gas} reservoir.
        As shown in \refsec{sec:correldarkgas}, the fact that clear variations
        of $G_\sms{dust}$ are correlated with $\Sigma_\sms{dust}$, but not with
        $\Sigma_\sms{gas}$, suggests that the \hi\ and CO traced gas is
        blind to actual variations of $G_\sms{dust}$.
        Assuming that our $X_\sms{CO}$ is correct, the decrease of 
        $G_\sms{dust}$ with $\Sigma_\sms{dust}$ could be partly due to the 
        presence of a dark gas component more massive than what was determined 
        in \refsec{sec:correlin}.
        The latter could have been underestimated because it was lost in the 
        scatter of the relation.
  \item[{\bf Dust-to-metal ratio.}] 
        Variations of $G_\sms{dust}$ could be due to local 
        variations of the dust-to-metal mass ratio \refeqp{eq:D2M}.
        In principle, the gas-to-dust mass ratio could go down to 
        $G_\sms{dust}^\sms{lim.}$ \refeqp{eq:Glim} and go up to infinity.
        Assuming an efficient mixing of freshly formed dust grains  
        throughout the ISM, an increase of $G_\sms{dust}$ would be the sign of 
        an enhanced destruction, mainly by SN blastwaves.
        This is probably the case at low surface density, although it is
        degenerate with biases 1 and 2.
        On the other hand, a decrease of $G_\sms{dust}$ would then be the sign 
        of an increase of the condensation efficiency of metals onto dust.
        This effect is degenerate with the dark gas component.
        Assuming there is no dark gas, for this scenario to be realistic, the 
        metal condensation onto dust 
        is expected to be enhanced in regions where the gas density is higher.
        However, the left panels of \reffig{fig:correldarkgas} show that there
        is no clear correlation between $G_\sms{dust}$ and the gas mass surface 
        denstity.
        This effect should therefore be minor.
        Although the dust-to-metal mass ratio may vary in the LMC, our
        data indicate that this potential variation is not at the origin of 
        the trend, except maybe for low surface density.
  \item[{\bf Emissivity.}] 
        The variations of $G_\sms{dust}$ could be due to variations of the
        emissivity of the grains in different regions.
        This emissivity variation would lead us to systematically 
        misestimate the dust mass in regions where it would differ from 
        the assumed opacity.
        These variations are poorly known and more difficult to constrain.
        The most well-known of these processes, grain-grain coagulation, leads 
        to an increase of the far-IR opacity \citep[e.g.][]{stepnik03}. 
        It happens preferentially in dense clouds.
        However, the left panels of \reffig{fig:correldarkgas} do not show 
        a clear correlation between $G_\sms{dust}$ and $\Sigma_\sms{gas}$.
        Therefore, this effect \modif{does} not seem to be at the origin of the 
        overall 
        variations of $G_\sms{dust}$.
        Another effect is the increase of the emissivity with grain temperature
        \citep[e.g.][]{meny07}.
        This effect would lead us to overestimate the dust mass in regions where
        the dust is hot.
        It would therefore lead to an anticorrelation between the observed 
        $G_\sms{dust}$ and $\langle U\rangle$.
        Therefore, this effect can not be at the origin of our trend.
        Although this review is not complete, we will assume that {\it local}
        variations of the grain emissivity are not responsible for the observed
        variations of $G_\sms{dust}$.
\end{description}

Considering these previous effects, we now make several simple assumptions to
interpret the trends of $G_\sms{dust}$ with $\langle U\rangle$ 
(right panels of \reffig{fig:correldarkgas}).
We assume that we can classify our pixels within the three following regimes, characterized by their average starlight intensity.
\renewcommand{\labelenumi}{\Alph{enumi}.}
\begin{enumerate}
  \item We assume that there are embedded regions, characterized by low 
        starlight intensities ($1\lesssim\langle U\rangle$), where the observed 
        gas-to-dust mass ratio is underestimated due to the presence of an
        undetected gas phase.
        The proportion of this overlooked gas rises, when $\langle U\rangle$
        decreases.
  \item We assume that there are regions, characterized by moderate 
        starlight intensities ($1\lesssim\langle U\rangle\lesssim10$),
        corresponding to the \hi\ dominated regime, where there is no dark gas.
        The observed gas-to-dust mass ratio is assumed to be correct in these 
        regions, and can be used as a reference.
  \item We assume that the gas-to-dust mass ratio may be higher in
        very diffuse regions, where shocks might have recently destroyed the
        grains. 
        In these regions ($\langle U\rangle\gtrsim10$), the biases of the 
        modelling can also be significant,
        as demonstrated in \refapp{ap:MC} and \reffig{fig:bias_Md}.
        An artificial correlation between $\langle U\rangle$ and $G_\sms{dust}$ 
        is expected.
\end{enumerate}
\renewcommand{\labelenumi}{\theenumi.}

\begin{figure}[h!tbp]
  \centering
  \includegraphics[width=0.95\linewidth]{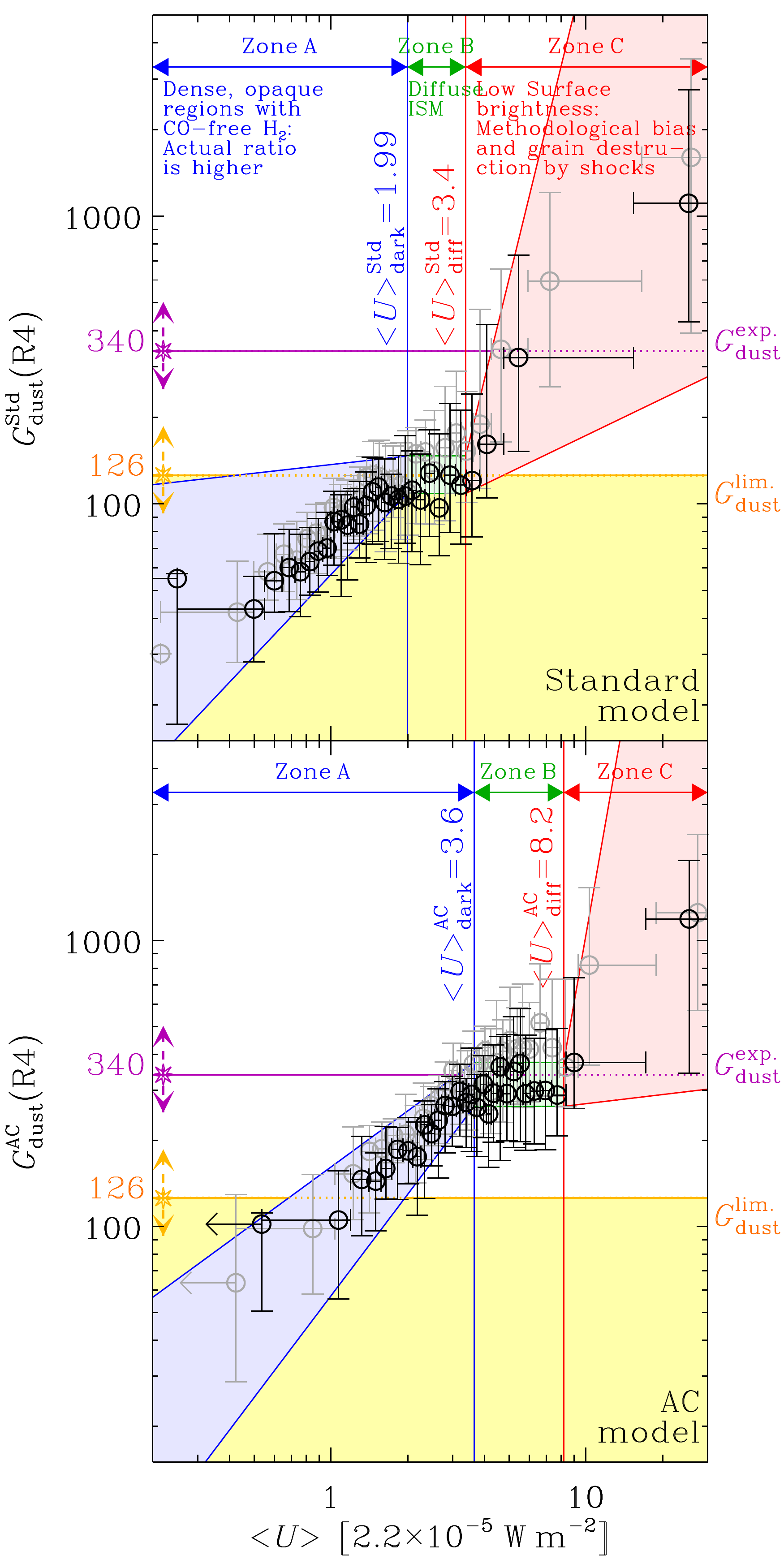}
  \caption{{\sl Definition of the various zones for the two models, for R4 
            (54~pc).}
           Each panel shows the binned trends of the right panels of 
           \reffig{fig:correldarkgas} (grey circles with error bars).
           The fitted trend (black symbols with error bars) is the 
           grey trend corrected for the offset $\Sigma_\sms{gas}^\sms{off}$
           (\reftab{tab:correlin}).
           We define the three zones depending on their starlight intensities.
           Each zone corresponds to a different regime of the observed 
           gas-to-dust mass ratio.
           The blue-green-red stripes are the envelopes of the fits of
           \refeq{eq:zonedef} to the trend.
           The yellow and purple dashed error bars, with the star symbol,
           represent the $90\;\%$ confidence uncertainities on 
           $G_\sms{dust}^\sms{lim.}$ and $G_\sms{dust}^\sms{exp.}$, 
           respectively (\reftab{tab:Z}).
           This complex figure is explained in details in 
           \refsec{sec:interpretation}.}
  \label{fig:dark_main}
\end{figure}
\reffig{fig:dark_main} illustrates this scenario.
The two panels show the binned trends of \reffig{fig:correldarkgas} with the
starlight intensity, for the two models.
The grey circles with error bars show the original trend, while the black circles with error bars show the trend corrected for the offset $\Sigma_\sms{gas}^\sms{off}$ derived in \reftab{tab:correlin}.
We fit the latter.
We decompose the trends into the three zones described above, by fitting the following function:
\begin{equation}
  \log G_\sms{dust}^\sms{fit} =
  \left\{
  \begin{array}{l}
    a_\sms{dark}\log\frac{\displaystyle\langle U\rangle}{\displaystyle\langle U\rangle_\sms{dark}} + \log G_\sms{dust}^\sms{ref}
    \mbox{ for } 
    \frac{\displaystyle\langle U\rangle}{\displaystyle\langle U\rangle_\sms{dark}}<1
    \\
    \\
    \log G_\sms{dust}^\sms{ref} 
    \mbox{ for } 
    \langle U\rangle_\sms{dark}\leq \langle U\rangle 
    \leq \langle U\rangle_\sms{diff}
    \\
    \\
    a_\sms{diff}\log\frac{\displaystyle\langle U\rangle}{\displaystyle\langle U\rangle_\sms{diff}} 
    + \log G_\sms{dust}^\sms{ref} 
    \mbox{ for } 
    \frac{\displaystyle\langle U\rangle}{\displaystyle\langle U\rangle_\sms{diff}} > 1,
  \end{array}
  \right.
  \label{eq:zonedef}
\end{equation}
where $a_\sms{dark}$, $a_\sms{diff}$, $G_\sms{dust}^\sms{ref}$, 
$\langle U\rangle_\sms{dark}$ and $\langle U\rangle_\sms{diff}$ are
free parameters.
$G_\sms{dust}^\sms{ref}$ is the \citengl{reference} gas-to-dust mass ratio, 
\ie\ the gas-to-dust mass ratio in the diffuse ISM,
and $\langle U\rangle_\sms{dark}$ and $\langle U\rangle_\sms{diff}$ are
the starlight intensities defining these zones (\reffig{fig:dark_main}).
\reftab{tab:zonedef} gives the main parameters of these fits for each model.

\begin{table}[h!tbp]
\centering
\begin{tabular}{lrrr}
\hline\hline
  & \citengl{Standard model} & \citengl{AC model} & Units\\
  \hline
    $G_\sms{dust}^\sms{ref}$ & $114_{-14}^{+25}$ & $306_{-38}^{+34}$ 
      & \\
    $\langle U\rangle_\sms{dark}$ & $1.99_{-0.21}^{+0.01}$ & $3.6_{-0.4}^{+0.5}$ 
      & $[2.2\E{-5}\;\rm W\,m^{-2}]$ \\
    $\langle U\rangle_\sms{diff}$ & $3.4_{-0.4}^{+0.6}$ & $8.2_{-0.9}^{+0.9}$ 
      & $[2.2\E{-5}\;\rm W\,m^{-2}]$ \\
  \hline
\end{tabular}
\caption{{\sl Fitted parameters defining the three zones of 
          \reffig{fig:dark_main}.}
          These values come from the best fit of the trends of 
          \reffig{fig:dark_main} with \refeq{eq:zonedef}, 
          and the subsequent error propagation.}
\label{tab:zonedef}
\end{table}
This point of view leads to the same conclusion than in \refsec{sec:correlin}, namely that the \citengl{standard model} is unphysical.
Indeed, the derived reference gas-to-dust mass ratio, $G_\sms{dust}^\sms{ref}\simeq114_{-14}^{+25}$ (\reftab{tab:zonedef}), in regions where dark gas is unlikely, is on average below the hard limit, $G_\sms{dust}^\sms{lim.}\simeq117$.
On the contrary, the \citengl{AC model} gives a reference $G_\sms{dust}^\sms{ref}\simeq306_{-38}^{+34}$ close to the expected value, based on the metallicity of the LMC, $G_\sms{dust}^\sms{exp.}\simeq339$.
More importantly, this reference ratio does not violate the elemental abundances, with the \citengl{AC model}.

In summary, now that we have taken into account the various competing processes, we can safely conclude that the \citengl{standard model}, which 
works for the Milky Way, does not apply to the LMC.
It means that the chemical composition of the grains in the LMC is {\it systematically} different than that of the Galaxy.
The LMC grains have on average a larger submm opacity.
Although the possible compositions are numerous, this work gives a plausible one, respecting the elemental abundances, based on optical properties of grains that have been observed in the laboratory \citep[ACAR amorphous carbons in \modif{lieu} of graphite;][]{zubko96}.

  \subsubsection{Remarks on the Submillimiter Emissivity Index}

First, as detailed in \refapp{ap:mie}, our \citengl{AC model} has a lower far-IR 
emissivity index ($\beta\simeq1.7$) than the \citengl{standard model} 
($\beta=2$).
In our case, $\beta$ does not vary much with wavelength in the \hersc\ range.
However, in general, $\beta$ is a function of the wavelength, and
there are grain species showing strong variations in this range 
\citep[e.g.][]{quinten02}.

Second, the emissivity of our grain model is what we call the {\it intrinsic} 
emissivity.
On the other hand, the isothermal fit of an SED with free emissivity index provides
the {\it effective} emissivity index (noted $\beta_\sms{eff.}$ by us).
The two are similar ($\beta=\beta_\sms{eff.}$) only if 
the isothermal approximation is valid.
On the contrary, if the spatial resolution is such that there is likely a mix of temperatures within the beam, then the two values are going to differ.
The SED will be broadened by the temperature distribution and $\beta_\sms{eff.}$
will likely be lower than $\beta$.
In the LMC, we are mostly in the second case.
Due to this degeneracy, the \modif{shapes} of the SED do not provide direct constraints 
on $\beta$.
Other constraints have to be invoked.
In our case, we used the constraints on the gas-to-dust mass ratio to discriminate between different $\beta$.

We note that \citet{aguirre03}, using COBE observations of the LMC, reached the same qualitative conclusion as \modif{derived here}. 
They concluded that the grain opacities were different in the LMC than in the Milky Way, although they adopted a different set of grain species.
However, these authors based their analysis only on isothermal fits, and did not use the gas-to-dust mass ratio as a constraint.
Therefore their demonstration was incomplete.
Similarly, \citet[][using \hersc\ data]{gordon10} and 
\citet[][using {\it Planck} data]{planck-collaboration11} showed that the {\it effective} $\beta$ of the LMC was around 1.5.
However, they did not constrain the {\it intrinsic} $\beta$.

Thus, in light of these precisions, we have shown, for the first time, that the {\it intrinsic far-IR/submm} emissivity index of the grains in the LMC \modif{is} lower than in the Galaxy.

  \subsubsection{Constraints on the Dark Gas Content}

Using the point of view of \reffig{fig:dark_main}, we can put an upper limit
on the dark gas content.
Since we have shown in \refsec{sec:interpretation} that most of the variation of 
the gas-to-dust mass ratio could originate in an overlooked gas component, we 
can estimate the mass of this component, from this point of view.

\begin{table}[h!tbp]
  \centering
  \begin{tabular}{lrr}
    \hline\hline
      & \multicolumn{2}{c}{Mass fraction $M_\sms{gas}^\sms{DG up}/(M_\sms{gas}^\sms{\hi}+M_\sms{gas}^\sms{\hmol})$} \\   
    \hline
                 & \multicolumn{1}{c}{\citengl{Standard model}} 
                 & \multicolumn{1}{c}{\citengl{AC model}} \\
    \hline
 R4 & $87_{-28}^{+52}$ $\%$ & $52_{-33}^{+49}$ $\%$ \\
 R5 & $87_{-28}^{+51}$ $\%$ & $50_{-31}^{+49}$ $\%$ \\
 R6 & $85_{-27}^{+49}$ $\%$ & $47_{-29}^{+48}$ $\%$ \\
 R7 & $76_{-23}^{+48}$ $\%$ & $42_{-28}^{+48}$ $\%$ \\
 R8 & $69_{-22}^{+40}$ $\%$ & $38_{-23}^{+52}$ $\%$ \\
 R9 & $71_{-23}^{+43}$ $\%$ & $37_{-24}^{+60}$ $\%$ \\
 R10 & $64_{-17}^{+40}$ $\%$ & $41_{-28}^{+62}$ $\%$ \\
    \hline
  \end{tabular}
  \caption{{\sl Properties of the overlooked gas component.}
           This is the mass fraction of the optimistic estimate of the dark gas
           mass, defined in \refeq{eq:Mdark}.
           This quantity is given as a function of spatial resolution for both 
           models.}
  \label{tab:missgas}
\end{table}
To provide an optimistic estimate, we define the mass of overlooked gas $M_\sms{gas}^\sms{/DG up}$ by the mass added to the observed gas mass $M_\sms{gas}^\sms{\hi}+M_\sms{gas}^\sms{\hmol}$,
in order to match the expected gas-to-dust mass ratio $G_\sms{dust}^\sms{exp.}$
in regions where the uncorrected $G_\sms{dust}$ is lower than 
$G_\sms{dust}^\sms{exp.}$:
\begin{equation}
  M_\sms{gas}^\sms{DG up} =
  \left\{
  \begin{array}{ll}
    \left(G_\sms{dust}^\sms{exp.}-G_\sms{dust}\right)\times M_\sms{dust}
      & \mbox{if } G_\sms{dust} \leq G_\sms{dust}^\sms{exp.} \\
      & \\
      0 & \mbox{if } G_\sms{dust} > G_\sms{dust}^\sms{exp.}.
  \end{array}
  \right.
  \label{eq:Mdark}
\end{equation}
Technically, this dark gas includes the CO-free \hmol, as well as a fraction of the CO-associated \hmol\ where the $X_\sms{CO}$ factor is higher than our 
adopted value.
This definition implies that this dark component covers most of the strip with the \citengl{standard model},
while it is going to cover only a fraction of it with the \citengl{AC model}.
\reftab{tab:missgas} gives the mass fraction of this component, for both models, at each spatial resolution.
The most reliable value is probably the value at R4 of the \citengl{AC model}. 

Therefore, comparing the conservative estimate of \reftab{tab:correlin} and the optimistic estimate of \reftab{tab:missgas}, with their respective uncertainties, we can bracket the dark gas mass fraction $f_\sms{gas}^\sms{DG}=M_\sms{gas}^\sms{DG}/(M_\sms{gas}^\sms{\hi}+M_\sms{gas}^\sms{\hmol})\simeq10-100\,\%$ in the strip, with the \citengl{AC model}.

For comparison, \citet{bernard08} found that this fraction was 182$\,\%$. 
The reason \modif{for} this discrepancy might be due to the fact that the region we have modelled here is relatively poor in dark gas, compared to the entire LMC.

  \subsection{What the \SPIREiii\ Excess Is Not}
  \label{sec:r500}

As mentioned in \refsec{sec:dale}, we have not fitted the \SPIREiii\ excess present in our data.
We have excluded this waveband, in order to avoid being biased by this effect, 
since its origin is still unknown.
This excess can, {\it a priori}, affect our previous result concerning the grain properties, for two different reasons.
First, this excess could originate in colder dust that our model has not accounted for. 
Our dust mass would then have been biased.
Second, if the origin of this excess is an unkown mechanism, not related to cold dust, it could also affect shorter wavelength bands, and bias our dust mass.

Although \SPIREiii\ was not used as a constraint, for our dust SED model, it is 
possible to study the behaviour of its excess with the physical conditions.

  \subsubsection{Systematic Analysis of the Excess}

\begin{figure}[h!tbp]
  \centering
  \includegraphics[width=0.95\linewidth]{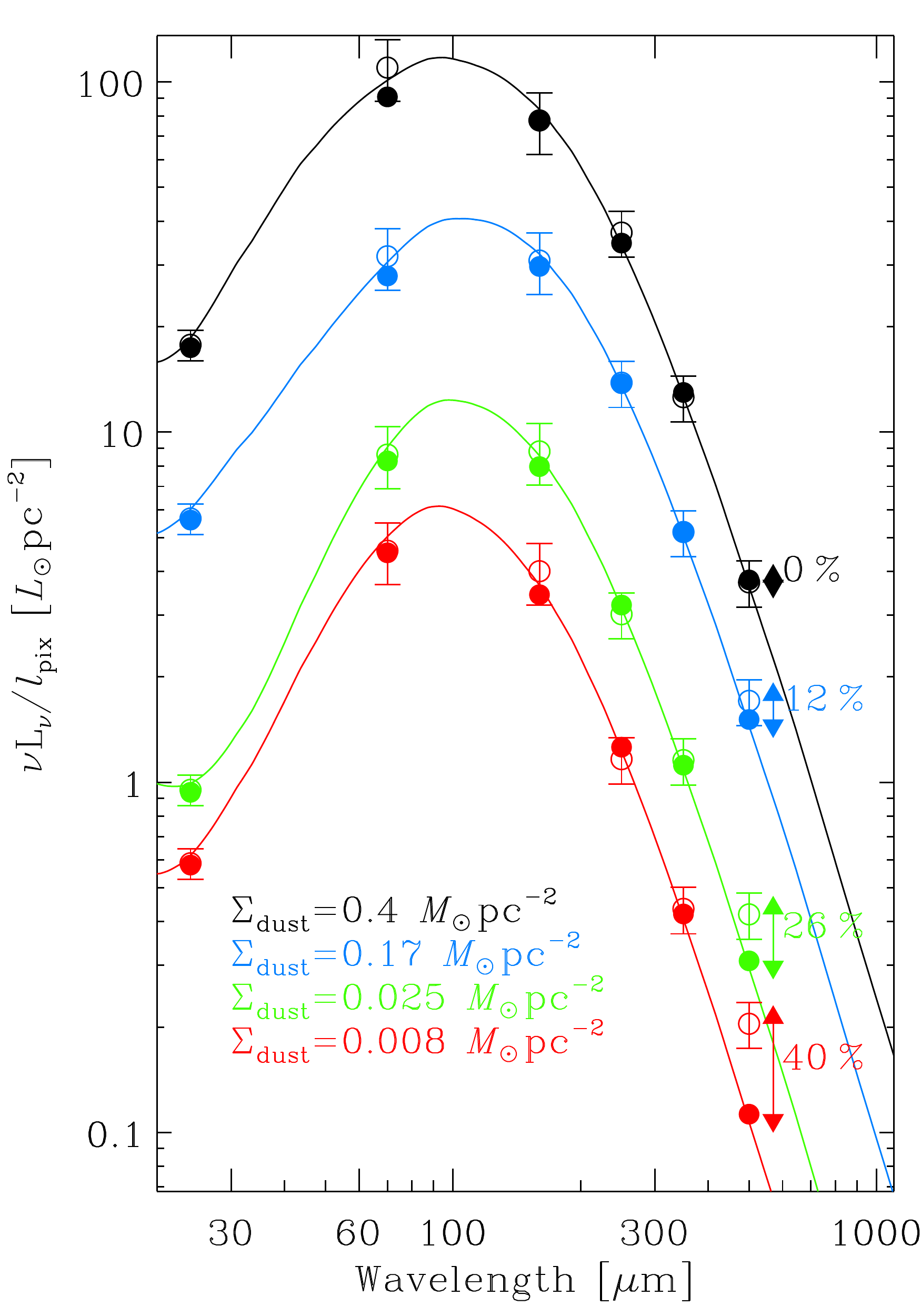}
  \caption{\modif{{\sl Different SEDs spanning the \SPIREiii\ excess range.}
            These SEDs correspond to four pixels (at resolution R4; 54~pc), 
            arbitrarily chosen.
            For each pixel the observations (open circle and error bar), 
            the model (solid line), and the model integrated in the broadband 
            filters (filled circle) are displayed.
            For each SED, the corresponding dust mass surface density 
            and the value of $r_{500}$ are given.}}
  \label{fig:r500_sed}
\end{figure}
Let's define the absolute and relative excesses of the \SPIREiii\ 
band, \modif{respectively} by:
\begin{eqnarray}
  \Delta_{500} & = & \nu L_\nu^\sms{obs}(\mbox{\SPIREiii})-\nu L_\nu^\sms{mod}(\mbox{\SPIREiii}) \label{eq:D500} \\
  r_{500} & = & \frac{L_\nu^\sms{obs}(\mbox{\SPIREiii})-L_\nu^\sms{mod}(\mbox{\SPIREiii})}{L_\nu^\sms{obs}(\mbox{\SPIREiii})}.
  \label{eq:r500}
\end{eqnarray}
\modif{\reffig{fig:r500_sed} shows four SEDs with different values of the excess.}

\begin{table}[h!tbp]
  \centering
  \begin{tabular}{l|rr|rr}
    \hline\hline
      Resolution  &  \multicolumn{2}{c}{\citengl{Standard model}}   
                  &  \multicolumn{2}{c}{\citengl{AC model}}         \\
    \hline
          & \multicolumn{4}{c}{Relative \SPIREiii\ excess $r_{500}$ $[\%]$} \\
    \hline
    R1 & $12.5_{-9.1}^{+7.2}$ & $[$$-13$,$30$$]_{90\,\%}$ &$10.5_{-9.1}^{+6.5}$ & $[$$-16$,$26$$]_{90\,\%}$ \\
    R2 & $12.6_{-9.7}^{+7.2}$ & $[$$-13$,$29$$]_{90\,\%}$ &$10.6_{-9.0}^{+6.5}$ & $[$$-16$,$26$$]_{90\,\%}$ \\
    R3 & $12.7_{-10.1}^{+7.2}$ & $[$$-13$,$30$$]_{90\,\%}$ &$10.8_{-9.0}^{+6.2}$ & $[$$-15$,$26$$]_{90\,\%}$ \\
    R4 & $13.1_{-10.1}^{+6.8}$ & $[$$-13$,$29$$]_{90\,\%}$ &$11.3_{-9.2}^{+5.7}$ & $[$$-15$,$26$$]_{90\,\%}$ \\
    R5 & $13.6_{-10.6}^{+6.2}$ & $[$$-14$,$29$$]_{90\,\%}$ &$11.8_{-9.5}^{+5.0}$ & $[$$-15$,$26$$]_{90\,\%}$ \\
    R6 & $14.2_{-11.0}^{+5.4}$ & $[$$-14$,$29$$]_{90\,\%}$ &$12.5_{-10.4}^{+4.3}$ & $[$$-15$,$26$$]_{90\,\%}$ \\
    R7 & $15.2_{-8.2}^{+7.2}$ & $[$$-10$,$31$$]_{90\,\%}$ &$12.9_{-8.1}^{+6.1}$ & $[$$-12$,$28$$]_{90\,\%}$ \\
    R8 & $15.9_{-8.2}^{+7.2}$ & $[$$-9$,$32$$]_{90\,\%}$ &$13.4_{-7.7}^{+5.9}$ & $[$$-11$,$28$$]_{90\,\%}$ \\
    R9 & $15.0_{-8.5}^{+7.5}$ & $[$$-10$,$31$$]_{90\,\%}$ &$13.1_{-7.7}^{+5.6}$ & $[$$-13$,$27$$]_{90\,\%}$ \\
    R10 & $16.1_{-7.8}^{+6.4}$ & $[$$-7$,$31$$]_{90\,\%}$ &$14.1_{-6.9}^{+6.1}$ & $[$$-9$,$29$$]_{90\,\%}$ \\
    \hline
       & \multicolumn{4}{c}{Absolute \SPIREiii\ excess 
                            $\Delta_{500}$ $[10^{-2}\,L_\odot\,\rm pc^{-2}]$} \\
    \hline
    R1 & $7.6_{-5.8}^{+4.6}$ & $[$$-7$,$18$$]_{90\,\%}$ &$6.4_{-5.4}^{+4.1}$ & $[$$-9$,$16$$]_{90\,\%}$ \\
    R2 & $7.6_{-6.0}^{+4.1}$ & $[$$-8$,$18$$]_{90\,\%}$ &$6.4_{-5.4}^{+3.8}$ & $[$$-8$,$16$$]_{90\,\%}$ \\
    R3 & $7.6_{-5.9}^{+4.2}$ & $[$$-8$,$18$$]_{90\,\%}$ &$6.5_{-5.5}^{+3.8}$ & $[$$-9$,$16$$]_{90\,\%}$ \\
    R4 & $7.6_{-5.9}^{+3.9}$ & $[$$-8$,$17$$]_{90\,\%}$ &$6.6_{-5.2}^{+3.2}$ & $[$$-9$,$15$$]_{90\,\%}$ \\
    R5 & $7.5_{-5.9}^{+3.2}$ & $[$$-7$,$16$$]_{90\,\%}$ &$6.5_{-5.3}^{+2.7}$ & $[$$-8$,$14$$]_{90\,\%}$ \\
    R6 & $7.6_{-5.9}^{+2.6}$ & $[$$-7$,$15$$]_{90\,\%}$ &$6.7_{-5.6}^{+2.0}$ & $[$$-8$,$13$$]_{90\,\%}$ \\
    R7 & $6.4_{-3.5}^{+3.0}$ & $[$$-4$,$13$$]_{90\,\%}$ &$5.4_{-3.4}^{+2.6}$ & $[$$-5$,$12$$]_{90\,\%}$ \\
    R8 & $5.8_{-3.0}^{+2.6}$ & $[$$-3$,$11$$]_{90\,\%}$ &$4.9_{-2.8}^{+2.2}$ & $[$$-4$,$10$$]_{90\,\%}$ \\
    R9 & $5.1_{-2.9}^{+2.5}$ & $[$$-3$,$11$$]_{90\,\%}$ &$4.5_{-2.6}^{+1.9}$ & $[$$-4$,$9$$]_{90\,\%}$ \\
    R10 & $4.7_{-2.3}^{+1.9}$ & $[$$-1.9$,$9$$]_{90\,\%}$ &$4.1_{-2.1}^{+1.8}$ & $[$$-2.8$,$8$$]_{90\,\%}$ \\
    \hline
  \end{tabular}
  \caption{{\sl Absolute and relative \SPIREiii\ excesses as a function of 
            spatial resolution.}
            This is the cumulative excess.
            The observations are identical at all resolutions.
            Only the model and the error bars change slightly from one 
            resolution to the other.
            The convention for error display is defined in \refeqs{eq:median} 
            and (\ref{eq:sup}).}
  \label{tab:r500}
\end{table}
\reftab{tab:r500} shows the value of the excesses, as a function of the spatial resolution.
In general, the excess tends to be slightly smaller with the \citengl{AC model}, since it has a slightly flatter submm slope.
The \modif{relative} excess tends to rise slightly when the spatial resolution decreases, but
this trend is not statistically significant.
We emphasize that $r_{500}$ being a relative quantity, most of the calibration error cancels, and the errors quoted in \reftab{tab:r500} are uncorrelated
between different spatial resolution.
Most of the error in this excess comes from the noise in the diffuse emission.
The most reliable estimates of the excess are therefore those of R10 (integrated 
strip):
$r_{500}^\sms{Std}=16.1_{-7.8}^{+6.4}\;\%$ and 
$r_{500}^\sms{AC}=14.1_{-6.9}^{+6.1}\;\%$.
This excess is only marginally detected in the global SED.
Since, the excess is relatively similar for both models, and that we have
previously shown that the \citengl{AC model} is more realistic, we will discuss the excess only for the \citengl{AC model}, in the rest of this section.

\begin{figure}[h!tbp]
  \centering
  \includegraphics[width=0.95\linewidth]{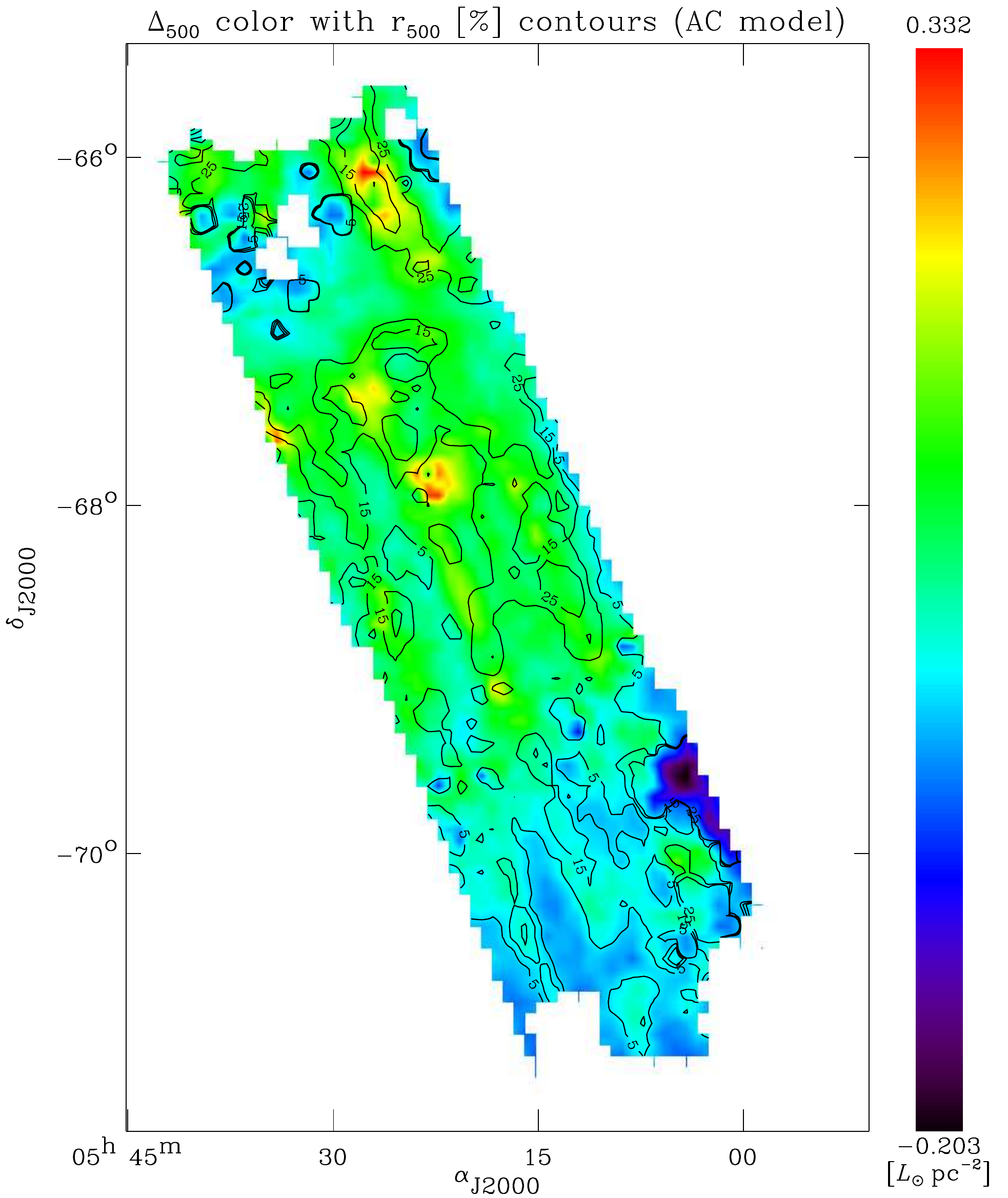}
  \caption{{\sl Spatial distributions of the \SPIREiii\ excesses.}
            The spatial resolution is R4 (54~pc).
            The color image is the absolute excess $\Delta_{500}$, and the
            contours are the relative excess $r_{500}$ (the contours are labeled 
            in percent).}
  \label{fig:im_delta}
\end{figure}
\reffig{fig:im_delta} compares the spatial distribution of the absolute
\SPIREiii\ excess (\refeqnp{eq:D500}, in colors), with the relative excess
(\refeqnp{eq:r500}, with contours).
We first note that the absolute excess is not homogeneous, and contains structures associated with the known features of the LMC, in particular N$\,44$, and the north superbubble.
We also note that the absolute value of the excess is much larger than the typical noise (\reftab{tab:rms}).
We emphasize that the map \modif{has} been background subtracted in such a way that the fluxes on the north and south of the strip are zero.
These arguments show that this excess can not be attributed to emission external to the LMC, like CMB fluctuations or cirrus foreground.
\modif{The distribution of the relative excess $r_{500}$ also shows structures}, that appear to be roughly anticorrelated with the IR luminosity.

\begin{figure*}[h!tbp]
  \centering
  \includegraphics[width=0.95\linewidth]{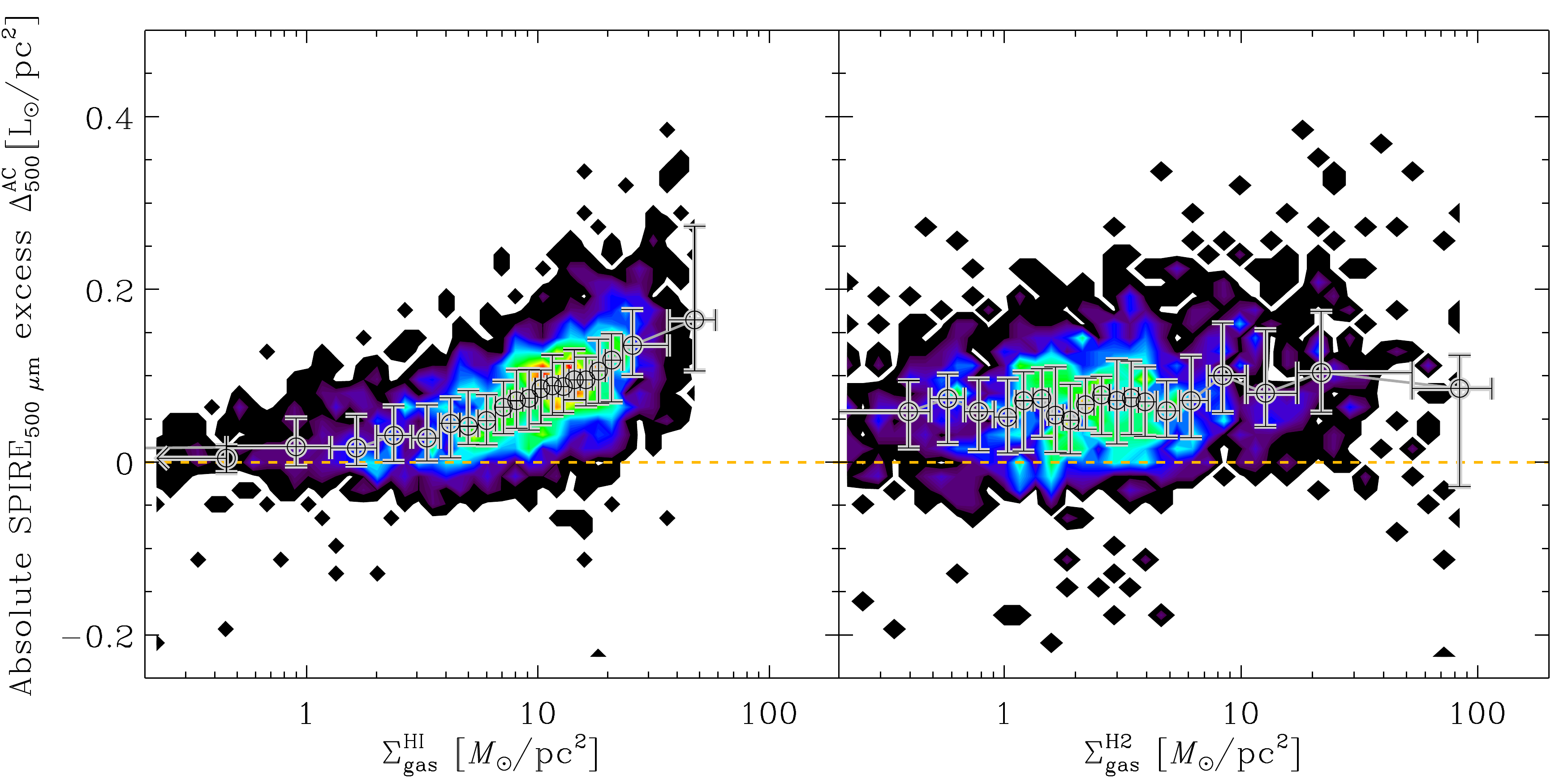}
  \caption{{\sl Pixel-to-pixel \SPIREiii\ absolute excess as a function of
            the column density of the atomic, and detected molecular phase.}
            The spatial resolution is R4, the dust model is \citengl{AC}.
            The color scale is identical to \reffig{fig:correldarkgas}.
            The error bars correspond to the binned trends.
            $\Sigma_\sms{\hmol}$ is the column density of molecular gas derived 
            from the CO line observations, but does not account for the
            dark component detected in \refsec{sec:darkCO}.
            The yellow dashed line shows the zero value (no excess).}
  \label{fig:delta}
\end{figure*}
Let's look in more details at the correlation of the excess with different quantities, in order to decipher its origin.
\reffig{fig:delta} shows the relation of the absolute excess to the gas phases.
It shows that the absolute excess is not correlated with the molecular gas, but is correlated with the \hi\ column density.
It therefore suggests that the \SPIREiii\ excess is associated with the atomic medium but not with dense phases.
To go further, we need to look into the various interpretations that have been proposed, for submm excesses in galaxies.

  \subsubsection{Consistency Test of Different Interpretations}
  \label{sec:testr500}

A submm emission excess has been reported in different systems.
We warn the reader that this excess extends up to cm wavelengths.
It is still uncertain if the same physical process is responsible for the entire 
wavelength range, or if it is the combination of several phenomena.

\citet{reach95} first reported a long wavelength excess in the {\it COBE}/FIRAS
spectrum of the Galaxy.
This excess could be fit with a very cold dust component (4-7~K).
However, the authors rejected this solution, since it was located at high Galactic latitudes.
This excess was fit by \citet{li01} invoking a long wavelength 
enhancement of the opacity of their silicate grains (see also \refapp{ap:mie}).
\citet{galliano03} discovered a submm excess in the SCUBA and MAMBO
observations of the blue compact galaxy \ngc{1569}.
Then several studies reporting similar excesses in galaxies, mainly 
low-metallicity dwarf galaxies were published, \eg: \citet{dumke04}, 
\citet{galliano05}, \citet{bendo06}, \citet{galametz09,galametz10}, 
\citet{ohalloran10}.
The relative intensity of this excess appears to decrease with metallicity
(\eg\ the sequence SMC, LMC, MW).
\citet{planck-collaboration11} presented the global excess at long 
wavelength.

The main interpretations of this excess, found in the literature, are the following.
\begin{enumerate}
  \item \citet{galliano03,galliano05} discussed the consequences of a very cold 
        dust component, showing that it was not inconsistent if the emissivity 
        index of these grains was $\beta\simeq1$.
        This scenario would imply a small number (at most a few hundreds) of 
        very dense, parsec-size clumps, containing a large fraction 
        ($40-80\,\%$) of the ISM mass.
  \item \citet{meny07} showed that the excess was successfully fit with a 
        physical model of temperature dependent grain emissivity. 
        Their model implies that the emissivity index decreases, 
        when the temperature of the grain increases.
  \item \citet{bot10} have shown that the long wavelength spectrum of the 
        excess renders the sole very cold hypothesis impossible.
        It has to be combined with another process to be realistic: 
        \eg\ very cold dust and spinning dust grains \citep{draine98b}.
        Alternatively, \citet{planck-collaboration11} proposed that the excess 
        in the SMC could be a combination of the \citet{meny07} grain model
        and of spinning grains.
        However, \citet{planck-collaboration11} attributed the excess in the 
        LMC to CMB fluctuations.
        \citet{bot10} explained the excess in the LMC and SMC with a combination 
        of 12~K dust and spinning grains \citep{ysard10}.
\end{enumerate}
Here, we discuss $r_{500}$ in light of these scenarios.

\begin{figure*}[h!tbp]
  \centering
  \begin{tabular}{cc}
    \includegraphics[width=0.47\linewidth]{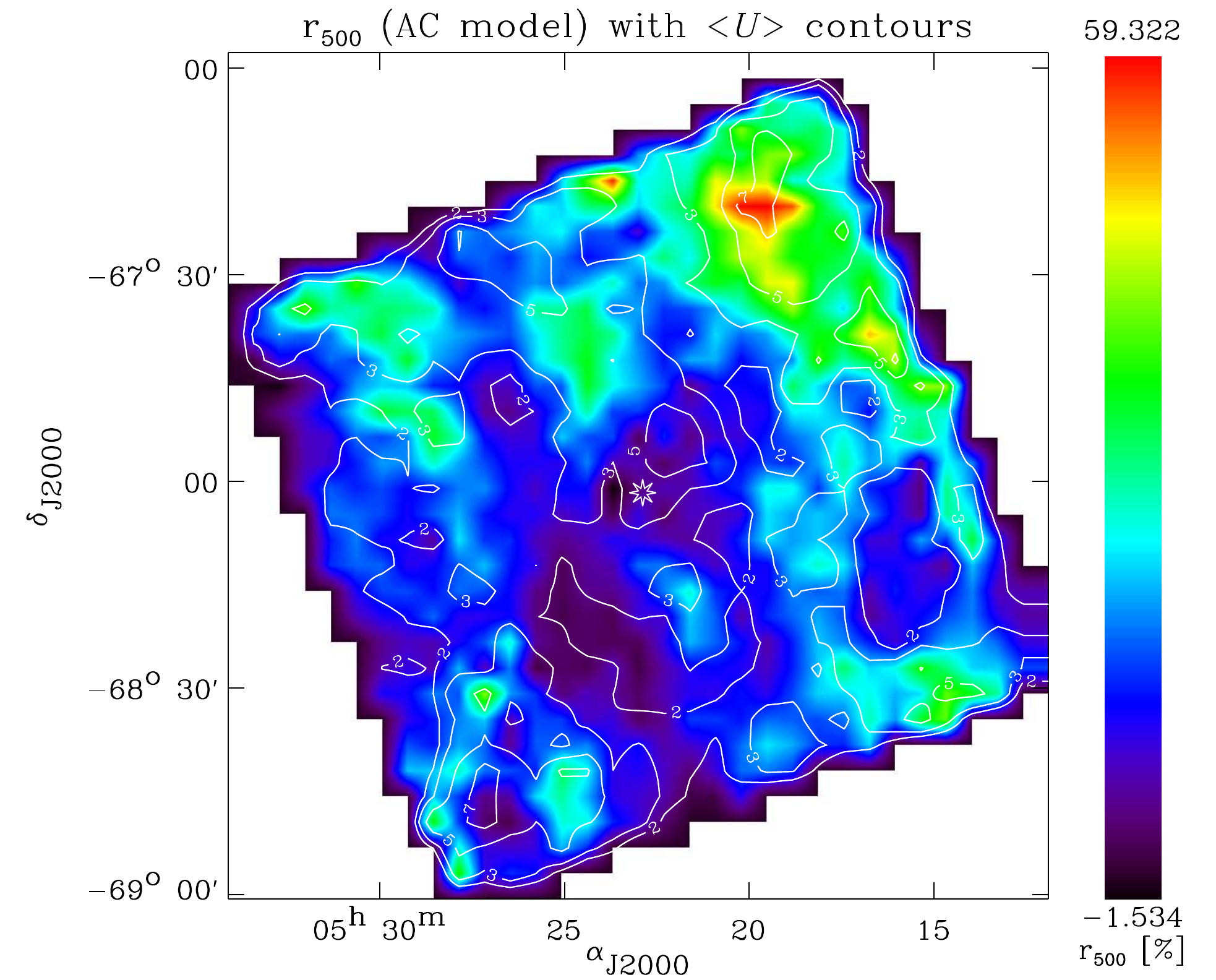} &
    \includegraphics[width=0.47\linewidth]{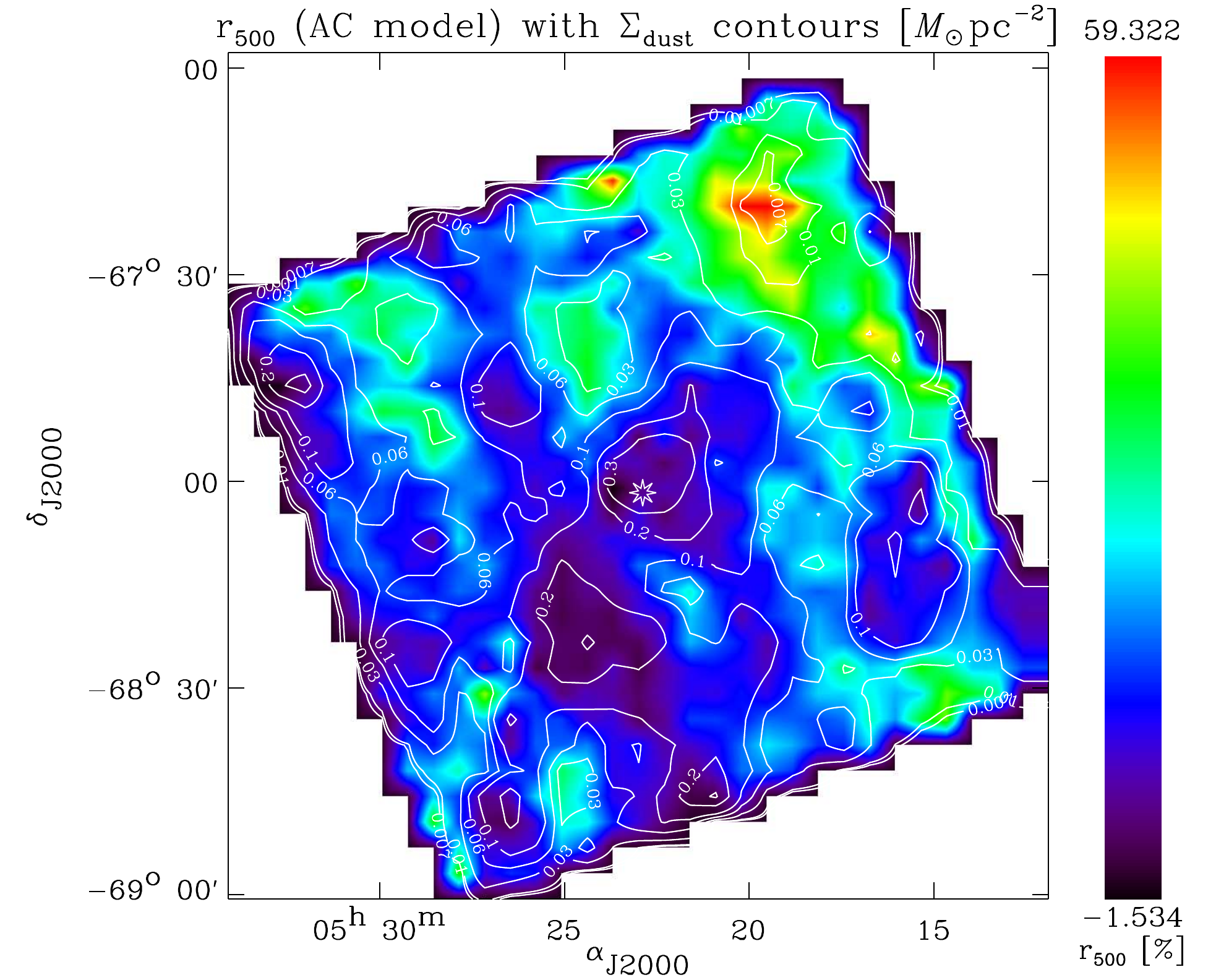} \\
  \end{tabular}
  \caption{{\sl Zoom on N$\,$44 and its surroundings.}
           The color image is $r_{500}^\sms{AC}$ at R4, with the \citengl{AC 
           model}, 
           for both panels.
           The contours are $\langle U\rangle^\sms{AC}(R4)$ for the left panel,
           and $\Sigma_\sms{dust}^\sms{AC}(R4)$ for the right panel.
           The white central star shows the location of the IR peak emission.
           The distributions are very similar, with the two models.}
  \label{fig:imn44}
\end{figure*}
\begin{figure*}[h!tbp]
  \centering
  \includegraphics[width=0.95\linewidth]{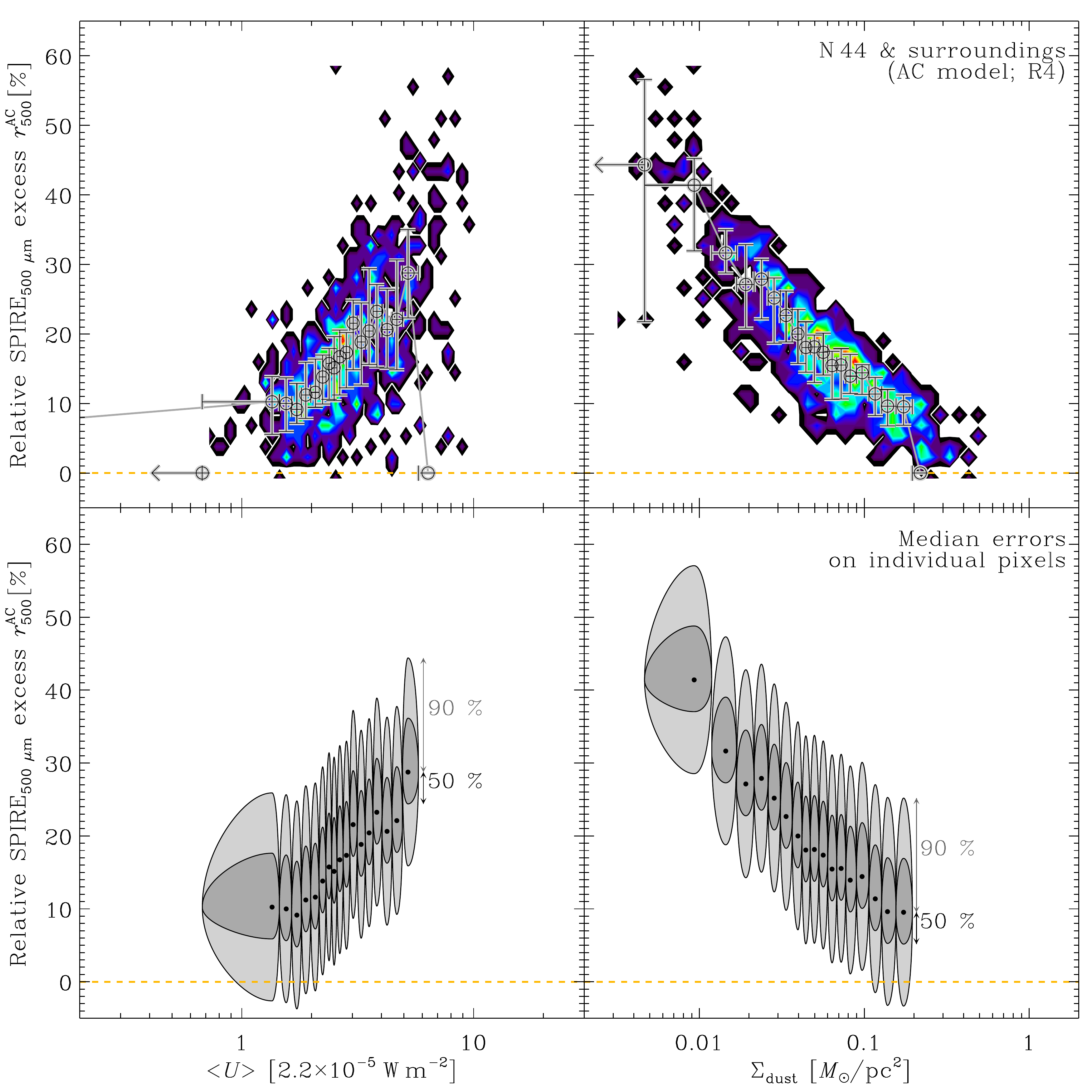}
  \caption{{\sl Pixel-to-pixel variations of $r_{500}$ around N$\,$44, for the 
           \citengl{AC model}.}
           This relation includes only the pixels of the window centered around
           N$\,$44 and displayed in \reffig{fig:imn44}.
           The spatial resolution is R4.
           The two top panels show the pixel density number correlations 
           (colors; scale indentical to \reffig{fig:correldarkgas}).
           The error bars represent the stacked trends in different bins.
           The central positions and the error bars are the median of the pixel 
           distribution.
           They account for the scattering of the relation, but not for the
           error on individual pixels.
           On the contrary, the two bottom panels show the same trends, but the
           ellipses are the median error bars on individual pixels 
           (dark: 50$\,\%$; light: 90$\,\%$).
           The excess is significant when looking at resolved scales.
           The yellow dashed line shows the zero value (no excess).}
  \label{fig:r500_n44}
\end{figure*}
Since the actual value of this excess is very sensitive to the noise, let's
zoom into the brightest star forming region of our observations: N$\,$44 and its surroundings.
\reffig{fig:imn44} shows the central distribution of the excess around N$\,$44,
compared to the starlight intensity and the dust mass column density.
\reffig{fig:r500_n44} shows the corresponding pixel-to-pixel variations.
It also demonstrates the typical error bars on individual pixels, showing that
the excess is significant when looking at spatially resolved regions.
Both $\langle U\rangle$ and $\Sigma_\sms{dust}$ are correlated with $r_{500}$.
However, the correlation is better with $\Sigma_\sms{dust}$.
This is confirmed by the spatial distribution: although, on average the excess
corresponds to high starlight intensity regions, there is one spot north of N$\,$44, with no excess and high starlight intensities.
Therefore, our observations indicate that the excess is primarily
associated with diffuse regions.
As a consequence, this excess can not come from contamination by the \COvtoiv\
line.

In regard to these trends, the very cold dust hypothesis is very unlikely.
Indeed, we would expect very dense clumps to be associated with dense regions.
On the contrary, the observations show that most of these clumps would be present in the diffuse ISM, and very few or none of them in dense regions.
However, this is only a qualitative argument showing that very cold dust is unlikely.
To rule out this component, we develop the following, more quantitative point of view.
In order to reach temperatures below 10~K, very cold dust has to be efficiently shielded. 
And, the dust responsible for the shielding is going to absorb the ambient radiation and reach warm temperatures, emitting in the wavelength range that has been modelled in the previous sections.
We can therefore estimate the mass of shielding dust necessary to allow the very
cold dust to reach low temperatures.
By comparing the order of magnitude of the minimum mass of shielding dust to the observed mass in each pixel, \refapp{ap:VCD} shows that very cold dust can reasonably be ruled out.

Incidentally, by showing:\textlist{\thetextlist~that very cold dust is unlikely to contribute significantly to \SPIREiii\ and 
\thetextlist~that the \SPIREiii\ excess is found mainly in diffuse regions,} we 
have demonstrated that this excess does not have any impact on our main discussion about dust mass (\refsec{sec:darkCO}).


\section{Summary and Conclusion}
\label{sec:conclusion}

In this paper, we have presented the modelling of the spatially resolved \spitz/IRAC, \spitz/MIPS and \hersc/SPIRE data of a strip covering one quarter of the LMC.
The purposes of this work \modif{was} 
to:\textlist{\thetextlist~systematically study all the effects leading to 
     inaccuracies or biases affecting the dust mass estimate of a galaxy, in 
     the \hersc\ era;
   \thetextlist~explore the peculiar ISM properties of the LMC.}Our main results 
are the following.
\begin{enumerate}
  \item We have presented an empricial model to fit IR/submm SEDs.
        This model adopts realistic grain properties, and accounts for a 
        possible distribution of starlight intensity in the region studied.
  	    We have described in detail the general propagation of 
        the observational errors (noise and calibration) through the SED 
        fitting process.
        We have shown that, even when the signal-to-noise ratio is high, the 
        errors on the dust mass are important and strongly asymmetric 
        (typically $\simeq_{-25\,\%}^{+40\,\%}$).
        However, \emph{we have shown that relative quantities (ratio of two 
        parameters) can have small error bars} 
        (typically $\simeq_{-7\,\%}^{+10\,\%}$) due to the partial cancelation 
        of the correlated calibration errors.
  \item By modelling the same set of maps, but with different pixel sizes,
        we have shown that the lack of spatial resolution can lead to a 
        systematic underestimate of the dust mass by $\simeq50\;\%$.
        This bias could be the result of the veiling of cold components by the
        emission from warmer regions. 
        Although the amplitude of this bias is probably specific to the type 
        of environment found in the LMC and to the wavelength coverage of our 
        data set, we believe that this trend with spatial resolution 
        is general.
        \emph{Modelling the integrated SED of a galaxy leads
        to underestimating the dust mass.}
  \item We have performed our SED fitting with two sets of grain composition,
        in order to explore the sensitivity of the dust mass estimate to the
        submm emissivity of the grains.
        We have shown that both compositions give equally good fits, but that
        the \citengl{standard} grain composition (graphite \&\ silicate), that 
        works for the Milky Way, fails to give physically realistic results in 
        the LMC.
        It violates the elemental abundances.
        We discuss the fact that this discrepancy could either be due to 
        modified submillimeter grain opacities (our AC model: amorphous carbon 
        \&\ silicate), or to the presence of an unaccounted for gas reservoir.
        In particular, we have shown that \emph{there is a degeneracy between 
        the intrinsic grain emissivity and their temperature distribution, 
        and that considerations on the mass are a powerful way to remove some of
        this degeneracy.}
  \item The detailed spatial analysis of the variations of the observed 
        gas-to-dust mass ratio has proven the need for grain 
        opacities different than those of the Galaxy.
        More precisely, our analysis has demonstrated that \emph{grains in the 
        LMC have on average a larger far-IR/submm opacity}.
        We propose a physical dust model that is consistent with these 
        properties (emissivity index $\beta\simeq1.7$, and 
        $\kappa_\sms{abs}(160\mic)=1.6\;\rm m^2\,kg^{-1}$), although we insist 
        that this particular composition is not a unique solution.
  \item We have shown that the mass averaged starlight intensity is a better
        tracer of regions where the observed gas-to-dust mass ratio is depleted.
        Comparing different methods, we have constrained the amount of 
        \citengl{dark gas} (i.e. unaccounted for by \hiline, and \COio).
        \emph{This reservoir accounts for $\simeq10-100\,\%$ of the 
        gas mass} of the strip.
        Simply correcting the standard CO-to-\hmol\ conversion factor is not 
        sufficient to account for this dark gas reservoir, since it is abundant
        in regions where no CO detection is reported.
  \item We have analyzed the excess emission in the \SPIREiii\ waveband,
        which has been reported before in the LMC and other low-metallicity 
        galaxies.
  	    On average the amplitude of the relative excess is $\simeq15\,\%$, and
	    it can vary spatially between 0 and $\simeq40\,\%$.
        By scrutinizing its behaviour, we show that \emph{this excess is 
        mainly anticorrelated with the density of the ISM.}
        We show that this excess can not be due to CMB fluctuations, nor 
        to massive amounts of very cold dust.
        However, the nature of this excess remains unknown.
\end{enumerate}

Considering the uncertainties in 
grain optical properties, our study demonstrates that, when modelling the IR/submm SED of a galaxy, \emph{the derived dust mass can be considered as a constraint rather than a result}.
Indeed, we have shown that standard Galactic grain properties were 
leading to unphysical masses.
We therefore have selected another model which was giving reasonable results.
However, going from this conclusion to providing the reader with
an actual dust mass or gas-to-dust mass ratio for the LMC would be
a circular process.
It is now clear that grain properties can vary significantly from a galaxy
to another.
The range of gas-to-dust mass ratios that results from using different grain 
species makes the estimate of the absolute dust mass irrelevant,
without independent constraints.
Such independent constraints could be the elemental depletions, or 
spectroscopic information on the precise chemical composition of the grains, or 
simultaneous extinction measurements.


\appendix
\section{The Grain Properties of Our Models}
\label{ap:mie}
  
  \subsection{The Submillimetre Opacities}

\begin{figure}[h!tbp]
  \centering
  \includegraphics[width=0.95\linewidth]{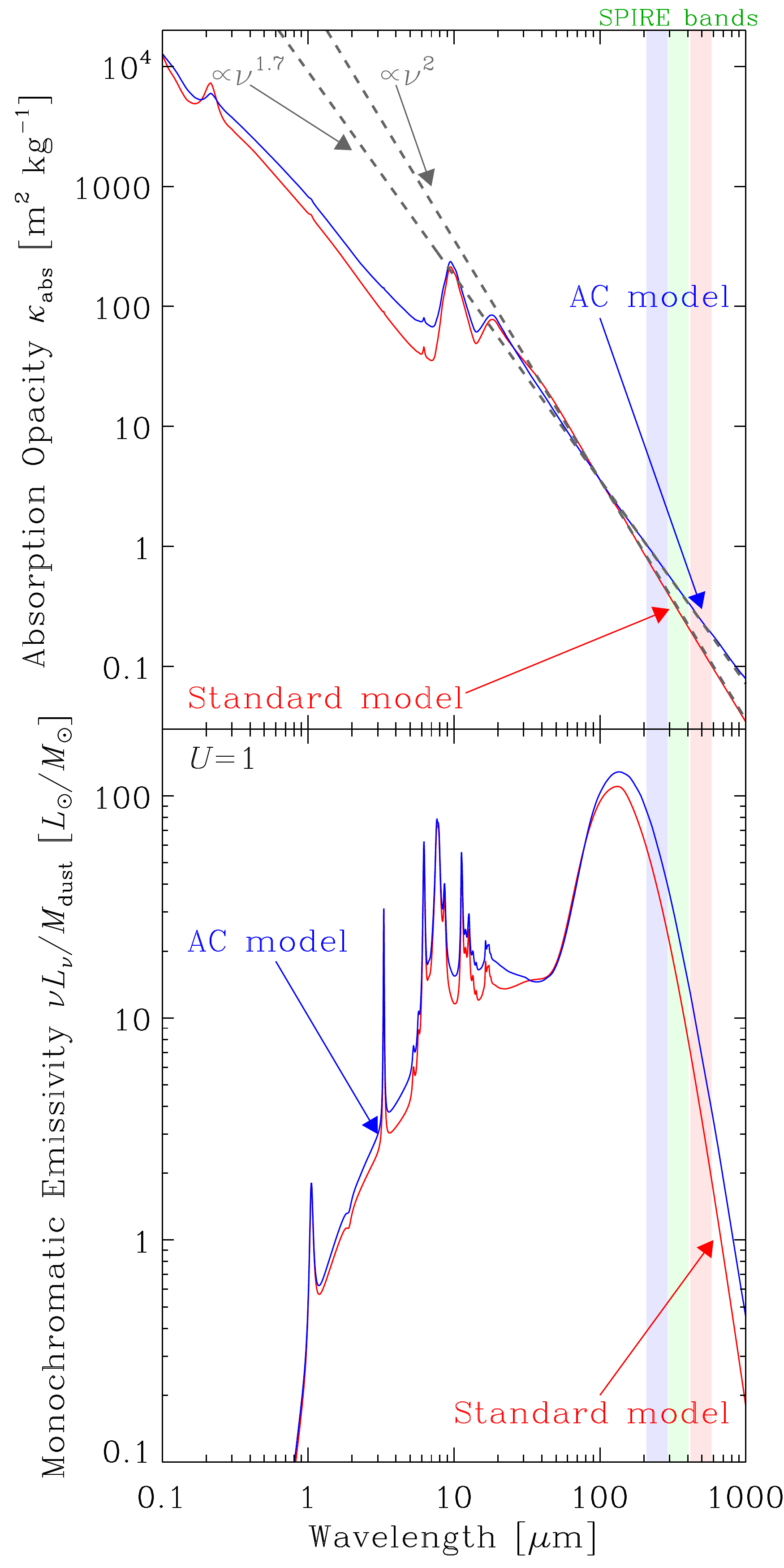}
  \caption{{\sl Comparison of the grain properties of our two models.}
           The top panels show the opacity of the two mixtures.
           The grey dash lines show long wavelength fits of these opacities,
           with empirical laws $\propto\nu^\beta$.
           The bottom panel compares the SEDs of the two grain mixtures 
           illuminated by the ISRF of the diffuse Galactic ISM ($U=1$).}
  \label{fig:kabs}
\end{figure}
The top panel of \reffig{fig:kabs} compares the absorption opacities
of the grain mixtures of our two models.
These opacities are the sum of PAHs, carbon grains and silicates.
The effective submillimeter emissivity index $\beta$ is defined by the logarithmic index of the opacity:
\begin{equation}
  \kappa_\sms{abs}(\lambda) \propto \lambda^{-\beta}.
  \label{eq:beta} 
\end{equation}
It is $\beta\simeq2$ for 
the \citengl{standard model} and $\beta\simeq1.7$ for the \citengl{AC model}, although $\beta$ for 
amorphous carbons alone is even lower.
\reftab{tab:kabs} gives the submilimeter properties of our two models approximated by \refeq{eq:beta}.
\begin{table}[h!tbp]
  \centering
  \begin{tabular}{lrr}
    \hline\hline
       & \citengl{Standard model} & \citengl{AC model} \\
    \hline
      $\beta$                     & 2      & 1.7 \\
      $\kappa_\sms{abs}(160\mic)$ & $1.4\;\rm m^2\,kg^{-1}$  
                                  & $1.6\;\rm m^2\,kg^{-1}$ \\
    \hline
  \end{tabular}
  \caption{{\sl Submillimeter properties of our dust compositions.}
           The two parameters $\beta$ and $\kappa_\sms{abs}(160\mic)$ are
           the parameters to approximate our dust opacities (\reffig{fig:kabs})
           with \refeq{eq:beta}.}
  \label{tab:kabs}
\end{table}

The bottom panel compares the infrared emission of the compositions, for the
same starlight intensity.
It demonstrates that the \citengl{AC model} has more emissivity, especially in the submm.
It therefore allows us to fit the same observed SED with less mass, and slightly hotter grains.

Fitting the {\it COBE}/FIRAS high latitude Galactic \modif{cirrus}, \citet{li01} had 
to modify the imaginary part of the dielectric function of their silicate grains, at wavelengths greater than $250\mic$.
This modification would have a very limited effect on our conclusion, since
our constraints go up to $350\mic$ only, where this increase is very limited (it is $\pm 12\;\%$ in the $250\mic\leq\lambda\leq1000\mic$ range).
This modification actually lowers the emissivity between 250 and 850~\mmic, and
increases it, at $\lambda>850\mic$.
For our dust mixture, the amplitude of this modification is even lower, since the
contribution of silicates to the far-IR is lower than for the \citet{li01} model.
The difference between the \citengl{standard model} and the one using the modified silicate of \citet{li01} is invisible on \reffig{fig:kabs}.
The physical origin of this excess may be similar to the one we find here.
However, its amplitude is much larger in the LMC, and manifests at shorter wavelengths ($\lambda\gtrsim100\mic$).

  \subsection{Size Distribution Considerations}

It is important to note that the size distribution used here does not include
grains larger than $0.35\mic$, and the contribution to the emission of grains larger than $a\gtrsim 0.1\mic$ is negligible.
Large grains tend to have a flat UV-visible opacity, and therefore, to have
a lower equilibrium temperature than smaller equilibrum grains, exposed
to the same ISRF.
Therefore, adding larger grains would increase the submillimetre emissivity of 
the model, and allow us to fit the SPIRE fluxes, without having to go to low
starlight intensities.
However, having large grains or having colder dust would give the same 
discrepancies in terms of gas-to-dust mass ratios.
Moreover, a significant excess of large grains would flatten the UV rise of the extinction curve, contradicting the observations of several lines of sight within the LMC \citep{gordon03}.

Another feature of our model is that we have been forced to lower the abundance
of non-PAH small grains ($a<10$~nm; both carbon and silicate grains) by a factor of 2.
Without this modification, we were not able to get good fits of the \MIPSi\ of the diffuse regions.
This modification has a very minor effect on the dust mass ($\lesssim10\;\%$), and it is systematic, thus it has no impact on our conclusions.
However, this is puzzling since the fit of the extinction curves of the LMC indicates a larger fraction of small grains \citep[{\eg}][]{weingartner01}.
The fact is that the \citet[][BARE-GR-S]{zubko04} model has a higher small grain contribution in \MIPSi\ than the \cite{draine07}.
On the other hand, replacing graphite by amorphous carbon, as shown in this paper, allows us to get rid of the $30\mic$ graphite feature, and decrease the contribution of small grains in the \MIPSi\ band, making the $24\mic$ fit better without having to alter the size distribution.
This is another indirect consistency check of the conclusion of this paper.

\subsection{Overview of the Derived PAH Properties}
\label{ap:PAH}

Although the scope of our paper was not to discuss the PAH properties, their abundance is a natural output of our model.
In this section, we summarize these results.

\begin{table}[h!tbp]
  \centering
  \begin{tabular}{l|rr|rr}
    \hline\hline
       & \multicolumn{2}{c}{\citengl{Standard model}} 
                 & \multicolumn{2}{c}{\citengl{AC model}} \\
    \hline
R1 & $0.67_{-0.03}^{+0.03}$ & $[$$0.57$,$0.75$$]_{90\,\%}$ & $0.78_{-0.03}^{+0.04}$ & $[$$0.71$,$0.89$$]_{90\,\%}$ \\
R2 & $0.67_{-0.03}^{+0.03}$ & $[$$0.57$,$0.75$$]_{90\,\%}$ & $0.78_{-0.03}^{+0.03}$ & $[$$0.71$,$0.88$$]_{90\,\%}$ \\
R3 & $0.66_{-0.03}^{+0.03}$ & $[$$0.59$,$0.74$$]_{90\,\%}$ & $0.78_{-0.03}^{+0.03}$ & $[$$0.71$,$0.88$$]_{90\,\%}$ \\
R4 & $0.66_{-0.03}^{+0.03}$ & $[$$0.59$,$0.74$$]_{90\,\%}$ & $0.78_{-0.03}^{+0.03}$ & $[$$0.70$,$0.87$$]_{90\,\%}$ \\
R5 & $0.66_{-0.03}^{+0.03}$ & $[$$0.60$,$0.73$$]_{90\,\%}$ & $0.78_{-0.03}^{+0.03}$ & $[$$0.71$,$0.87$$]_{90\,\%}$ \\
R6 & $0.65_{-0.03}^{+0.03}$ & $[$$0.59$,$0.73$$]_{90\,\%}$ & $0.77_{-0.03}^{+0.03}$ & $[$$0.69$,$0.86$$]_{90\,\%}$ \\
R7 & $0.64_{-0.02}^{+0.03}$ & $[$$0.59$,$0.72$$]_{90\,\%}$ & $0.77_{-0.03}^{+0.03}$ & $[$$0.70$,$0.86$$]_{90\,\%}$ \\
R8 & $0.64_{-0.03}^{+0.02}$ & $[$$0.59$,$0.71$$]_{90\,\%}$ & $0.79_{-0.03}^{+0.03}$ & $[$$0.72$,$0.89$$]_{90\,\%}$ \\
R9 & $0.62_{-0.03}^{+0.03}$ & $[$$0.57$,$0.69$$]_{90\,\%}$ & $0.78_{-0.03}^{+0.03}$ & $[$$0.71$,$0.88$$]_{90\,\%}$ \\
R10 & $0.62_{-0.03}^{+0.03}$ & $[$$0.57$,$0.69$$]_{90\,\%}$ & $0.77_{-0.03}^{+0.03}$ & $[$$0.71$,$0.86$$]_{90\,\%}$ \\
    \hline
  \end{tabular}
  \caption{{\sl PAH mass fraction as a function of spatial resolution.}
            The quantity $f_\sms{PAH}$ (\refeqnp{eq:lnu}; \reftab{tab:parm})
            is the PAH-to-total-dust mass ratio, divided by the Galactic value 
            ($4.6\,\%$).
            In other words, $f_\sms{PAH}=1$ in the Galaxy.
            The convention for error display is defined in \refeqs{eq:median} 
            and (\ref{eq:sup}).}
  \label{tab:PAH}
\end{table}
\reftab{tab:PAH} shows the PAH mass fractions for the two models.
This parameter is relatively well constrained and does not vary significantly with spatial resolution.
Indeed, it depends essentially on the \IRACiv-to-total-IR luminosity ratio.
The mass fraction for the \citengl{AC model} is systematically higher, since the bulk of the dust is more emissive.

\begin{figure}[h!tbp]
  \includegraphics[width=0.95\linewidth]{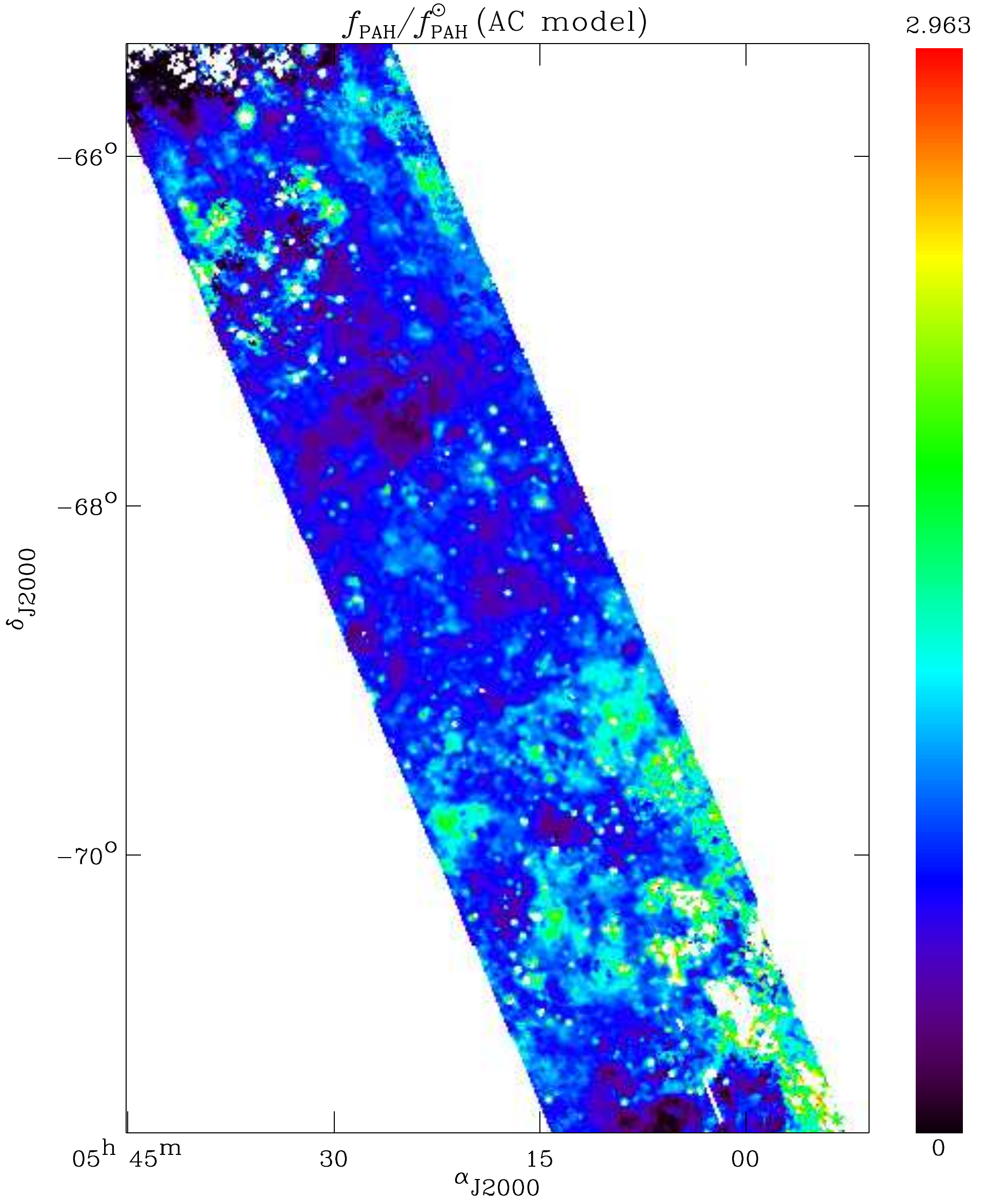}
  \caption{{\sl PAH mass fraction, $f_\sms{PAH}$.}
           The map is shown for the \citengl{AC model}, at spatial resolution 
           R1 (54~pc).}
  \label{fig:mapah}
\end{figure}
\reffig{fig:mapah} shows the spatial distribution of the mass fraction of PAHs,
$f_\sms{PAH}$ \refeqp{eq:lnu}.
We confirm the \citet{paradis09} results showing an excess of PAH emission 
toward the stellar bar.
However, $f_\sms{PAH}$ is biased by the fact that, in absence of detailed mid-IR spectrum, we have arbitrarily fixed the charge fraction to $1/2$.
The charge fraction controls the emissivity of the C-C modes \citep{galliano08b}.
Therefore, it is difficult to uniquely interpret this excess emissivity by a local increase of the PAH abundance.

\section{Details Concerning the Error Analysis}
\label{ap:MC}

  \subsection{Classes of Observed SEDs}

The error analysis presented in \refsec{sec:error} would imply having to fit
the $156\,577$ pixels (\reftab{tab:dolls}) 300 times with the two models, leading to
a total of $\simeq 10^8$ fits.
However, all these SEDs have a lot of similarities, and it is not necessary to perform the Monte Carlo iterations on each pixels.
Here we describe the approximation method we demonstrated in \reffig{fig:dist_MC}.

\begin{figure}[h!tbp]
  \centering
  \includegraphics[width=0.95\linewidth]{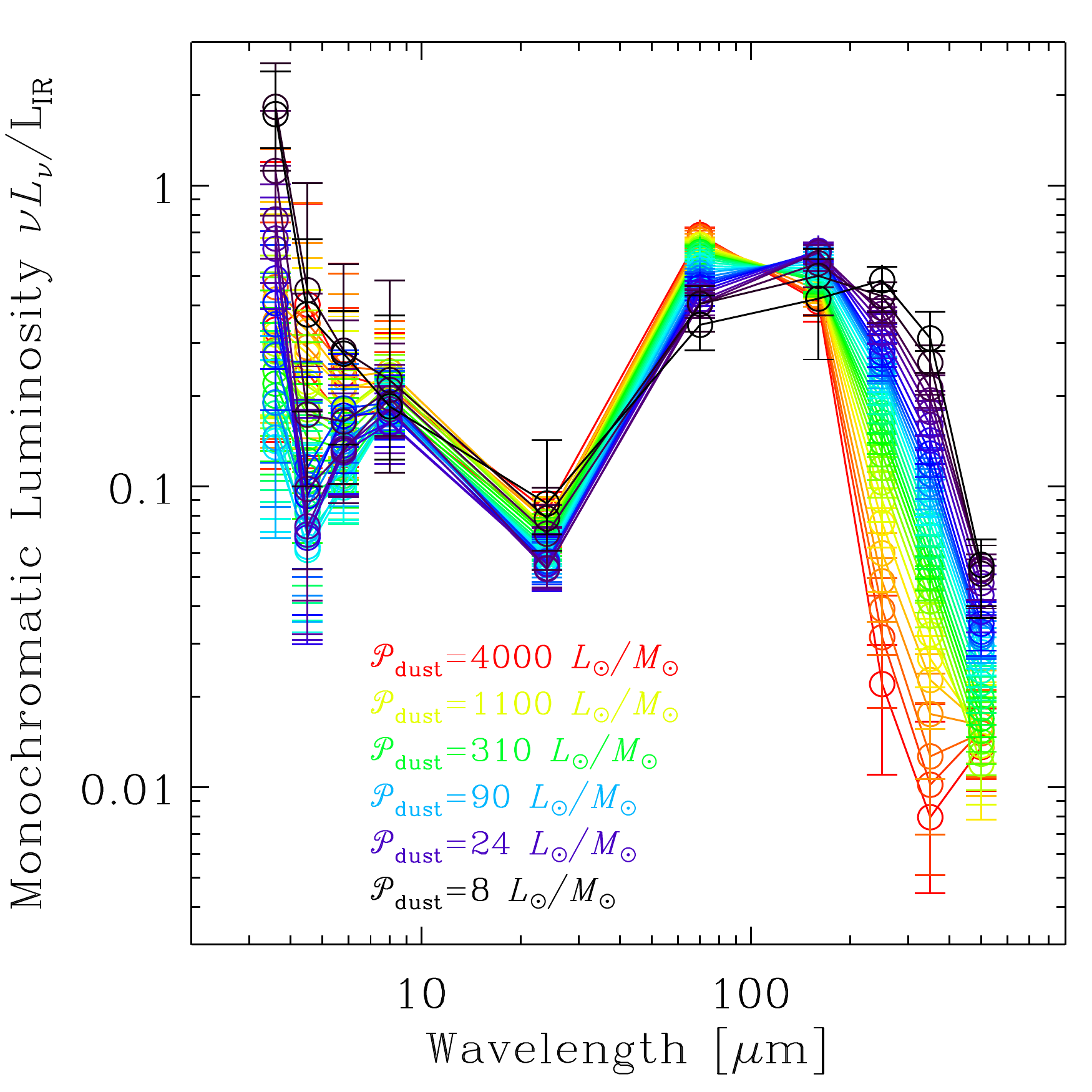}
  \caption{{\sl Classes of observed SEDs for the \citengl{standard model}.}
           The SEDs are normalised by their integrated IR luminosity, 
           $L_\sms{IR}$.
           The classes are ordered according to their specific power 
           $\mathcal{P}_\sms{dust}$.
           The classes derived with the \citengl{AC model} are very similar.
           Only the specific power is systematically scaled up by a factor of
           $\simeq1/0.38_{-0.02}^{+0.04}$ (\reffig{fig:comparison}).}
  \label{fig:sedclass}
\end{figure}
We start from the actual fit of the unperturbed SED of each pixel of \reftab{tab:dolls}.
We order these SEDs according to their specific emitted power $\mathcal{P}_\sms{dust}=L_\sms{IR}/M_\sms{dust}$, making 30 logarithmic bins.
\reffig{fig:sedclass} shows the classes of observed SEDs, for the \citengl{standard model}.
At each waveband $\lambda_0$, the central value is the median of the 
normalized monochromatic power $\lambda_0L_{\lambda_0}/L_\sms{IR}$ of each
pixel within the considered bin of specific power.
The error shows the dispersion of the pixels within the bin.
On first approximation, the specific power is proportional to the mass averaged starlight intensity $\langle U\rangle$.
That is the reason why the SEDs of \reffig{fig:sedclass} are nicely ordered according to the wavelength peak of the dust emission.
Notice that the dispersion at most wavebands is small compared to the variation spread by the different classes.
The only wavelength range where it is not true is the near-IR ($\lambda\lesssim5\mic$), where the stellar emission dominates.

  \subsection{Class Interpolation and Error Estimate}

We perform the Monte Carlo error analysis of \refsec{sec:MC} on each SED class, for 6 different noise levels, spanning the whole range of observed signal-to-noise ratios, S/N.
We perform this study with the two models.
Then, for each pixel of \reftab{tab:dolls}, the errors on the parameters are determined by logarithmically interpolating the precomputed errors in $\mathcal{P}_\sms{dust}$ and S/N.

The validity of this interpolation method is demonstrated in \reffig{fig:dist_MC}.
We show that this method reproduces well the central value and error bars of each parameter, since it reproduces well the skewness of the probability distribution.

\subsection{Biases of the SED Fits}
\label{ap:biases}

\begin{figure}[h!tbp]
  \centering
  \includegraphics[width=0.95\linewidth]{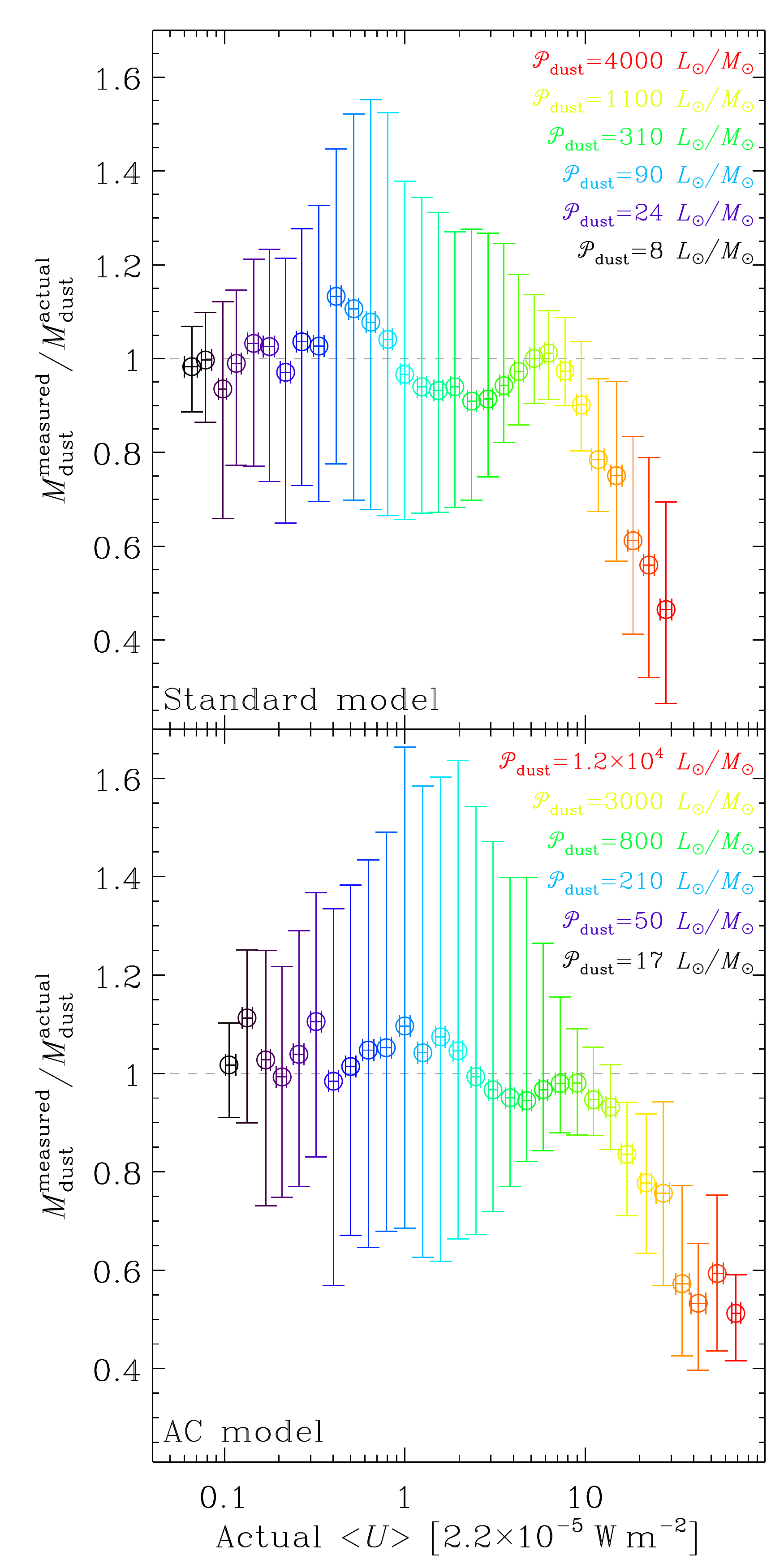}
  \caption{{\sl Bias in the dust mass estimate as a function of the starlight 
            intensity, for both models.}
            In each panel, we plot the ratio of the median of the dust masses
            measured by fitting the randomly perturbed SEDs of 
            \reffig{fig:sedclass} ($M_\sms{dust}^\sms{measured}$) to the 
            actual dust mass of the unperturbed SED 
            ($M_\sms{dust}^\sms{actual}$).
            This quantity is plotted as a function of the actual average 
            starlight intensity $\langle U\rangle$.
            The color code of the classes is identical to 
            \reffig{fig:sedclass}.
            We clearly see that for high starlight intensities, the dust mass
            is systematically underestimated by a factor up to 2.}
  \label{fig:bias_Md}
\end{figure}
The large database of Monte-Carlo fits of the SEDs of \reffig{fig:sedclass}
can be used to study the systematic effects \modif{of} our method.
\reffig{fig:bias_Md} shows the bias on the dust mass, as a function of the mean equilibrium grain temperature.

The two panels of this figure correspond to the two models.
The $x$-axis is the \citengl{actual} $\langle U\rangle$, \ie\ the value of $\langle U\rangle$ derived from the unperturbed SEDs of \reffig{fig:sedclass}.
Each value of $\langle U\rangle$ defines one of the 30 classes, since $\langle U\rangle\propto\mathcal{P}_\sms{dust}$.
The $y$-axes show the ratio between: $M_\sms{dust}^\sms{measured}$, which is the
median of the mass estimates derived from the fitting of the 300 Monte-Carlo perturbed SEDs of each class; and $M_\sms{dust}^\sms{actual}$ which is the dust
mass corresponding to the unperturbed SED.
In other words, $M_\sms{dust}^\sms{measured}/M_\sms{dust}^\sms{actual}$ quantifies the deviation from the dust mass derived from an SED
fit, to its actual value, as a function of the starlight intensity.

\reffig{fig:bias_Md} shows that up to $\langle U\rangle^\sms{Std}\lesssim 6$ 
and $\langle U\rangle^\sms{AC}\lesssim9$, the dust masses derived from SED
fits are not significantly biased (\ie\ the deviation is smaller than the error bars).
However, for starlight intensities higher than these values, the SED fit tends to systematically underestimate the dust masses (\ie\ overestimate the gas-to-dust mass ratio).
The amplitude of this effect can go up to a factor of $\simeq2$, for high 
$\langle U\rangle$.

This is a demonstration of the effect invoked to explain the low gas-to-dust mass ratio, at high starlight intensities, in \reffig{fig:dark_main}.
Fortunately, this effect concerns only a regime containing a small fraction of 
the pixels.
Moreover, these pixels are the less massive, since they correspond to hot and 
diffuse regions.
Therefore, this bias does not have a significant impact on our global dust mass estimate.

\section{Relevance of Our Starlight Intensity Distribution}

  \subsection{The Unnecessariness of Adding a Diffuse Field Component}
  \label{ap:compD07}

We note that our approach \modif{of} modelling IR SEDs with an empirical combination of starlight intensities, is common in the literature.
In particular, \citet{draine07b} modelled the SEDs of the SINGS galaxies with
an extra, uniformly illuminated, component:
\begin{eqnarray}
  \frac{\dd M_\sms{dust}^\sms{extra}}{\dd U} & = &
    \gamma M_\sms{dust}\times\frac{(\alpha-1)}{U_\sms{min}^{1-\alpha}-(U_\sms{min}+\Delta U)^{1-\alpha}}U^{-\alpha} \nonumber\\
    & + & (1-\gamma)M_\sms{dust}\times\delta(U-U_\sms{min}),
   \label{eq:DL07}
\end{eqnarray}
where $\gamma$ is an extra parameter controlling the mass fraction of the 
$U^{-\alpha}$ component.
This formulation implicitly assumes that there are no massive quantities of dust colder than the diffuse ISM.
One of the advantages of this extra component is to avoid the dust mass to diverge, in the absence of submm constraints.
Indeed, as demonstrated by \citet{galametz11}, modelling the SEDs of galaxies
without submm data, using \refeq{eq:dale}, leads to gross errors on the 
dust mass. 
However, in our case, we have the valuable submm constraints provided by \hersc.
The relevance of this extra component can be tested using the Monte-Carlo method, described in \refsec{sec:MC}.

\begin{figure}[h!tbp]
  \includegraphics[width=0.95\linewidth]{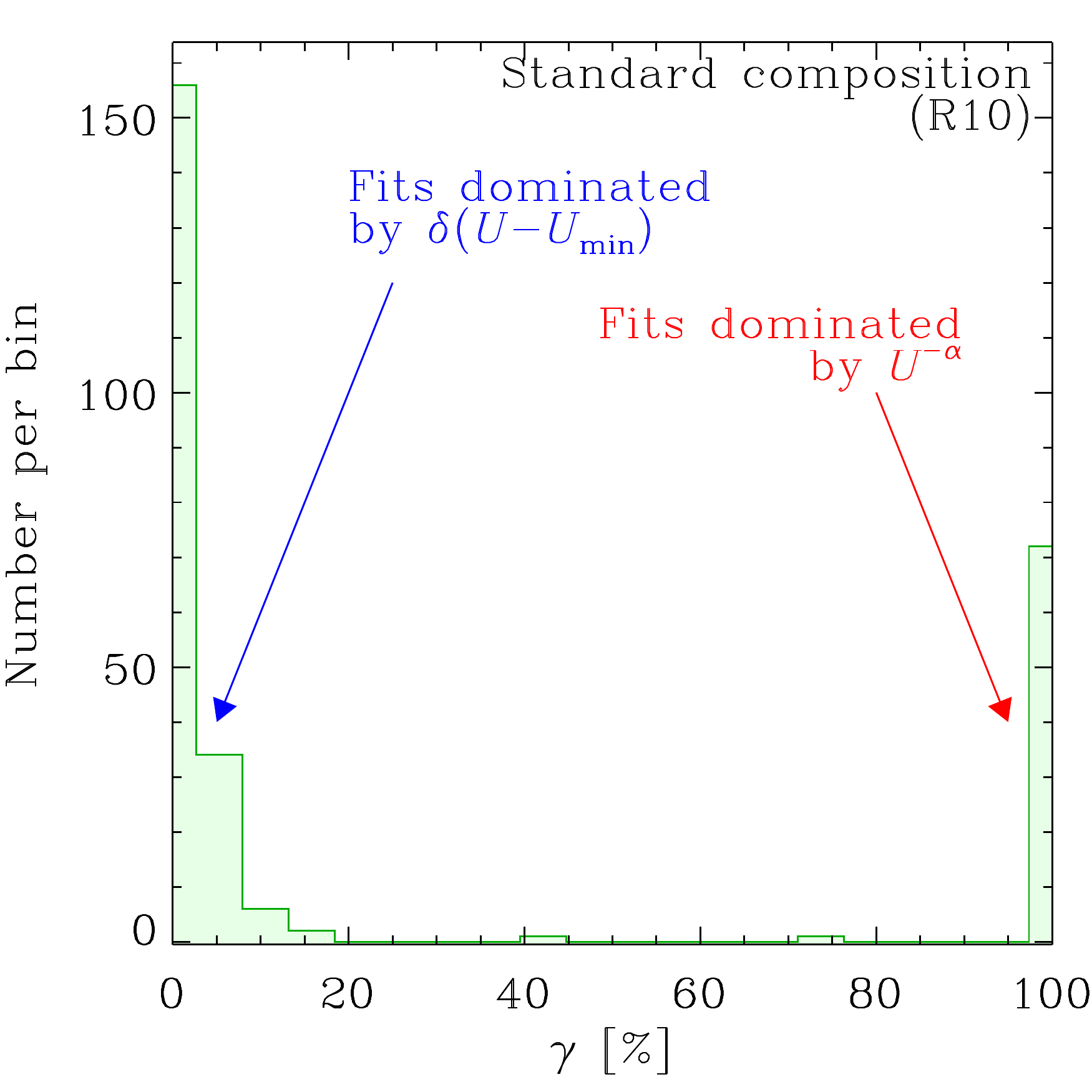}
  \caption{{\sl Test of the stability of the starlight intensity distribution of 
            \refeq{eq:DL07}.}
            The parameter $\gamma$ is the mass fraction of the component in
            $U^{-\alpha}$.
            The histogram is the distribution of the Monte-Carlo iterations of 
            the fit
            of \refeq{eq:DL07} to the integrated strip (R10), with the grain 
            composition of the \citengl{standard model}.
            The bimodality of the distribution is the sign of the instability of 
            this formalism.}
  \label{fig:test_DL07}
\end{figure}
First, we tested the stability of this extra component.
\reffig{fig:test_DL07} shows the distribution of the Monte-Carlo values 
of $\gamma$, when fitting the integrated SED strip (R10), with \refeq{eq:DL07}
and the \citengl{standard} grain composition.
This distribution is drastically bimodal.
Most of the perturbed SEDs are fit either without the extra 
$\delta(U-U_\sms{min})$ components ($\gamma\simeq1$) --~ which is equivalent to our model~-- or with a sole uniformly illuminated SED ($\gamma\simeq0$).
In other words, the value of the parameter $\gamma$ is totally uncertain.

Second, we tested the statistical relevance of adding this extra component.
To quantify this aspect, we performed a F-test.
We computed the statistic $F_\chi$ \citep{bevington03}: 
\begin{equation}
  F_\chi = 
  \frac{\chi^2_\sms{Std}-\chi^2_\sms{extra}}{\chi^2_\sms{extra}/(n-m_\sms{extra}-1)},
\end{equation}
where $\chi^2_\sms{Std}$ is the $\chi^2$ \refeqp{eq:chi2} of the fit of a given perturbed SED of the integrated strip with the \citengl{standard model}, and
$\chi^2_\sms{extra}$ is the corresponding value for \refeq{eq:DL07};
$n=10$ is the number of wavebands, and $m_\sms{extra}=7$ is the number of free parameters for the model of \refeq{eq:DL07} plus the stellar component of \refeq{eq:model}.
$F_\chi$ can be seen as a measure of how much the additional term has improved the value of the reduced $\chi^2$.
Statistically, $F_\chi\simeq 0.052_{-0.027}^{+0.023}\ll 0.67$ is lower than the value of the F-distribution, with $n-m_\sms{extra}-1=2$ degrees of freedom, with a probability of exceeding $F_\chi$ of $50\,\%$ \citep[or $P_F(0.67,1,n-m_\sms{extra}-1)=50\,\%$;][]{bevington03}.
In other words, adding the extra component of \refeq{eq:DL07} does not significantly improve the $\chi^2$. 
Our starlight intensity distribution \refeqp{eq:dale} is therefore statistically more significant.

Finally, we tested the conservativeness of the dust mass estimate with
\refeq{eq:DL07}.
We have performed the fits of exactly the same perturbed SEDs that were used 
in \reffig{fig:test_DL07}, with our \citengl{standard model} \refeqp{eq:dale}.
The ratio of the mass of our model to the mass obtained with the extra component is 
$M_\sms{dust}^\sms{Std}/M_\sms{dust}^\sms{extra}\simeq 1.000_{-0.045}^{+0.010}$.
In other words, the two starlight intensity distributions of \refeqs{eq:dale} and (\ref{eq:DL07}) give statistically identical dust masses, within a few percents.

In summary, for the analysis performed in this paper, the starlight intensity distribution of \refeq{eq:DL07} is not physically motivated, neither statistically stable nor relevant, and gives identical results to \refeq{eq:dale}.
Our model has an appropriate balance between free parameters and observational constraints.
It is flexible enough, but it does not lead to overinterpreting the data.

  \subsection{Comparison of Our Model with the Isothermal Approximation}
  \label{ap:isoth}

As justified in \refsec{sec:model}, our model accounts for a distribution of equilibrium dust temperatures within each pixel \refeqp{eq:dale}.
This formalism allows us to fit the submm slope more accurately than with a 
single temperature.
As a consequence, our model predicts more mass than a single black body fit having the same dust opacity.
Since the single black body approach is still widely used, even in the \hersc\ era, we have performed a systematic comparison in order to quantify the biases of such an approach.

To perform our comparison, we have fitted the R4 (54~pc) map with a single modified black body having the same grain opacity as our \citengl{standard model}: $\kappa_\sms{abs}(160\mic)=1.4\;\rm m^2\,kg^{-1}$, and $\beta=2$ (\reffig{fig:kabs}).
For each pixel, the mass and the temperature are free to vary, but $\beta$ is kept fixed.
We constrain this model with the \MIPSii, \MIPSiii, \SPIREi\ and \SPIREiii\ fluxes, weighted the same way as our complete model \refeqp{eq:chi2}, except that the weight of the \MIPSii\ flux is divided by 100.
The purpose of this trick is the 
following:\textlist{\thetextlist~when the peak of the SED is well constrained by 
\MIPSiii, \SPIREi\ and \SPIREii\ (low temperatures), the \MIPSii\ flux has 
almost no impact on the $\chi^2$.
\thetextlist~\modif{when} the peak is not well constrained (high temperatures),
  the \MIPSii\ becomes important \modif{and temperature divergence is avoided.}}
The mass for each pixel is noted $M_\sms{dust}^\sms{1BB($\beta=2$)}$.

\begin{figure}[h!tbp]
  \includegraphics[width=0.95\linewidth]{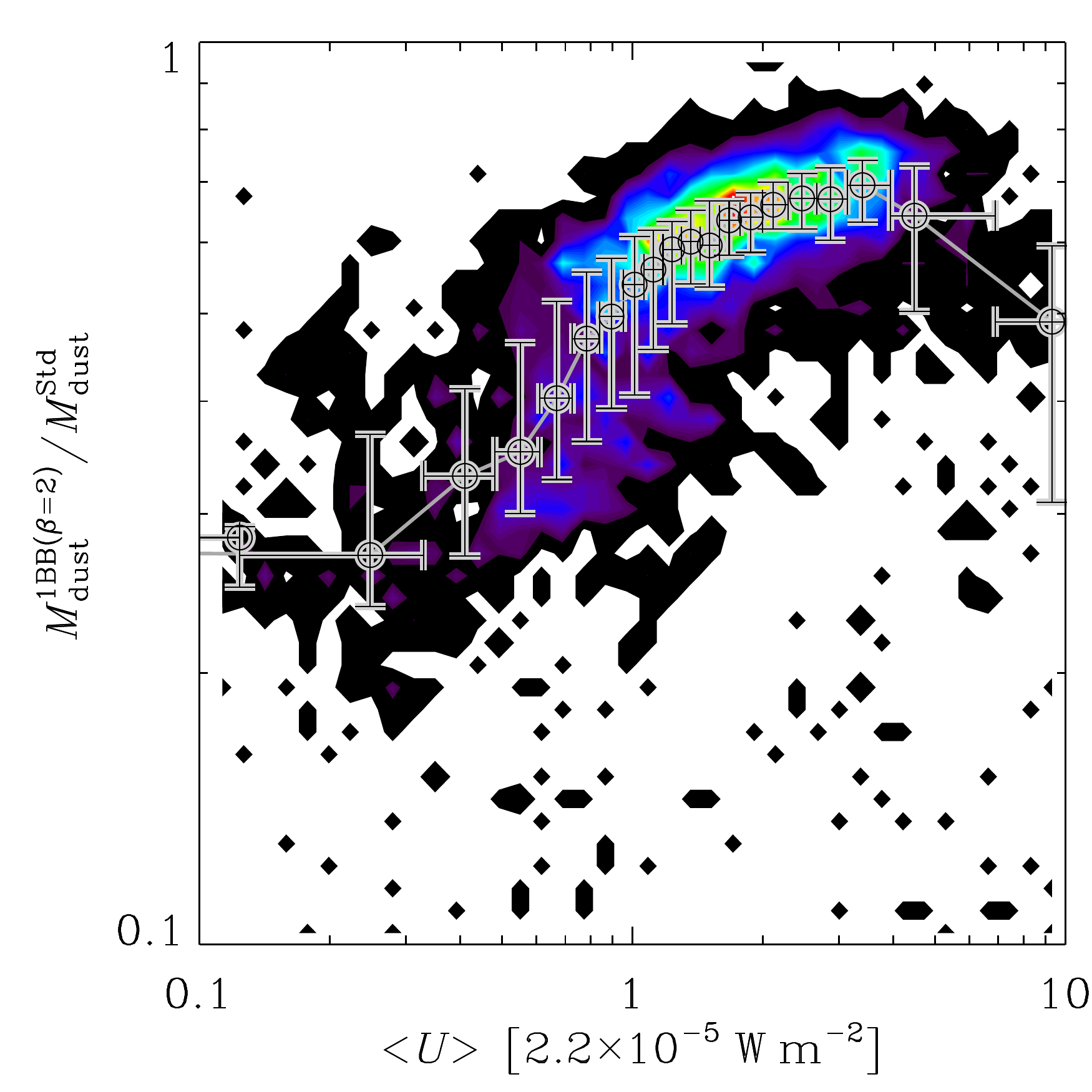}
  \caption{{\sl Comparison between our \citengl{standard model} and a single
            black body fit.}
            The spatial resolution is R4 (54~pc).
            The $x$ axis is the mass starlight averaged intensity 
            $\langle U\rangle$ of the \citengl{standard model}.
            The $y$ axis is the ratio between the dust derived with a single
            black body fit $\left(M_\sms{dust}^\sms{1BB($\beta=2$)}\right)$
            and with the \citengl{standard model} 
            $\left(M_\sms{dust}^\sms{Std}\right)$, for the same pixel.
            The color code of the pixel density and the binning of the trend 
            are similar to \reffig{fig:correldarkgas}.}
  \label{fig:comp_isoth}
\end{figure}
\reffig{fig:comp_isoth} shows the ratio between the dust masses obtained with this isothermal fit and with our \citengl{standard model}, as a function of the mass averaged starlight intensity $\langle U\rangle$.
It shows that for $\langle U\rangle\gtrsim1$, the mass ratio is roughly constant between $60$ and $80\,\%$, but for $\langle U\rangle\lesssim1$, the ratio drops down to less than $30\,\%$.
This simulation demonstrates that an isothermal fit will induce a bias in dust masses with average temperature.
The regions with high starlight intensities ($\langle U\rangle\gtrsim1$) are generally diffuse regions and their SEDs \modif{are} almost isothermal.
This explains why the mass ratio does not vary much with starlight intensity in
this range.
The average value of the ratio is around $70\,\%$ in these regions, which is reasonable considering the crudeness of the approximation.
On the contrary, regions with low starlight intensities ($\langle U\rangle\lesssim1$)
are generally dense regions. 
Since the mean free path of photons is much shorter in these regions, there is
a significant mix of cold and hot components within each pixel.
The isothermal approximation is not valid anymore in this range of 
$\langle U\rangle$, and the mass obtained with a single black body fit is 
biased.
Namely, it drops by a factor of $\simeq 2$ compared to its diffuse ISM value, down to $\simeq 30\,\%$ of the value of the \citengl{standard model}.
We obtain a similar trend comparing the \citengl{AC model} with a single black body fit having $\kappa_\sms{abs}(160\mic)=1.6\;\rm m^2\,kg^{-1}$, and 
$\beta=1.7$, except that the values of $\langle U\rangle$ are systematically shifted by a factor of $\simeq2$ (as demonstrated in \reffig{fig:comparison}).

Finally, we note that the increase of dust mass at low starlight intensities compared to the isothermal approximation is unlikely a bias induced by the submm excess extending down to the \SPIREii\ band.
Indeed, this excess is negligible in regions with low $\langle U\rangle$ 
(\refsec{sec:r500}).
It is prominent only in regions with high $\langle U\rangle$, where the trend of 
\reffig{fig:comp_isoth} is not significant.

\section{Inconsistency of the Very Cold Dust Hypothesis}
\label{ap:VCD}

As discussed by \citet{galliano03,galliano05} and \citet{galametz09,galametz10},
the \SPIREiii\ excess could be attributed to very cold dust (VCD; $T_\sms{eq}^\sms{VCD}\lesssim 10\; \rm K$). 
Simply considering the shape of the SED, this explanation is not unlikely.
Such a component would produce a change in the submm slope of the SED.
Indeed, warm and cold dust are heated by the transmitted stellar light, in different environments, at different optical depths.
Their temperature distribution is therefore continuous.
On the other hand, in order to reach \emph{very} cold temperatures ($T_\sms{dust}\lesssim10\;\rm K$), the dust has to be fully shielded from stellar radiation.
In these conditions, the dominant heating sources are the collisions and the IR radiation (galaxy and CMB) which do not depend on the optical depth.
\modif{It could reach temperatures lower than $T_\sms{dust}\simeq 5\;\rm K$ \citep{galliano03} only with difficulty.}
Very cold dust therefore corresponds to a change of regime in the dust heating.
If it exists, its temperature should not significantly vary with the optical depth in the cloud. 
Consequently its emission would be seen as a roughly isothermal component at 
$5\;{\rm K}\lesssim T_\sms{dust}\lesssim 10\;\rm K$.
With sufficient mass, it would produce an excess emission, with a change of slope, at wavelengths $300\mic\lesssim\lambda\lesssim 1\;\rm mm$.

Until now, we did not have the spatial resolution to explore this hypothesis.
In this section, we do an order of magnitude estimate to test the likeliness of the VCD model.
What follows is an update of the discussion at the end of \citet{galliano03}.
This is a complement to \refsec{sec:testr500}.

Let's first assume that the excess originates in equilibrium grains
having the UV-to-IR cross-section of silicates:
\begin{equation}
  \begin{array}{ll}
  \mbox{V band opacity:} & \kappa_\sms{abs}(\lambda_\sms{V})\simeq130\;{\rm m^2\,kg^{-1}} \\
  \mbox{\SPIREiii\ opacity:} & \kappa_\sms{abs}(\mbox{\SPIREiii})\simeq0.23\;{\rm m^2\,kg^{-1}}, \\ 
  \end{array}
\end{equation}
and a submillimeter opacity index $\beta^\sms{VCD}=1$, and an equilibrium temperature of $T_\sms{eq}^\sms{VCD}\simeq10$~K.
Those are the most optimistic values. 
If we can invalidate them, then the VCD hypothesis will be unrealistic for lower temperatures, steeper submm opacities, and larger UV cross-sections.

Although the relative excess is stronger in diffuse regions, we find an excess
in almost every pixel of the strip (\reffig{fig:im_delta}).
Let's assume that VCD lies in the core of very dense spherical clumps in the diffuse ISM.
Then, the ISRF they have to be shielded from is the typical ISRF of the diffuse ISM.
For the more realistic model (\citengl{AC}), this is basically $\langle U\rangle\gtrsim3$.
The optical depth, assuming a slab extinction, to shield this dust, and to allow it to reach $T_\sms{eq}^\sms{VCD}$ is $A_\sms{V}^\sms{VCD}\simeq2.9$.
Since these clumps have to be small, and that we see the excess everywhere, it
means that each clump is unresolved, even at R1.
Therefore, the diameter of these clumps has to be lower than the pixel size of R1, which is $l_\sms{pix}(R1)\simeq 10$~pc.

This maximum size translates into a minimum density of $n_\sms{H}^\sms{min}$, in order to reach the required optical depth of $A_\sms{V}^\sms{VCD}\simeq2.9$:
\begin{equation}
 n_\sms{H}^\sms{min}\gtrsim
 \frac{\displaystyle\frac{A_\sms{V}^\sms{VCD}}{1.086}\times 
       G_\sms{dust}^\sms{exp.}}
 {\displaystyle m_\sms{H}\times\kappa_\sms{abs}(\lambda_\sms{V})\times 
  \frac{l_\sms{pix}(R1)}{2}}
 \simeq2.5\E{4}\;\rm H\,cm^{-3},
\end{equation}
$m_\sms{H}$ being the mass of an H atom.
Notice that, since we are performing only an order of magnitude estimate, we
use the slab extinction for a sphere.

With $\langle U\rangle\simeq 3$, the temperature of the shielding dust, around the core, is $T_\sms{dust}^\sms{shield}\simeq 21\;\rm K$. 
Thus, a typical excess of $r_\sms{500}\simeq 15\,\%$ corresponds to a mass ratio, between the VCD core and the shielding ISM dust around, of:
\begin{equation}
  \frac{M_\sms{dust}^\sms{VCD}}{M_\sms{dust}^\sms{shield}}
  \simeq \frac{r_\sms{500}}{1-r_\sms{500}}
  \frac{l_\nu^\sms{VCD}(\mbox{\SPIREiii})}{l_\nu^\sms{shield}(\mbox{\SPIREiii})}
  \simeq 0.5,
\end{equation}
$l_\nu$ \modif{being} the specific monochromatic power of the components.
Therefore, for this picture to be correct, there should be at least
$\Sigma_\sms{dust}^\sms{shield}$ of shielding ISM dust surface density, such 
that:
\begin{equation}
  \Sigma_\sms{dust}^\sms{shield}
  \gtrsim \frac{4\pi}{3}\left(\frac{l_\sms{pix}(R1)}{2}\right)^3
  \frac{n_\sms{H}^\sms{min}m_\sms{H}}{G_\sms{dust}^\sms{exp.}}
  \frac{1}{l_\sms{pix}(R1)^2}
  \simeq10\msun\,\rm pc^{-2}
\end{equation}
This value is 2 orders of magnitude higher than the typical dust mass surface density, and one order of magnitude higher than the highest values
(\reffig{fig:correldarkgas}).
Moreover, this argument is conservative, since the highest excesses are found in lowest surface density regions (\refsec{sec:r500}).

In summary, we have shown that the \SPIREiii\ excess could not be accounted for
by very cold dust, since it would require a minimum mass of shielding dust,
within each pixel, too large compared to the observed surface density.
It is equivalent to note that, the excess being present in most pixels of R1, we would need at least $10^5$ clumps of cold dust.
It is more efficient to hide very cold dust in a small number of clumps.
For example, \citet{galliano03} estimated that this number could be less than a few hundreds in the Magellanic dwarf galaxy \ngc{1569}.

\begin{acknowledgements}
We would like to thank Anthony Jones, Vincent Guillet, Laurent Verstraete, Bruce Draine and Karl Gordon for stimulating discussions on the matter of this paper.
We thank the anonymous referee, for his report, and the editor, Malcolm Walmsley, for his careful reading of the paper.
We acknowledge extensive use of the Levenberg-Marquardt $\chi^2$ minimization routine written by Craig Markwardt.

SPIRE has been developed by a consortium of institutes led
by Cardiff Univ.\ (UK) and including Univ.\ Lethbridge (Canada);
NAOC (China); CEA, LAM (France); IFSI, Univ.\ Padua (Italy);
IAC (Spain); Stockholm Observatory (Sweden); Imperial College
London, RAL, UCL-MSSL, UKATC, Univ.\ Sussex (UK); Caltech,
JPL, NHSC, Univ.\ Colorado (USA). This development has been
supported by national funding agencies: CSA (Canada); NAOC
(China); CEA, CNES, CNRS (France); ASI (Italy); MCINN (Spain);
SNSB (Sweden); STFC, UKSA (UK); and NASA (USA).
\end{acknowledgements}

\bibliographystyle{/Users/fgallian/Astro/TeXstyle/bib_notes}
\bibliography{/Users/fgallian/Astro/TeXstyle/references}

\end{document}